\documentclass{jpp}
\usepackage{graphicx}

\usepackage[utf8]{inputenc}
\usepackage[T1]{fontenc}
\usepackage{amsmath}
\usepackage{color}
\usepackage{mathtools}
\usepackage{amssymb}
\usepackage{hyperref}
\usepackage{natbib}
\usepackage{subcaption}

\newcommand{\dd}[1]{\mathrm{d}#1 \,}

\newcommand{\pdev}[2]{\frac{\partial #1}{\partial #2}}
\newcommand{\pdevn}[3]{\frac{\partial^{#3} #1}{\partial #2^{#3}}}

\newcommand{\dev}[2]{\frac{\mathrm{d}#1}{\mathrm{d}#2}}
\renewcommand{\v}[1]{\boldsymbol{#1}}

\newcommand{\crl}[1]{\langle #1 \rangle}
\newcommand{\delgam}{\Delta\Gamma}

\newcommand{\iden}{\mathbb{I}}
\newcommand{\ee}{\mathrm{e}}
\newcommand{\ii}{\mathrm{i}}
\newcommand{\eci}{\mathrm{ei}}
\newcommand{\ice}{\mathrm{ie}}
\newcommand{\ISO}{\mathrm{I}}
\newcommand{\ANI}{\mathrm{A}}

\shorttitle{Collisionless relaxation of Lynden-Bell plasma}
\shortauthor{R. J. Ewart et al.}

\title{TUTORIAL\\ ~ \\ Collisionless relaxation of a Lynden-Bell plasma}

\author{R. J. Ewart\aff{1,2} \corresp{\email{robert.ewart@physics.ox.ac.uk}}, A. Brown\aff{3},
  T. Adkins\aff{1,4} \and A. A. Schekochihin\aff{1,4}}

\affiliation{\aff{1} Rudolf Peierls Centre for Theoretical Physics, University of Oxford, Oxford, OX1 3PU, UK
\aff{2}Balliol College, Oxford, OX1 3BJ
\aff{3}Exeter College, Oxford, OX1 3DP
\aff{4}Merton College, Oxford, OX1 4JD
}

\begin{document}

\maketitle

\begin{abstract}
Plasmas whose Coulomb-collision rates are very small may relax on shorter time scales to non-Maxwellian quasi-equilibria, which, nevertheless, have a universal form, with dependence on initial conditions retained only via an infinite set of Casimir invariants enforcing phase-volume conservation. These are distributions derived by \cite{LyndenBell67} via a statistical-mechanical entropy-maximisation procedure, assuming perfect mixing of phase-space elements. To show that these equilibria are reached dynamically, one must derive an effective `collisionless collision integral' for which they are fixed points---unique and inevitable provided the integral has an appropriate H-theorem. We describe how such collision integrals are derived and what assumptions are required for them to have a closed form, how to prove the H-theorems for them, and why, for a system carrying sufficiently large electric-fluctuation energy, collisionless relaxation should be fast. It is suggested that collisionless dynamics may favour maximising entropy locally in phase space before converging to global maximum-entropy states. Relaxation due to interspecies interaction is examined, leading, \textit{inter alia}, to spontaneous transient generation of electron currents. The formalism also allows efficient recovery of `true' collision integrals for both classical and quantum plasmas.
\end{abstract}
\section{Introduction}
\label{Section:Intro}

The problem of the relaxation of a gas to equilibrium is an old one. It was argued by \citet{Maxwell1860} that the distribution function of neutral particles undergoing elastic collisions would relax to what is now known as the Maxwellian. Maxwell's insight was driven by statistical assumptions about the neutral gas, but it was not until Boltzmann used the \textit{Stosszahlansatz} \citep{Boltzmann} to derive his collision integral that Maxwell's prediction was put on dynamical footing. Neutral particles, of course, greatly simplify the problem due to their short interaction length. When the interactions are long range, this substantially alters the system. In plasmas, the long-range Coulomb potentials are screened by the presence of positive and negative charges, meaning that `collisions' in the system are divided into encounters that are closer or more distant than the Debye length. At distances greater than the Debye length, particles experience a mean field of many particles. Within a few Debye lengths, the unshielded nature of the charge enables `true collisions' to occur. Owing to these true collisions, a homogeneous plasma, like a neutral gas, is doomed to a Maxwellian equilibrium \citep{Landau1936,Lenard60,Balescu60}. The frequency of these true collisions in real plasmas can, however, be very low, and interactions with the mean field may lead to considerably more circuitous routes towards stability. Instabilities may flare up, saturate, and decay, significantly altering the initial distribution function, usually rendering it stable long before the onset of collisional physics. If the rate of true collisions could be set artificially to zero, there would then be certain non-Maxwellian distributions that would not evolve. It is, therefore, natural to ask whether, if collisions are made arbitrarily weak, the system will preferentially evolve to one of these `quasi-stable' distributions on timescales much shorter than the collision timescale. 

\citet{LyndenBell67} first addressed an analogous question in the context of gravitationally interacting stellar systems, coining the term `violent relaxation' to describe this rapid collisionless process. Lynden-Bell argued that this relaxation would be approximately described by the collisionless Vlasov equation, which conserves phase-space volumes. He proposed that the system would reach a quasi-stable distribution function that maximised a certain entropy function subject to the conditions that the total particle number, momentum, energy, and phase-space volumes were conserved. This entropy-maximisation procedure led to statistics very similar to Fermi-Dirac statistics, a natural one for a system consisting of elementary objects (phase-space elements) that excluded the same phase volume. Soon after Lynden-Bell's paper, \citet{Kadomtsev_Pogutse70} derived a collision integral for the evolution of a plasma that evolved certain simplified intial conditions towards Fermi-Dirac-like distributions predicted by Lynden-Bell. Kadomtsev and Pogutse's calculation followed the ethos of the Balescu-Lenard collision integral, but deviated by assuming that the underlying exact distribution function was a piecewise constant function that was everywhere equal either to 0 or to a single constant, a so-called `waterbag distribution'. While taking a step towards verifying that Lynden-Bell's stable distributions were relevant entities, Kadomtsev and Pogutse's assumption of the distribution function only taking two possible values was a very restrictive one. 

In this paper, we will show how Kadomtsev and Pogutse's collision integral can be generalised to `multi-waterbag distributions', which capture all conceivable initial conditions and lead to the full range of Lynden-Bell's stable states, which are distributions that are monotonically decreasing functions solely of the particle energy (cf. \citealt{Gardner63}) that preserve the `waterbag content' of the initial conditions. Our results recover, or generalise, the previous extensions of the Kadomtsev-Pogutse result to multiple waterbags for gravitational and fluid-dynamical kinetic systems, due to \cite{Severne1980} and \cite{Chavanis2004,Chavanis2005}. They also allow one to recover very straightforwardly the Balescu-Lenard integral for collisional plasmas as well as Coulomb collision integrals for quantum plasmas consisting of fermions and bosons.

This paper is structured as follows. In section \ref{Section:2}, we outline a quasilinear derivation of a `collisionless collision integral', which mimics that of \citet{Kadomtsev_Pogutse70}. This will lead us to a general collision integral in terms of an unknown second-order correlator of the exact distribution function, for which a closure is required. In section~\ref{Section:2.5}, we make the first step towards this closure by assuming short correlation lengths in phase space (the `microgranulation ansatz', the collisionless version of Boltzmann's \textit{Stosszahlansatz}). In section~\ref{Section:3}, we will make the `waterbag closure' that will lead to a closed collision integral. To do this, we will first discuss, in section~\ref{Section:3.1}, the `single-waterbag' closure used by Kadomtsev and Pogutse, and the difficulty of generalising it to multiple waterbags; then, in section~\ref{Section:3.2}, we will show how one can use one's incomplete knowledge of the system to find a `multi-waterbag' closure (cf. \citealt{Chavanis2005}). Having derived a multi-waterbag collision integral, we will show in section~\ref{Section:3.3.1} that its fixed points are the equilibria predicted by Lynden-Bell and briefly review the thermodynamic properties of these fixed points. In section \ref{Section:3.3.2}, we will show that these fixed points are stable by proving that the multi-waterbag collision integral has an H-theorem, viz., that the system will increase Lynden-Bell's entropy. In section~\ref{Section:4}, we will show that the resolution of the closure problem in section~\ref{Section:3.2} is actually unnecessary if one is willing to consider the kinetics of a different set of objects: not the distribution function of particles, but the distribution function of waterbags. This `hyperkinetics' treats the waterbags as the fundamental objects and describes the evolution of a `waterbag distribution function' in a seven-dimensional phase space, as opposed to the usual six-dimensional phase space of conventional kinetics. In section \ref{Section:4.1}, we derive the `hyperkinetic collision integral' (cf. \citealt{Severne1980}), which is pleasingly similar in form to the previous collision integral but has no need for its thermodynamically motivated closure. In section~\ref{Section:4.2}, we prove an H-theorem for the hyperkinetic collision integral, whence it follows that the latter also pushes the distribution function towards the Lynden-Bell equilibria. In section~\ref{Section:5}, we relate the two collision integrals by proving their equivalence over a broad range of initial conditions. Their equivalence however, is not guaranteed for all initial conditions, as the thermodynamically motivated closure can sometimes be too restrictive. In section~\ref{Section:6}, we compare the `effective collisions' described by the hyperkinetic collision integral to the `true' collisions described by the Balescu-Lenard collision integral. Namely, we show that, should the microgranulation ansatz hold, the effective collision rate and the energy stored in the fluctuating electric fields will be generically much larger than for true collisions, which gives a method by which the microgranulation ansatz can be tested. A number of caveats with regard to the existence of Lynden-Bell plasmas are discussed in section~\ref{Section:5.4}. Inter-species interactions in Lynden-Bell plasmas are studied in section~\ref{Section:6}: while isotropisation of the electrons and the relaxation of the temperatures of the Lynden-Bell equilibria are a relatively straightforward generalisation of what happens due to standard collisional relaxation (sections~\ref{Section:Zeroth Order} and~\ref{Section:ion-electron}), the relaxation of the distributions' mean momentum turn out to contain a somewhat surprising effect of spontaneous generation of electron current (section~\ref{Section:First order}). In section~\ref{Section:7}, we summarise the narrative presented in this paper and discuss its limitations, uncertainties, and further steps.

\section{Derivation of collision integrals from quasilinear theory}
\label{Section:2}
The calculation contained in this section is fundamentally a textbook one, although in practice it may be difficult to find a textbook where it is presented in quite this form, which we consider to be the most transparent.

To understand the relaxation of a distribution function to equillibrium, we begin by considering the evolution of a plasma consisting of multiple species of particles, indexed by $\alpha$, with mass $m_{\alpha}$ and charge $q_{\alpha}$. The distribution function $f_{\alpha}$ of these particles evolves according to the collisionless Vlasov equation
\begin{equation}
\label{eqn:S1:E1}
\pdev{f_{\alpha}}{t} + \v{v}\cdot\v{\nabla} f_{\alpha} - \frac{q_{\alpha}}{m_{\alpha}}(\v{\nabla} \varphi)\cdot \pdev{f_{\alpha}}{\v{v}} = 0,
\end{equation}
where the potential $\varphi$ is determined by Poisson's equation
\begin{equation}
\label{eqn:S1:E2}
-\nabla^{2}\varphi = 4\pi\sum_{\alpha}q_{\alpha}\int \dd{\v{v}}f_{\alpha}.
\end{equation}
For simplicity, we shall limit ourselves to the consideration of the electrostatic case only, forbidding the plasma to host any magnetic field.

In principle, (\ref{eqn:S1:E1}) and (\ref{eqn:S1:E2}) already contain the information necessary to evolve the distribution towards equilibrium. Of course, Michelangelo's David was wholly contained within a block of marble, which did not, however, provide great insight into what could lie beneath \citep{coonin2014marble}. The aim of this calculation is then to discern what information can be cut away from from (\ref{eqn:S1:E1}) and (\ref{eqn:S1:E2}) to leave only that which is necessary to understand the relaxation of the mean distribution. If we wished to answer the question of collisionless relaxation with complete generality, then the most likely answer is that no information can be cut away. We will therefore specialise to  the following simplified, but important, case: a system that is on average uniform in space and for which deviations from homogeneity occur only as small perturbations. In such a regime, it is natural to write~$f_{\alpha}$ as a sum of Fourier modes
\begin{equation}
\label{eqn:S1:E3}
f_{\alpha}(\v{r},\v{v}) = \sum_{\v{k}}f_{\v{k}\alpha}(\v{v})e^{i\v{k}\cdot\v{r}}.
\end{equation}
The evolution of the mean part of the $\v{k} = 0$ mode of the distribution function,~$f_{0\alpha} = ~\crl{f_{\v{k}=0,\alpha}}$, is then
\begin{equation}
\label{eqn:S1:E4}
\pdev{f_{0\alpha}}{t} = \pdev{}{\v{v}}\cdot\left[\frac{q_{\alpha}}{m_{\alpha}}\sum_{\v{k}}\v{k}\mathrm{Im}\crl{\varphi^{*}_{\v{k}}f_{\v{k}\alpha}} \right].
\end{equation}
Here the averages can be rationalised by having many copies of the system, which only differ from one another in microscopic detail. After these copies are allowed to evolve forward to reach the present time, they will each have different values of $f_{\v{k}\alpha}$ owing to the initial differences. The angle brackets therefore represent ensemble averages of the system, and the restriction of statistical homogeneity is that the average of any $f_{\v{k}\alpha}$ is zero.

From (\ref{eqn:S1:E4}), we see that to work out the evolution of $f_{0\alpha}$, we must know the remaining fluctuating part of the distribution~$f_{\v{k}\alpha}$. Therefore, we consider the linearised Vlasov equation for $f_{\v{k}\alpha}$:
\begin{equation}
\label{eqn:S1:E5}
\pdev{f_{\v{k}\alpha}}{t} + i\v{k}\cdot\v{v}f_{\v{k}\alpha} = i\frac{q_{\alpha}}{m_{\alpha}}\varphi_{\v{k}}\v{k}\cdot\pdev{f_{0\alpha}}{\v{v}},
\end{equation} 
where Poisson's equation (\ref{eqn:S1:E2}) becomes
\begin{equation}
\label{eqn:S1:E6}
\varphi_{\v{k}} = \sum_{\alpha}\frac{4\pi q_{\alpha}}{k^{2}}\int\dd{\v{v}}f_{\v{k}\alpha}.
\end{equation}
Note that in a homogeneous system, there can be no mean electric field. Provided that the fluctuations have much smaller amplitudes than $f_{0\alpha}$, (\ref{eqn:S1:E4}) and (\ref{eqn:S1:E5}) imply that $f_{\v{k}\alpha}$ evolves much faster than the mean distribution. Therefore, the programme for deriving the evolution of $f_{0\alpha}$ becomes the linear one (e.g., \citealt{Kadomtsev65}): find the evolution of the fluctuating $f_{\v{k}\alpha}$ subject to a constant mean distribution function $f_{0\alpha}$, then evolve~$f_{0\alpha}$ using this $f_{\v{k}\alpha},$ with the knowledge that, as $f_{0\alpha}$ varies, the fluctuations will constantly adjust themselves. 
\subsection{Linear theory for $f_{\v{k}\alpha}$}
\label{Section:2.1}
To solve for the evolution of $f_{\v{k}\alpha}$ from (\ref{eqn:S1:E5}) and (\ref{eqn:S1:E6}), we follow \citet{Landau1936} and introduce the Laplace transform
\begin{equation}
\label{eqn:S1:E7}
\hat{\varphi}(p) = \int_{0}^{\infty}e^{-pt}\varphi(t)\dd{t},
\end{equation}
and similarly for $f_{\v{k}\alpha}(t)$. We now take the Laplace transform of (\ref{eqn:S1:E5}) and (\ref{eqn:S1:E6}) to get
\begin{equation}
\label{eqn:S1:E8}
\hat{f}_{\v{k}\alpha}(p) = i\frac{q_{\alpha}}{m_{\alpha}}\frac{\hat{\varphi}_{\v{k}}(p)}{p+i\v{k}\cdot\v{v}}\v{k}\cdot\pdev{f_{0\alpha}}{\v{v}} + \hat{h}_{\v{k}\alpha}(p),
\end{equation}
\begin{equation}
\label{eqn:S1:E9}
\hat{\varphi}_{\v{k}}(p) = \sum_{\alpha'}\frac{4\pi q_{\alpha'}}{k^{2}\epsilon_{\v{k}}(p)}\int\dd{\v{v}'} \hat{h}_{\v{k}\alpha'}(p),
\end{equation}
where the dielectric function has emerged, defined by
\begin{equation}
\label{eqn:S1:E10}
\epsilon_{\v{k}}(p) = 1 - i\sum_{\alpha'}\frac{4\pi q_{\alpha'}^{2}}{m_{\alpha'}k^{2}}\int\dd{\v{v}'}\frac{1}{p+i\v{k}\cdot\v{v}'}\v{k}\cdot\pdev{f_{0\alpha'}}{\v{v}'},
\end{equation}
while information about the initial condition enters via 
\begin{equation}
\label{eqn:S1:E11}
\hat{h}_{\v{k}\alpha}(p) = \frac{g_{\v{k}\alpha}(\v{v})}{p+i\v{k}\cdot\v{v}}.
\end{equation}
where $g_{\v{k}\alpha}(\v{v}) = f_{\v{k}\alpha}(t=0,\v{v})$.

The time-dependent solution is given by the inverse Laplace transforms:
\begin{equation}
\label{eqn:S1:E12}
\varphi_{\v{k}}(t) = \frac{1}{2\pi i }\int_{-i\infty + \sigma}^{i\infty + \sigma}\dd{p}e^{pt}\hat{\varphi}_{\v{k}}(p),
\end{equation}
\begin{equation}
\label{eqn:S1:E13}
f_{\v{k}\alpha}(t) = \int_{-i\infty + \sigma}^{i\infty + \sigma}\frac{\dd{p}}{2\pi i }e^{pt}\left[i\frac{q_{\alpha}}{m_{\alpha}}\frac{\hat{\varphi}_{\v{k}}(p)}{p+i\v{k}\cdot\v{v}}\v{k}\cdot\pdev{f_{0\alpha}}{\v{v}} + \hat{h}_{\v{k}\alpha}(p) \right].
\end{equation}
where $\sigma$ must be chosen so that for all $p$ with $\mathrm{Re}(p)> \sigma$ the integrands are analytic functions.
\subsection{Quasilinear evolution of $f_{0\alpha}$}
\label{Section:2.2}
Having computed $\varphi(t)$ and $f_{\v{k}\alpha}(t)$ we are now in a position to determine the evolution of $f_{0\alpha}$. Substituting (\ref{eqn:S1:E12}) and (\ref{eqn:S1:E13}) into (\ref{eqn:S1:E4}) we obtain the earliest form of our collision integral for the evolution of $f_{0\alpha}$:
\begin{multline}
\label{eqn:S1:E14}
\pdev{f_{0\alpha}}{t} = -\pdev{}{\v{v}}\cdot \frac{q_{\alpha}}{m_{\alpha}}\sum_{\v{k}}\v{k}\mathrm{Im}\iint\frac{\dd{p}\dd{p'}}{(2\pi)^{2}}e^{(p+p')t}\Big[i\frac{q_{\alpha}}{m_{\alpha}}\frac{\crl{\hat{\varphi}_{\v{k}}(p)\hat{\varphi}^{*}_{\v{k}}(p'^{*})}}{p + i\v{k}\cdot\v{v}}\v{k}\cdot\pdev{f_{0\alpha}}{\v{v}}\\ + \crl{\hat{h}_{\v{k}\alpha}(p,\v{v})\hat{\varphi}^{*}_{\v{k}}(p'^{*})} \Big].
\end{multline}
Note that, to avoid confusion in (\ref{eqn:S1:E14}), both inverse Laplace transforms are taken along the contour running from $-i\infty +\sigma$ to $i\infty + \sigma$. Since the $p'$ contour in (\ref{eqn:S1:E14}) comes from a complex conjugation, this results in an overall minus sign appearing. 

Presently, (\ref{eqn:S1:E14}) is still in need of information, as it still depends on the averages of $\hat{\varphi}_{\v{k}}$ and $\hat{h}_{\v{k}}$. The eventual aim is to `close' this collision integral: to replace these correlators with expressions involving only the mean distribution $f_{0\alpha}$. This will give a differential equation for each $f_{0\alpha}$ in terms of other $f_{0\alpha'}$ only, which, in principle, can then be solved. Such a closure will come from some model of the averages that we have yet to compute, and will be enabled by assumptions introduced in sections \ref{Section:2.5} and \ref{Section:3}. 

First we will manipulate (\ref{eqn:S1:E14}) into a more agreeable form by rewriting the averages~$\crl{\hat{\varphi}_{\v{k}}(p)\hat{\varphi}^{*}_{\v{k}}(p'^{*})}$ and~$\crl{\hat{h}_{\v{k}\alpha}(p,\v{v})\hat{\varphi}^{*}_{\v{k}}(p'^{*})}$ as correlators solely of $\hat{h}_{\v{k}\alpha}$. For the first term in ~(\ref{eqn:S1:E14}), we get, using ~(\ref{eqn:S1:E9}),
\begin{multline}
\label{eqn:S1:E15}
i\frac{q_{\alpha}^{2}}{m_{\alpha}^{2}}\frac{\crl{\varphi_{\v{k}}(p)\varphi^{*}_{\v{k}}(p'^{*})}}{p+ i\v{k}\cdot\v{v}}\v{k}\cdot\pdev{f_{0\alpha}}{\v{v}} = \\
i\frac{q_{\alpha}^{2}}{m_{\alpha}^{2}}\sum_{\alpha'\alpha''}\frac{16\pi^{2}q_{\alpha'}q_{\alpha''}}{k^{4}\epsilon_{\v{k}}(p)\epsilon^{*}_{\v{k}}(p'^{*})} \iint \dd{\v{v}'}\dd{\v{v}''}\frac{\crl{\hat{h}_{\v{k}\alpha''}(p,\v{v}'')\hat{h}^{*}_{\v{k}\alpha'}(p'^{*},\v{v}')}}{p+i\v{k}\cdot\v{v}}\v{k}\cdot\pdev{f_{0\alpha}}{\v{v}}.
\end{multline}
The second term in (\ref{eqn:S1:E14}), again via~(\ref{eqn:S1:E9}), becomes
\begin{equation}
\label{eqn:S1:E16}
\begin{split}
\frac{q_{\alpha}}{m_{\alpha}}\crl{\hat{h}_{\v{k}\alpha}(p)\varphi_{\v{k}}^{*}(p'^{*})} &= \sum_{\alpha'}\frac{4\pi q_{\alpha}q_{\alpha'}}{m_{\alpha}k^{2}\epsilon_{\v{k}}^{*}(p'^{*})}\int\dd{\v{v}}\crl{\hat{h}_{\v{k}\alpha}(p,\v{v})\hat{h}^{*}_{\v{k}\alpha'}(p'^{*},\v{v}')} \\ &= \sum_{\alpha'}\frac{4\pi q_{\alpha}q_{\alpha'}}{m_{\alpha}k^{2}\epsilon_{\v{k}}(p)\epsilon^{*}_{\v{k}}(p'^{*})}\int\dd{\v{v}'}\Bigg[ \crl{\hat{h}_{\v{k}\alpha}(p,\v{v})\hat{h}^{*}_{\v{k}\alpha'}(p'^{*},\v{v}')} \\ &- i \sum_{\alpha''}\frac{4\pi q_{\alpha''}^{2}}{m_{\alpha''}k^{2}}\int\dd{\v{v}''}\frac{\crl{\hat{h}_{\v{k}\alpha}(p,\v{v})\hat{h}^{*}_{\v{k}\alpha'}(p'^{*},\v{v}')}}{p + i\v{k}\cdot\v{v}''}\v{k}\cdot\pdev{f_{0\alpha''}}{\v{v}''}\Bigg].
\end{split}
\end{equation}
At the last step, seemingly gratuitously, we multiplied and divided by the dielectric function (\ref{eqn:S1:E10}). This will prove to be a useful tactic, as it separates this correlator in two parts: the second, which bears strong resemblance to (\ref{eqn:S1:E15}), and the first, which will later vanish under certain assumptions. Since this first term is destined to vanish, we will continue as though it has already done so and confirm its disappearance at the end of section~\ref{Section:2.5}. The full evolution equation for the mean distribution function of species $\alpha$ is then 
\begin{multline}
\label{eqn:S1:E17}
\pdev{f_{0\alpha}}{t} = -\frac{q_{\alpha}}{m_{\alpha}}\pdev{}{\v{v}}\cdot\sum_{\v{k}}\frac{\v{k}\v{k}}{k^{4}}\cdot\mathrm{Re}\iint\frac{\dd{p}\dd{p'}}{(2\pi)^{2}}e^{(p+p')t}\sum_{\alpha'\alpha''}\frac{16\pi^{2}q_{\alpha'}q_{\alpha''}}{\epsilon_{\v{k}}(p)\epsilon^{*}_{\v{k}}(p'^{*})} \iint \dd{\v{v}'}\dd{\v{v}''}\\ \Bigg[\frac{\crl{\hat{h}_{\v{k}\alpha''}(p,\v{v}'')\hat{h}^{*}_{\v{k}\alpha'}(p'^{*},\v{v}')}}{p+i\v{k}\cdot\v{v}} \frac{q_{\alpha}}{m_{\alpha}}\pdev{f_{0\alpha}}{\v{v}}  - \frac{\crl{\hat{h}_{\v{k}\alpha}(p,\v{v})\hat{h}^{*}_{\v{k}\alpha'}(p'^{*},\v{v}')}}{p+i\v{k}\cdot\v{v}''} \frac{q_{\alpha''}}{m_{\alpha''}}\pdev{f_{0\alpha''}}{\v{v}''}  \Bigg].
\end{multline}
This can be written more compactly as  
\begin{equation}
\label{eqn:S1:E18}
\pdev{f_{0\alpha}}{t} = \pdev{}{\v{v}}\cdot\sum_{\alpha''}\int\dd{\v{v}''}\left[\mathsf{D}^{\alpha}_{\alpha\alpha''}(\v{v},\v{v}'')\cdot\pdev{f_{0\alpha}}{\v{v}} - \mathsf{D}^{\alpha}_{\alpha''\alpha}(\v{v}'',\v{v})\cdot\pdev{f_{0\alpha''}}{\v{v}''}\right],
\end{equation}
where the `diffusion kernel' is
\begin{multline}
\label{eqn:S1:E19}
 \mathsf{D}^{\alpha}_{\mu \nu}(\v{w},\v{v}) =\\ -\sum_{\nu'}\frac{16\pi^{2}q_{\mu}^{2}q_{\nu}q_{\nu'}}{m_{\alpha}m_{\mu}}\mathrm{Re}\sum_{\v{k}}\frac{\v{k}\v{k}}{k^{4}}\iint\frac{\dd{p}\dd{p'}}{(2\pi)^{2}}\frac{e^{(p + p')t}}{\epsilon_{\v{k}}(p)\epsilon^{*}_{\v{k}}(p'^{*})}\int\dd{\v{v}'}\frac{\crl{\hat{h}_{\v{k}\nu}(p,\v{v})\hat{h}^{*}_{\v{k}\nu'}(p'^{*},\v{v}')}}{p + i\v{k}\cdot\v{w}}.
\end{multline}
\subsection{Simplification of the diffusion kernel}
\label{Section:2.3}
\begin{figure}
\centering
\includegraphics[width=0.75\textwidth]{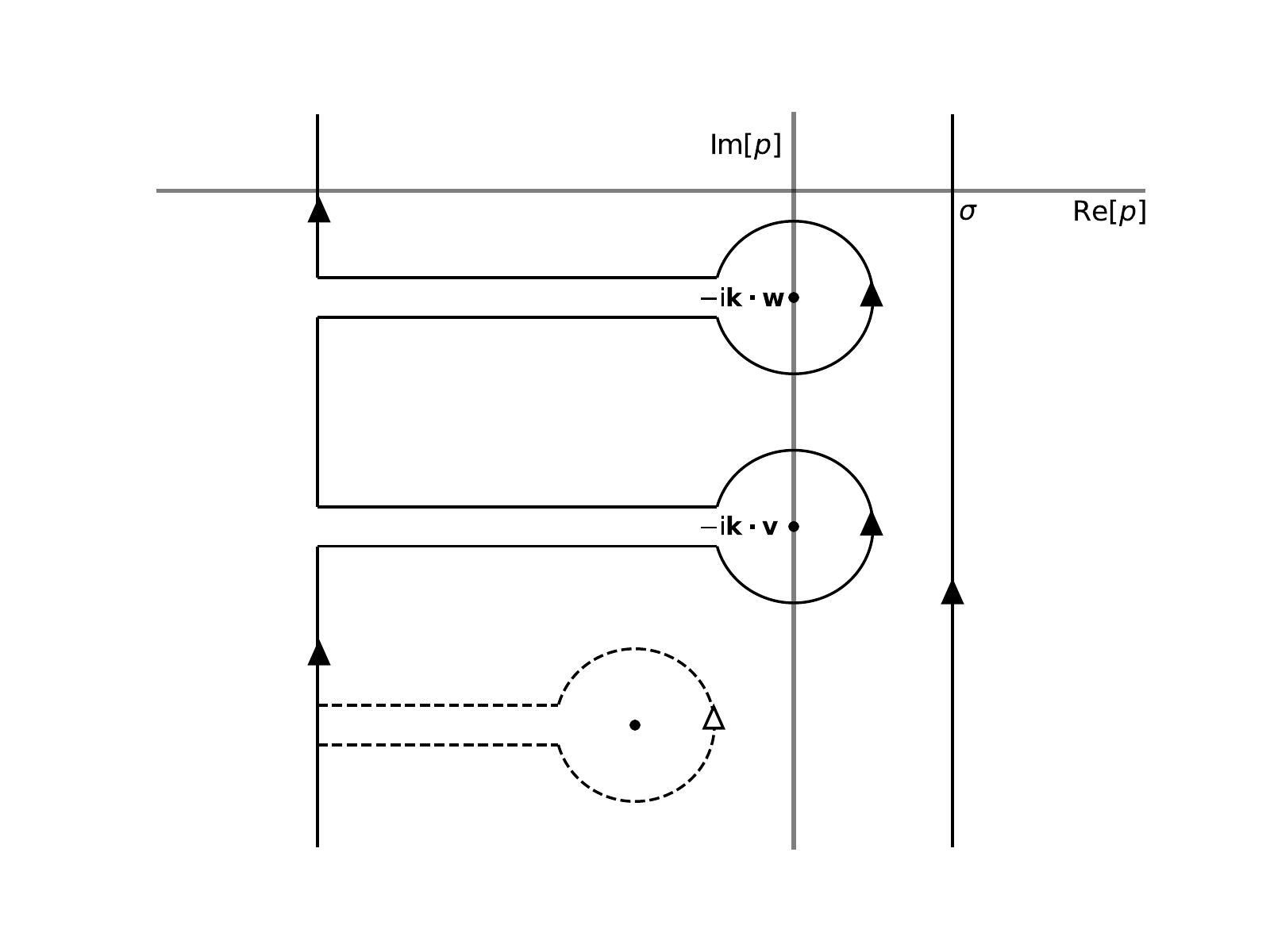}
\caption{The contours used for the $p$ integration in (\ref{eqn:S1:E21}). The right-hand contour is the original contour, which is calculated by deforming it around the poles of (\ref{eqn:S1:E22}) onto the left-hand contour. Poles of the dielectric function would correspond to decaying modes but are ignored (shown in the dotted line).}
\label{Figure 1}
\end{figure}

Even before we make any assumptions about the nature of the initial condition to decompose the averages, it is possible to simplify the diffusion kernel by appealing to the separation of timescales between the mean and the fluctuations. To do so, we will carry out the $p$ and $p'$ integrals. First we rewrite (\ref{eqn:S1:E19}) as follows:
\begin{equation}
\label{eqn:S1:E20}
\mathsf{D}^{\alpha}_{\mu\nu}(\v{w},\v{v}) = \sum_{\nu'}\frac{16\pi^{2}q_{\mu}^{2}q_{\nu}q_{\nu'}}{m_{\alpha}m_{\mu}}\mathrm{Re}\sum_{\v{k}}\frac{\v{k}\v{k}}{k^{4}}\int \dd{\v{v}}'\crl{g_{\v{k}\nu}(\v{v})g^{*}_{\v{k}\nu'}(\v{v}')}I_{\v{k}}(\v{v},\v{v}',\v{w}),
\end{equation}
where
\begin{equation}
\label{eqn:S1:E21}
I_{\v{k}}(\v{v},\v{v}',\v{w}) = -\iint\frac{\dd{p}\dd{p'}}{(2\pi)^{2}}\frac{e^{(p+p')t}}{\epsilon_{\v{k}}(p)\epsilon^{*}_{\v{k}}(p'^{*})}\frac{1}{(p+ i\v{k}\cdot\v{v})(p+i\v{k}\cdot\v{w})(p'-i\v{k}\cdot\v{v}')}.  
\end{equation}
Since this is a contour integral of a holomorphic function [note that $\epsilon_{\v{k}}^{*}(p'^{*})$ is a holomorphic function of $p'$], these integrals can be computed by deforming the $p$ and $p'$ contours far into the left-hand plane, where the real part of $e^{pt}$ will suppress the integral. All that will remain from this is the contribution from the poles of the integral. Generally, as well as the `ballistic poles' on the imaginary line, the dielectric function can have poles in the left and right halves of the complex plane. Poles of the dielectric function in the left half of the complex plane, however, correspond to decaying modes, which we will neglect (see figure~\ref{Figure 1}). Poles of the dielectric function in the right half of the complex plane correspond to linear instabilities of the distribution function and in principle cannot be neglected. However, we may restrict ourselves to linearly stable distribution functions with the proviso that we take our initial condition to be the distribution function after all instabilities have vanished and that the initial evolution with instabilities growing and saturating must be treated by a different theory. Within these assumptions the result of the contour integration is
\begin{equation}
\label{eqn:S1:E22}
I_{\v{k}}(\v{v},\v{v}',\v{w}) = \frac{i}{\epsilon_{\v{k}}(-i\v{k}\cdot\v{v})\epsilon^{*}_{\v{k}}(-i\v{k}\cdot\v{v}')}\frac{e^{-i\v{k}\cdot(\v{v}-\v{v}')t}}{\v{k}\cdot(\v{v}-\v{w})}\left[1 - \frac{\epsilon_{\v{k}}(-i\v{k}\cdot\v{v})}{\epsilon_{\v{k}}(-i\v{k}\cdot\v{w})}e^{i\v{k}\cdot(\v{v}-\v{w})t} \right].
\end{equation} 
We have previously assumed a separation of timescales between the mean distribution function and the perturbed distribution function (known generally as the Bogoliubov ansatz: see \citealt{SwansonKT}). Therefore, the perturbation can be allowed to evolve to large~$t$ before we consider its effect on the mean distribution function. Accordingly, we take the limit $t\to \infty$ in~(\ref{eqn:S1:E21}):
\begin{equation}
\label{eqn:S1:E23}
I_{\v{k}}(\v{v},\v{v}',\v{w}) = \pi\frac{e^{-i\v{k}\cdot(\v{v}-\v{v}')t}}{\epsilon_{\v{k},\v{k}\cdot\v{v}}\epsilon^{*}_{\v{k},\v{k}\cdot\v{v}'}}\delta\left(\v{k}\cdot(\v{v}-\v{w}) \right),
\end{equation}
where $\epsilon_{\v{k},\v{k}\cdot\v{v}} \equiv \epsilon_{\v{k}}(-i\v{k}\cdot\v{v})$. Substituting (\ref{eqn:S1:E23}) back into (\ref{eqn:S1:E20}) gives a simplified expression for the diffusion kernel:
\begin{multline}
\label{eqn:S1:E24}
\mathsf{D}^{\alpha}_{\mu\nu}(\v{w},\v{v}) =\\ \sum_{\nu'}\frac{16\pi^{3}q_{\mu}^{2}q_{\nu}q_{\nu'}}{m_{\alpha}m_{\mu}}\mathrm{Re}\sum_{\v{k}}\frac{\v{k}\v{k}}{k^{4}}\delta(\v{k}\cdot(\v{w}-\v{v}))\int\dd{\v{v}'}\frac{e^{-i\v{k}\cdot(\v{v}- \v{v}')t}\crl{g_{\v{k}\nu}(\v{v})g^{*}_{\v{k}\nu'}(\v{v}')}}{\epsilon_{\v{k},\v{k}\cdot\v{v}}\epsilon_{\v{k},\v{k}\cdot\v{v}'}^{*}}.
\end{multline}
Note that it is in obtaining this expression that the quasilinear approximation has first been truly used. In the more general expression~(\ref{eqn:S1:E19}), $\hat{h}_{\v{k}\nu}(p,\v{v})$, instead of being given by~(\ref{eqn:S1:E11}), can always be assumed to contain the nonlinear contributions neglected in~(\ref{eqn:S1:E5}).
\subsection{Conservation laws of the general quasilinear collision integral}
\label{Section:2.4}
Due to the symmetries that are possessed by (\ref{eqn:S1:E24}), it is now possible to show that the collision integral (\ref{eqn:S1:E18}) with the diffusion kernel (\ref{eqn:S1:E24}) conserves the particle number, total mean momentum and total mean energy between the species. 

The number of particles of species $\alpha$
\begin{equation}
\label{eqn:S1:E25}
N_{\alpha} = V\int\dd{\v{v}}f_{0\alpha},
\end{equation} 
is trivially conserved because the right-hand side of (\ref{eqn:S1:E18}) is a full derivative with respect to $\v{v}$. 

The total momentum of the system is given by
\begin{equation}
\label{eqn:S1:E26}
P = V\sum_{\alpha}\int\dd{\v{v}}m_{\alpha}\v{v}f_{0\alpha}.
\end{equation}
Taking the time derivative of (\ref{eqn:S1:E26}) using (\ref{eqn:S1:E18}), we get
\begin{equation}
\label{eqn:S1:E27}
\begin{split}
\dev{P}{t}  &= V\sum_{\alpha\alpha''}\iint\dd{\v{v}}\dd{\v{v}''}m_{\alpha}\v{v} \pdev{}{\v{v}}\cdot\left[\mathsf{D}^{\alpha}_{\alpha\alpha''}(\v{v},\v{v}'')\cdot\pdev{f_{0\alpha}}{\v{v}}- \mathsf{D}^{\alpha}_{\alpha''\alpha}(\v{v}'',\v{v})\cdot\pdev{f_{0\alpha''}}{\v{v}''} \right] \\
& = -V\sum_{\alpha\alpha''}\iint\dd{\v{v}}\dd{\v{v}''}\left[m_{\alpha}\mathsf{D}^{\alpha}_{\alpha\alpha''}(\v{v},\v{v}'')- m_{\alpha''}\mathsf{D}^{\alpha''}_{\alpha\alpha''}(\v{v},\v{v}'') \right]\cdot\pdev{f_{0\alpha}}{\v{v}}  \\& = 0.
\end{split}
\end{equation}
In the second equality, we integrated by parts and swapped the $\alpha$, $\alpha''$ indices, as well as the arguments $\v{v}$ and $\v{v}''$ in the second term of the integral. This expresses the condition for momentum conservation as the symmetry 
\begin{equation}
\label{eqn:S1:E28}
m_{\alpha}\mathsf{D}^{\alpha}_{\alpha\alpha''}(\v{v},\v{v}'')= m_{\alpha''}\mathsf{D}^{\alpha''}_{\alpha\alpha''}(\v{v},\v{v}'').
\end{equation}
From (\ref{eqn:S1:E24}), and indeed already from (\ref{eqn:S1:E20}), we see that this is manifestly satisfied, hence the total momentum is conserved but particles of different species may exchange momentum. 

We apply much the same procedure to the total energy 
\begin{equation}
\label{eqn:S1:E29}
E = V\sum_{\alpha}\int \dd{\v{v}}\frac{1}{2}m_{\alpha}|\v{v}|^{2}f_{0\alpha}(\v{v}).
\end{equation}
Taking the time derivative of (\ref{eqn:S1:E29}), we get
\begin{equation}
\label{eqn:S1:E30}
\begin{split}
\dev{E}{t} &= V\sum_{\alpha\alpha''}\iint\dd{\v{v}}\dd{\v{v}''}\frac{1}{2}m_{\alpha}|\v{v}|^{2}\pdev{}{\v{v}}\cdot\left[\mathsf{D}^{\alpha}_{\alpha\alpha''}(\v{v},\v{v}'')\cdot\pdev{f_{0\alpha}}{\v{v}}- \mathsf{D}^{\alpha}_{\alpha''\alpha}(\v{v}'',\v{v})\cdot\pdev{f_{0\alpha''}}{\v{v}''} \right] \\
& =  -V\sum_{\alpha\alpha''}\iint\dd{\v{v}}\dd{\v{v}''}\left[m_{\alpha}\v{v}\cdot \mathsf{D}^{\alpha}_{\alpha\alpha''}(\v{v},\v{v}'')- m_{\alpha''}\v{v}''\cdot \mathsf{D}^{\alpha''}_{\alpha\alpha''}(\v{v},\v{v}'') \right]\cdot\pdev{f_{0\alpha}}{\v{v}} \\ & = 0.
\end{split}
\end{equation}
Again the swapping of the indices and velocities allows the condition of total-energy conservation to be cast as the symmetry 
\begin{equation}
\label{eqn:S1:E31}
m_{\alpha}\v{v}\cdot \mathsf{D}^{\alpha}_{\alpha\alpha''}(\v{v},\v{v}'') = m_{\alpha''}\v{v}''\cdot \mathsf{D}^{\alpha''}_{\alpha\alpha''}(\v{v},\v{v}''),
\end{equation}
which is satisfied by (\ref{eqn:S1:E24}) due to the delta function that emerged from the approximation relating to the separation of timescales. Essentially what this means is that when the mean distribution function is considered to be linearly stable and to change slowly compared to the fluctuations, the energy $E$ of this distribution function cannot change.

While it is encouraging that we have a collision integral in a general form into which conservation laws are hard-wired, without a general form of the correlator $\crl{g_{\alpha\v{k}}(\v{v})g^{*}_{\nu\v{k}}(\v{v}')}$ it is still not possible to determine the evolution of $f_{0\alpha}$. We will now introduce the approximations necessary to arrive at collision integrals in a closed form. 
\subsection{Microgranulation ansatz}
\label{Section:2.5}
The collision integral (\ref{eqn:S1:E18}), with $\mathsf{D}$ given by (\ref{eqn:S1:E24}), expresses the evolution of the mean distribution function for each species in terms of correlators of the form $\crl{g_{\v{k}\alpha}(\v{v})g^{*}_{\v{k}\alpha'}(\v{v}')}$. Despite $g_{\v{k}\alpha}$ first entering the calculation as an initial condition, we must now interpret it in a slightly different way.

Suppose we begin at time $t_{0}$ with a fluctuation in the distribution function in the form of an initial condition $g_{\v{k}\alpha} = f_{\v{k}\alpha}(t = t_{0})$. We then evolve this perturbation,~$f_{\v{k}\alpha}(t)$, according to the linearised Vlasov equation (\ref{eqn:S1:E5}). This evolution is fast, and we assume that the mean distribution, $f_{0\alpha}(t_0)$, remains constant during it. But of course, the mean distribution does evolve, albeit on a longer time scale. So if we pick some later time~$t_{1}$, such that $t_{1}-t_{0}$ is comparable to that longer time scale, we will find a slightly altered mean distribution function, $f_{0\alpha}(t_{1})$. It would then make sense to evolve the perturbation~$f_{\v{k}\alpha}$ using this new mean distribution. We therefore `reset' the initial condition to $g_{\v{k}\alpha} = f_{\v{k}\alpha}(t=t_{1})$ and restart the evolution of $f_{\v{k}\alpha}(t)$ from this `new' initial state.

This is very similar to how one might do this numerically: we evolve linearly under a given $f_{0\alpha}$ and initial condition; then, after a small amount of time we, ratchet-like, restart the system, declaring the new initial conditions of this next iteration to be the final state of the previous iteration. The ideal model for the correlation function of $g_{\v{k}\alpha}$ would then be one that continually updated to be the correct one for a given mean distribution function. The determination of the steady state phase-space correlation function associated with a given $f_{0\alpha}$ is a difficult problem, not in general solved (cf. \citealt{adkins_schekochihin_2018}). Instead, following \citet{Kadomtsev_Pogutse70}, \citet{Balescu60} and \citet{Lenard60}, we will make some simplifying assumptions about the correlations of~$g_{\v{k}\alpha}$. The first step towards doing this is to assume that these correlations are very short-distanced in phase space, i.e., that the correlator $\crl{g_{\alpha}(\v{r},\v{v})g_{\alpha'}(\v{r}',\v{v}')}$ is zero unless the points $(\v{r},\v{v})$ and $(\v{r}',\v{v}')$ lie very close to each other. We also assume that the perturbed distribution functions of different species are uncorrelated. Mathematically, we express these assumptions in the form of the \textit{microgranulation ansatz}:
\begin{equation}
\label{eqn:S1:E32}
\crl{g_{\nu}(\v{r},\v{v})g_{\nu'}(\v{r}',\v{v}')} = \delgam_{\nu}\delta_{\nu\nu'}\delta(\v{r}-\v{r}')\delta(\v{v}-\v{v}')\crl{g_{\nu}^{2}}(\v{v}).
\end{equation}
Here the remaining correlator $\crl{g_{\nu}^{2}}(\v{v})$ is assumed to be spatially independent due to the statistical homogeneity of the system. The new parameter $\delgam_{\nu}$ is the `volume' of phase space over which the distribution function of a given species has correlated fluctuations. The ansatz (\ref{eqn:S1:E32}) does not yet constitute a closure as we have not specified the correlator~$\crl{g_{\nu}^{2}}(\v{v})$ in terms of $f_{0\nu}(\v{v})$. But before this is done, let us implement the ansatz (\ref{eqn:S1:E32}) to simplify the diffusion kernel ~(\ref{eqn:S1:E24}) again. 

The Fourier-space correlation function that appears in (\ref{eqn:S1:E24}) is easily computed from~(\ref{eqn:S1:E32}):
\begin{equation}
\label{eqn:S1:E33}
\begin{split}
\crl{g_{\v{k}\nu}(\v{v})g_{\v{k}\nu'}^{*}(\v{v}')} & = \iint\frac{\dd{\v{r}}\dd{\v{r}'}}{V^{2}}e^{-i\v{k}\cdot(\v{r}-\v{r}')}\crl{g_{\nu}(\v{r},\v{v})g_{\nu'}(\v{r}',\v{v}')} \\ &= \frac{\delgam_{\nu}}{V} \delta_{\nu \nu'}\crl{g_{\nu}^{2}}(\v{v})\delta(\v{v}-\v{v}').
\end{split}
\end{equation}
Thus, the microgranulation ansatz simplifies (\ref{eqn:S1:E24}) to
\begin{equation}
\label{eqn:S1:E34}
\mathsf{D}^{\alpha}_{\mu \nu}(\v{w},\v{v}) = \frac{16\pi^{3}q_{\mu}^{2}q_{\nu}^{2}\delgam_{\nu}}{m_{\alpha}m_{\mu}}\sum_{\v{k}}\frac{\v{k}\v{k}}{k^{4}}\frac{\delta(\v{k}\cdot(\v{w}-\v{v}))}{|\epsilon_{\v{k},\v{k}\cdot\v{v}}|^{2}}\crl{g_{\nu}^{2}}(\v{v}).
\end{equation}
This turns the collision integral (\ref{eqn:S1:E18}) into the following form, written in terms of an as yet undetermined correlator $\crl{g_{\alpha}^{2}}(\v{v})$:
\begin{multline}
\label{eqn:S1:E35}
\pdev{f_{0\alpha}}{t} = \sum_{\alpha''}\frac{16\pi^{3}q_{\alpha}^{2}q_{\alpha''}^{2}}{m_{\alpha}V}\pdev{}{\v{v}}\cdot\sum_{\v{k}}\frac{\v{k}\v{k}}{k^{4}}\cdot\int \dd{\v{v}''}\frac{\delta(\v{k}\cdot(\v{v}-\v{v}''))}{\left|\epsilon_{\v{k},\v{k}\cdot\v{v}} \right|^{2}} \\ \bigg[ \frac{\delgam_{\alpha''}}{m_{\alpha}}\crl{g_{\alpha''}^{2}}(\v{v}'')\pdev{f_{0\alpha}}{\v{v}} - \frac{\delgam_{\alpha}}{m_{\alpha''}}\crl{g_{\alpha}^{2}}(\v{v})\pdev{f_{0\alpha''}}{\v{v}''}\bigg].
\end{multline}
In the next section we will show that one can link $\crl{g_{\alpha}^{2}}(\v{v})$ to assumptions about the nature of the exact distribution function, leading finally to a closure in terms of $f_{0\alpha}$.

However, first, let us take care of a piece of unfinished business: we are now in a position to confirm that the first term in (\ref{eqn:S1:E16}) does indeed vanish. The contribution of that term to (\ref{eqn:S1:E14}) is
\begin{multline}
\label{eqn:S1:E36}
-\pdev{}{\v{v}}\cdot\sum_{\v{k}}\v{k}\mathrm{Im}\sum_{\alpha'}\frac{4\pi q_{\alpha}q_{\alpha'}}{m_{\alpha}k^{2}} \iint\frac{\dd{p}\dd{p'}}{(2\pi)^{2}}\frac{e^{(p+p')t}}{\epsilon_{\v{k}}(p)\epsilon_{\v{k}}^{*}(p'^{*})} \int\dd{\v{v}'}\crl{\hat{h}_{\v{k}\alpha}(p,\v{v})\hat{h}_{\v{k}\alpha'}^{*}(p'^{*},\v{v}')}
\end{multline}
Therefore, this term will be zero if the quantity
\begin{equation}
\label{eqn:S1:E37}
\sum_{\alpha'}\frac{4\pi q_{\alpha}q_{\alpha'}}{m_{\alpha}k^{2}} \iint\frac{\dd{p}\dd{p'}}{(2\pi)^{2}}\frac{e^{(p+p')t}}{\epsilon_{\v{k}}(p)\epsilon_{\v{k}}^{*}(p'^{*})} \int\dd{\v{v}'}\frac{\crl{g_{\v{k}\alpha}(\v{v})g^{*}_{\v{k}\alpha'}(\v{v}')}}{\left(p+i\v{k}\cdot\v{v}\right)\left(p'-i\v{k}\cdot\v{v}'\right)}
\end{equation}
is real. Taking the complex conjugate of (\ref{eqn:S1:E37}) and permuting $p\leftrightarrow p'$, we find
\begin{equation}
\label{eqn:S1:E38}
\sum_{\alpha'}\frac{4\pi q_{\alpha}q_{\alpha'}}{m_{\alpha}k^{2}}\iint\frac{\dd{p}\dd{p'}}{(2\pi)^{2}} \frac{e^{(p+p')t}}{\epsilon_{\v{k}}(p)\epsilon^{*}_{\v{k}}(p'^{*})}\int\dd{\v{v}'}\frac{\crl{g^{*}_{\v{k}\alpha}(\v{v})g_{\v{k}\alpha'}(\v{v}')}}{\left(p' - i\v{k}\cdot\v{v}\right)\left(p + i\v{k}\cdot\v{v}'\right)},
\end{equation}
which is the same as (\ref{eqn:S1:E37}) if the correlation function satisfies
\begin{equation}
\label{eqn:S1:E39}
\int\dd{\v{v}'}\frac{\crl{g_{\v{k}\alpha}(\v{v})g^{*}_{\v{k}\alpha'}(\v{v}')}}{\left(p+i\v{k}\cdot\v{v}\right)\left(p'-i\v{k}\cdot\v{v}'\right)} =\int\dd{\v{v}'}\frac{\crl{g^{*}_{\v{k}\alpha}(\v{v})g_{\v{k}\alpha'}(\v{v}')}}{\left(p' - i\v{k}\cdot\v{v}\right)\left(p + i\v{k}\cdot\v{v}'\right)}.
\end{equation}
The microgranulation ansatz (\ref{eqn:S1:E32}), manifestly does satisfy this, so our earlier neglect of the first term in (\ref{eqn:S1:E16}) is vindicated. Of course, the microgranulation ansatz is quite a simplification of the correlation function of the phase-space density. It is possible that the true correlation function would not have this symmetry and, consequently, give rise to a qualitatively different evolution of the mean distribution function. Such a possibility will be explored in a separate publication.
\section{Waterbag representation}
\label{Section:3}
The collisionless Vlasov equation has the property that it conserves integrals of all functions of solely the distribution function $f_{\alpha}(\v{v})$ over all phase space. Equivalently, under the Vlasov equation, `phase volume is incompressible', i.e., for any $\eta$, the quantity 
\begin{equation}
\label{eqn:S3:E1}
\Gamma_{\alpha}(\eta) = \iint\dd{\v{r}}\dd{\v{v}}H\left(f_{\alpha}(\v{r},\v{v}) - \eta \right),
\end{equation}
where $H(x)$ is the Heaviside function, is a constant of the motion. This implies that packets of phase-space density~$\eta$ travel in phase space and may deform but not rarefy or compress. This motivates us, as it did \cite{LyndenBell67} and \citet{Kadomtsev_Pogutse70}, to consider the concept of a waterbag distribution---a distribution for which the phase-space density is a piecewise constant function. A single-waterbag distribution is then one for which $f_{\alpha}(\v{r},\v{v})$ is piecewise constant and equal either to zero or to a single value $\eta$, while a multi-waterbag distribution function can take some countable set of values $\lbrace\eta_{i}\rbrace$. The name `waterbag' should conjure the image of these objects correctly: packets of phase `fluid' that can be distorted but not compressed or rarefied. To emphasise this fluid analogy, we shall henceforth refer to the distribution function $f_{\alpha}(\v{r},\v{v})$ as `phase-space density' and think of it as a random field, whose mean $f_{0\alpha}(\v{v})$ is our primary object of interest.

Before discussing the multi-waterbag case, we will extend Kadomtsev and Pogutse's single-waterbag model to multiple species, recovering their results and highlighting the analytical gain of the waterbag model.
\subsection{Single-waterbag closure}
\label{Section:3.1}
\begin{figure}
\centering
\begin{minipage}{0.49\textwidth}
\centering
\includegraphics[width=1.1\textwidth]{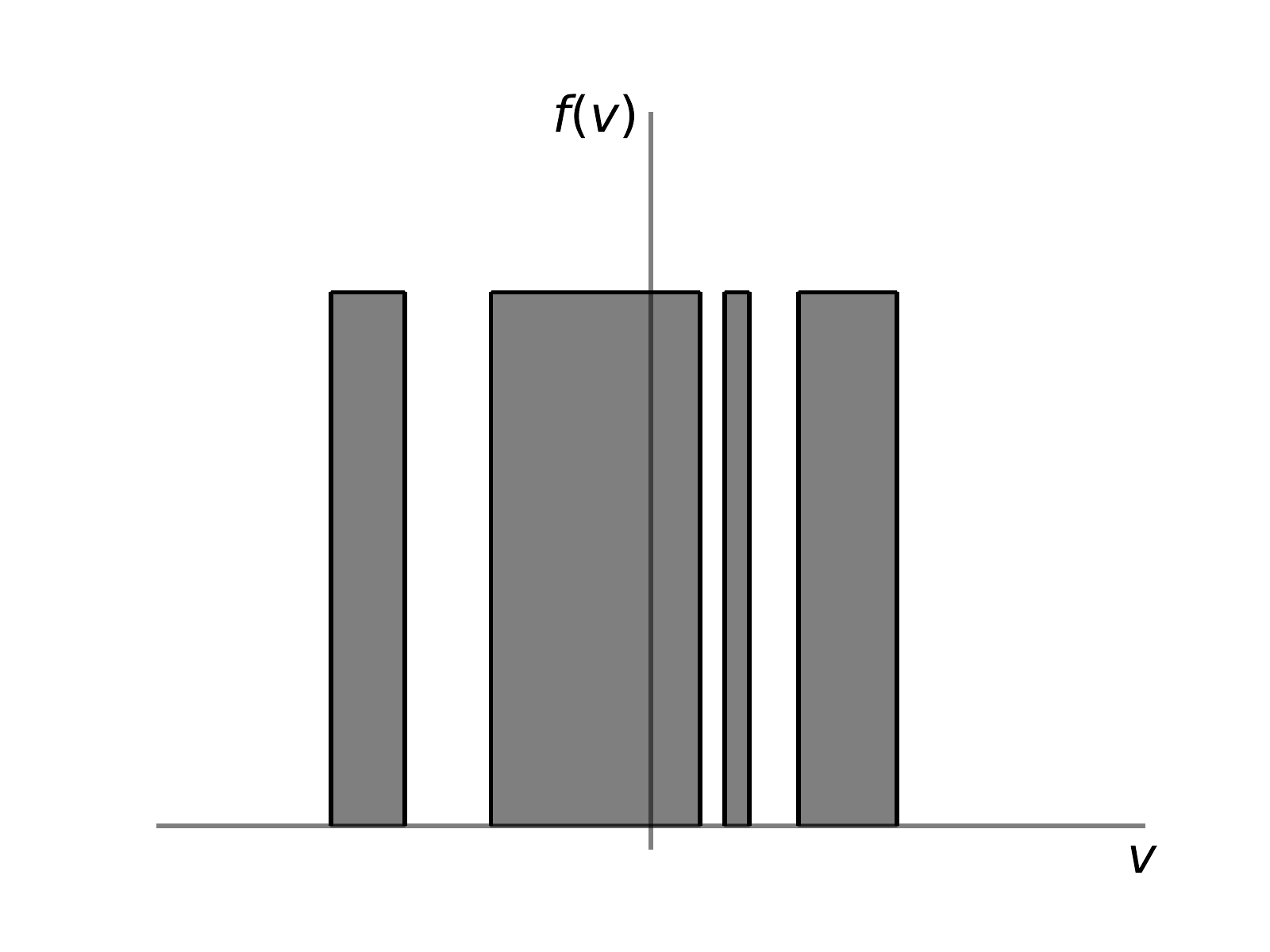}
\end{minipage}
\begin{minipage}{0.49\textwidth}
\centering
\includegraphics[width=1.1\textwidth]{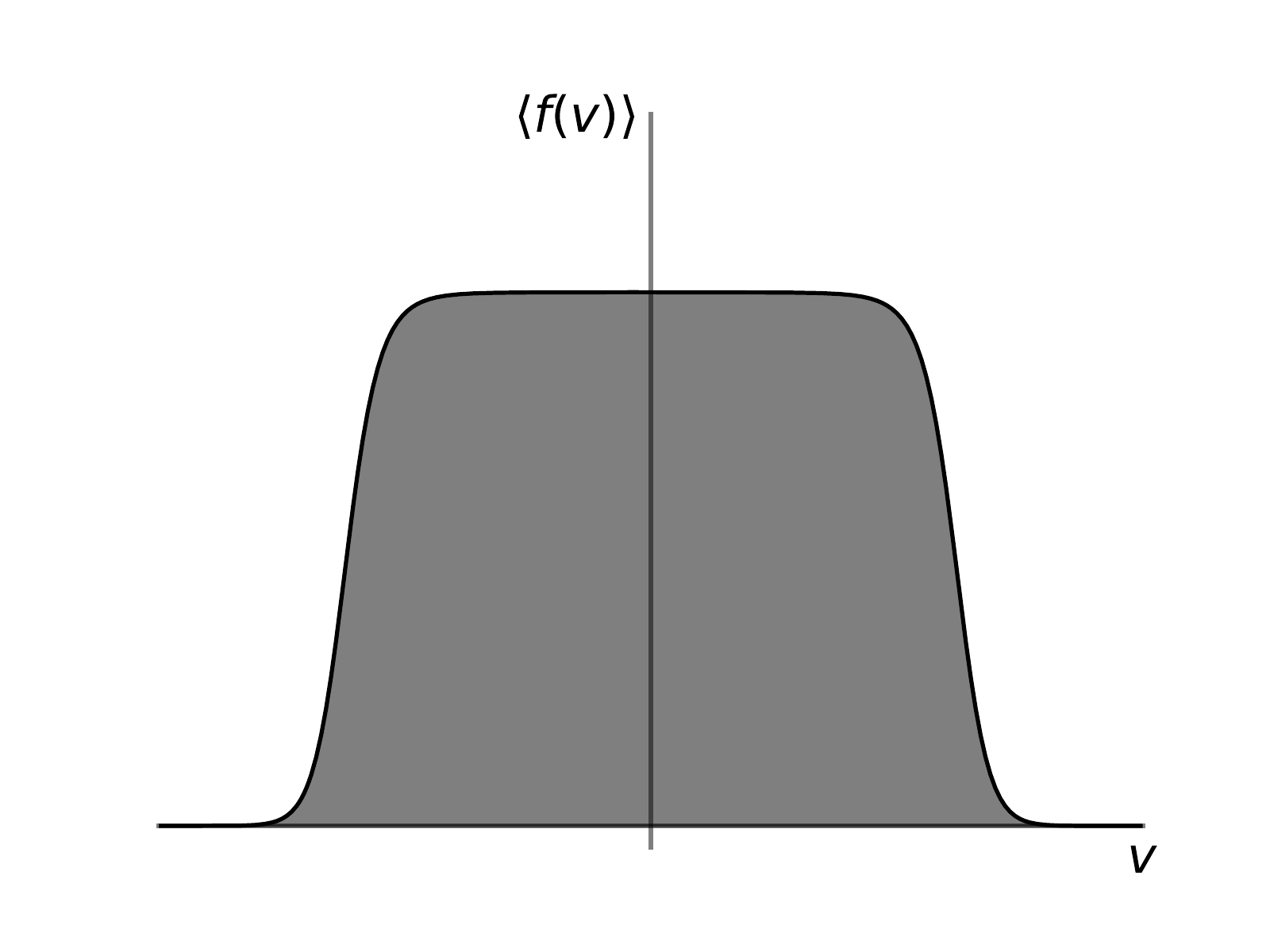}
\end{minipage}
\caption{A 1D cartoon of the exact and mean distribution functions (phase-space densities) for the single-waterbag model. As discussed in section \ref{Section:2.5}, the mean distribution should be considered in the ensemble-averaged sense, so that the many realisations of systems like the exact one given in the left panel evolve on average to look like the Fermi-Dirac distribution in the right panel.}
\label{Figure 2}
\end{figure}
When only one waterbag density $\eta_{\alpha}$ for each species $\alpha$ is assumed, this has immediate implications for the mean phase-space density. If the exact phase-space density $f_{\alpha}(\v{r},\v{v})$ is only ever $\eta_{\alpha}$ or zero, then its average value is directly related to the probability of a portion of phase space being occupied. We will therefore define $p_{\alpha}(\v{v})$ as the probability that, at a given point $\v{v}$, the phase-space density is $\eta_{\alpha}$ for particles of species $\alpha$. This can also be said as `$p_{\alpha}(\v{v})$ is the probability that there is a waterbag of species $\alpha$ at $\v{v}$' with the understanding that a waterbag is a patch of phase space of a certain density.

Written in terms of this $p_{\alpha}(\v{v})$, the mean phase-space density $f_{0\alpha}(\v{v})$ is then 
\begin{equation}
\label{eqn:S3:E3}
f_{0\alpha}(\v{v}) = \crl{f_{\alpha}(\v{r},\v{v})} =\eta_{\alpha} p_{\alpha}(\v{v}).
\end{equation}
Note that there is no $\v{r}$ dependence because our system is assumed to be statistically homogeneous in the position space. Likewise, the squared phase-space density will be $\eta_{\alpha}^{2}$ with probability $p_{\alpha}$ or zero otherwise. Therefore,
\begin{equation}
\label{eqn:S3:E4}
\crl{f_{\alpha}^{2}}(\v{v}) = \eta_{\alpha}^{2}p_{\alpha}(\v{v}) = \eta_{\alpha} f_{0\alpha}(\v{v}).
\end{equation}
The correlator $\crl{g_{\alpha}^{2}}(\v{v})$ can then be determined immediately:
\begin{equation}
\label{eqn:S3:E5}
\crl{g_{\alpha}^{2}}(\v{v}) = \crl{\left[f_{\alpha}(\v{r},\v{v}) - f_{0\alpha}(\v{v})\right]^{2}} = \crl{f_{\alpha}^{2}}(\v{v})-f_{0\alpha}^{2}(\v{v}) = \left[\eta_\alpha - f_{0\alpha}(\v{v}) \right]f_{0\alpha}(\v{v}).
\end{equation}
With this closure, (\ref{eqn:S1:E35}) becomes the multi-species generalisation of Kadomtsev and Pogutse's collision integral:
\begin{multline}
\label{eqn:S3:E6}
\pdev{f_{0\alpha}}{t} = \\ \sum_{\alpha''}\frac{16\pi^{3}q_{\alpha}^{2}q_{\alpha''}^{2}}{m_{\alpha}V}\pdev{}{\v{v}}\cdot\sum_{\v{k}}\frac{\v{k}\v{k}}{k^{4}}\cdot\int\dd{\v{v}''}\frac{\delta(\v{k}\cdot(\v{v}-\v{v}''))}{\left|\epsilon_{\v{k},\v{k}\cdot\v{v}} \right|^{2}}\Bigg\lbrace\frac{\delgam_{\alpha''}}{m_{\alpha}}\Big[\eta_{\alpha''}-f_{0\alpha''}(\v{v}'')\Big]f_{0\alpha''}(\v{v}'')\pdev{f_{0\alpha}}{\v{v}}\\ - \frac{\delgam_{\alpha}}{m_{\alpha''}}\Big[\eta_{\alpha} - f_{0\alpha}(\v{v})\Big]f_{0\alpha}(\v{v})\pdev{f_{0\alpha''}}{\v{v}''}\Bigg\rbrace.
\end{multline}

The fixed points of this collision integral are Fermi-Dirac distributions:
\begin{equation}
\label{eqn:S3:E7}
f_{0\alpha}(\v{v}) = \frac{\eta_{\alpha}}{1 + \exp\left[\beta\delgam_{\alpha}\eta_{\alpha}\left(\frac{1}{2}m_{\alpha}|\v{v}|^{2} - \mu_{\alpha}\right) \right]}.
\end{equation}
We can see that these are indeed fixed points by noting that, for (\ref{eqn:S3:E7}),
\begin{equation}
\label{eqn:S3:E8}
\frac{\delgam_{\alpha''}}{m_{\alpha}}\pdev{f_{0\alpha}}{\v{v}} = -\beta\delgam_{\alpha}\delgam_{\alpha''}\v{v}\Big[\eta_{\alpha} - f_{0\alpha}(\v{v}) \Big]f_{0\alpha}(\v{v}). 
\end{equation}
Thus, the two bracketed terms in (\ref{eqn:S3:E6}) become identical except for a factor of $\v{v}$ or $\v{v}''$. This difference vanishes when the bracket is dotted with $\delta\left(\v{k}\cdot(\v{v}-\v{v}'')\right)\v{k}$, setting the collision integral to zero. The demonstration that the fixed points (\ref{eqn:S3:E7}) are stable will come in section \ref{Section:3.3}, where we prove the general H-theorem for the multi-waterbag model, which reduces to the single-waterbag model in the appropriate limit. This H-theorem will guarantee that the mean phase-space density tends towards the maximum of a certain entropy (exactly the entropy proposed by \citealt{LyndenBell67}) subject to the conservation laws proved in section \ref{Section:2.4}. These conservation laws will therefore enforce a particular choice of the parameters $\beta$ and $\mu_{\alpha}$ in \ref{eqn:S3:E7}, which we call the `thermodynamic beta' and `chemical potentials', respectively. This is the correct number of free parameters, one for each conservation law. Note that we have, without loss of generality, moved into the frame where the plasma has zero net momentum. 
 
Physically, the Fermi-Dirac distribution has emerged because it is the maximiser of an entropy, subject to the conservation of energy, particle number and phase volume. Phase-volume conservation manifests as a Pauli-like exclusion effect: the waterbags cannot overlap, so they are forced to different points of phase space, removing the possibility of a Maxwellian equilibrium. Instead, the qualitative shape of the distribution function is set by the relation between the constant volume occupied by the waterbags and the constant energy of those waterbags. Clearly, for an initial condition with waterbags of species $\alpha$ occupying a phase volume $\Gamma_{\alpha}$ in phase space, there will be some finite minimum energy that the intial condition must possess. This minimum energy is non-zero because the possibility of all waterbags sinking to $\v{v} = 0$ is forbidden by the exclusion principle. Instead, some waterbags will be forced to higher energies, giving a constant mean phase-space density for all energies below a certain value, the Fermi energy $\epsilon_{\mathrm{F}\alpha}$. The Fermi energy is then set by the intial volume of the waterbags with
\begin{equation}
\label{eqn:S3:Fermi from volume}
\frac{\Gamma_{\alpha}}{V} = \frac{4\pi}{3}\left(\frac{2\epsilon_{\mathrm{F}\alpha}}{m_{\alpha}} \right)^{3/2},
\end{equation} 
where $V$ is the volume of the position space. The minimum possible energy $E_{\mathrm{min}}$ for such an initial condition is
\begin{equation}
\frac{E_{\mathrm{min}}}{V} = \sum_{\alpha}\frac{2\pi m_{\alpha} \eta_{\alpha}}{5}\left(\frac{2\epsilon_{\mathrm{F}\alpha}}{m_{\alpha}} \right)^{5/2} = \sum_{\alpha}\frac{2\pi m_{\alpha} \eta_{\alpha}}{5} \left(\frac{3\Gamma_{\alpha}}{4\pi V} \right)^{5/3}.
\end{equation}
When the energy of the system far exceeds this minimum,~${E \gg E_{\mathrm{min}}}$, the volume of phase space available to a given waterbag will be much greater than the volume occupied by other waterbags. In this limit, the exclusion effect will be unimportant and the distribution functions will be approximately Maxwellian with the parameters $\mu_{\alpha}$ and~$\beta$ in (\ref{eqn:S3:E7}) chosen to give the correct particle number for each species and the correct total energy. In the opposite limit, where the energy of the system begins very close to the minimum possible energy, $E-E_{\mathrm{min}} \ll E_{\mathrm{min}}$, the exclusion effect will be paramount, with waterbags tightly packed at low energies. This gives rise to a mean phase-space density that is a smoothed step function, taking a value nearly $\eta_{\alpha}$  at energies below~$\epsilon_{\mathrm{F}\alpha}$ and nearly zero above it. The values of $\beta$ and $\mu_{\alpha}$ can then be determined via a Sommerfeld expansion in the same way as it is done in standard statistical mechanics.
\subsection{Multi-waterbag closure}
\label{Section:3.2}
By assuming, as we did in section \ref{Section:3.1}, that the exact phase-space density was a single-waterbag distribution, we made the closure for $\crl{g_{\alpha}^{2}}(\v{v})$ very simple. The reason for this boils down to the fact that, for a single-waterbag distribution, knowledge of the mean phase-space density was sufficient to determine all the requisite statistical information about the exact one, namely the probabilities $p_{\alpha}(\v{v})$. In a similar vein, one can ask what information there is to be determined in an $N$-waterbag system, where the exact phase-space density can take the values $\lbrace \eta_{J\alpha} \rbrace_{J =1,2,...,N}$. As a direct generalisation of the single-waterbag model, we then define $p_{J \alpha}(\v{v})$ to be the probability that the phase-space density of species~$\alpha$ will be equal to $\eta_{J\alpha}$ at velocity $\v{v}$. In other words, `the $\eta_{J\alpha}$ waterbag of species~$\alpha$ has probability $p_{J\alpha}(\v{v})$ of being present at $\v{v}$'. Then we can write the mean phase-space density as 
\begin{equation}
\label{eqn:S3:E9}
f_{0\alpha}(\v{v}) = \crl{f_{\alpha}(\v{r},\v{v})} = \sum_{J}\eta_{J\alpha}p_{J\alpha}(\v{v}),
\end{equation}
and the mean square one as
\begin{equation}
\label{eqn:S3:E10}
\crl{f_{\alpha}^{2}}(\v{v}) = \sum_{J}\eta_{J\alpha}^{2}p_{J\alpha}(\v{v}).
\end{equation}
Now the complication introduced by multiple waterbags becomes obvious. When ${N>1}$, knowledge of just $f_{0\alpha}$ cannot uniquely determine the probabilities $p_{J\alpha}(\v{v})$ and so does not exactly determine $\crl{g_{\alpha}^{2}}$ in (\ref{eqn:S1:E35}). Therefore, it is not possible to close (\ref{eqn:S1:E35}) without further information. In principle, this information could be extracted. One could construct the evolution equations for the higher-order moments of the exact phase-space density~$\crl{f_{\alpha}^{2}}$, $\crl{f_{\alpha}^{3}}$,..., $\crl{f_{\alpha}^{N}}$. The evolution of the $i$-th such moment would generally depend on the $(i+1)$-st moment. In that way, one would have $N$ analogues of (\ref{eqn:S3:E9}) and (\ref{eqn:S3:E10}) for the $N$ unknowns $p_{J\alpha}(\v{v})$. This would then allow one to write $\crl{f_{\alpha}^{N+1}}$ as a function of all the lower-order moments and finally close the system, featuring $N$ coupled collision integrals. 

In section \ref{Section:4}, we will show that such a scheme can be made tractable by calculating all such moments in one fell swoop by increasing the dimension of the phase space. First, however, we consider a simple closure that will prove illuminating in understanding the relaxation of collisionless systems. Instead of hoping to gain any further knowledge of the system, we ask what is the most likely assignment of probabilities $p_{J\alpha}(\v{v})$ given that the mean phase-space density is $f_{0\alpha}(\v{v})$ and nothing else is known. This can be answered in the spirit of statistical inference by maximising the standard \citet{Shannon1948} entropy:
\begin{equation}
\label{eqn:S3:E11}
S_{\alpha} = -\int\dd{\v{v}}\sum_{J = 0}^{N}p_{J\alpha}(\v{v})\ln p_{J\alpha}(\v{v}),
\end{equation} 
where the sum now includes $J=0$ as `the empty waterbag', representing the probability that a given point in phase space is has zero density of particles. The Shannon entropy~(\ref{eqn:S3:E11}) must be maximised subject to the condition that the mean phase-space density is~$f_{0\alpha}(\v{v})$, i.e., that the $p_{J\alpha}(\v{v})$'s obey (\ref{eqn:S3:E9}), and that the probabilities $p_{J\alpha}(\v{v})$ sum to unity at each $\v{v}$. Another constraint on these probabilities is that the total number of particles per unit volume contained within each waterbag,
\begin{equation}
\label{eqn:S3:E12}
n_{J\alpha} = \eta_{J\alpha}\int\dd{\v{v}}p_{J\alpha}(\v{v}),
\end{equation}
is fixed for all $J \neq 0$, since the phase volume corresponding to each $\eta_{J\alpha}$ must be conserved by the evolution under the Vlasov equation. Such invariants are often called (or are equivalent to) Casimir invariants (cf. \citealt{Chavanis2004,Chavanis2005}, \citealt{Zhdankin2021}). 

Thus, we maximise the following functional for each species:
\begin{multline}
\label{eqn:S3:E13}
-\int\dd{\v{v}}\sum_{J}p_{J\alpha}(\v{v})\ln p_{J\alpha}(\v{v}) - \int\dd{\v{v}}\psi_{\alpha}(\v{v})\left[\sum_{J}\eta_{J\alpha}p_{J\alpha}(\v{v}) - f_{0\alpha}(\v{v}) \right] \\ - \sum_{J\neq 0}\gamma_{J\alpha}\left[\int \dd{\v{v}}p_{J\alpha}(\v{v}) - \frac{n_{J\alpha}}{\eta_{J\alpha}}\right]  - \int\dd{\v{v}}\lambda_{\alpha}(\v{v}) \left[\sum_{J}p_{J\alpha}(\v{v}) - 1 \right] \to \text{ max},
\end{multline}
where $\psi_{\alpha}(\v{v})$, $\gamma_{J\alpha}$, and $\lambda_{\alpha}(\v{v})$ are Lagrange multipliers. The result is
\begin{equation}
\label{eqn:S3:E14}
p_{J\alpha}(\v{v}) = \frac{1}{Z_{\alpha}\left(\psi_{\alpha}(\v{v})\right)}e^{-\psi_{\alpha}(\v{v})\eta_{J\alpha}- \gamma_{J\alpha}},
\end{equation}
where the `partition function' of species $\alpha$ is 
\begin{equation}
\label{eqn:S3:E15}
Z_{\alpha}(\psi_{\alpha}(\v{v})) = 1 + \sum_{J\neq 0} e^{-\psi_{\alpha}(\v{v})\eta_{J\alpha} - \gamma_{J\alpha}},
\end{equation}
and the Lagrange multipliers $\psi_{\alpha}(\v{v})$ and $\gamma_{J\alpha}$ must be chosen to enforce the constraints~(\ref{eqn:S3:E9}) and~(\ref{eqn:S3:E12}).
Analogously to the standard \citet{Gibbs1902} statistical mechanics, (\ref{eqn:S3:E9}) becomes 
\begin{equation}
\label{eqn:S3:E16}
f_{0\alpha}(\v{v})  = \frac{1}{Z_{\alpha}\left(\psi_{\alpha}(\v{v})\right)}\sum_{J}\eta_{J\alpha}e^{-\psi_{\alpha}(\v{v})\eta_{J\alpha} - \gamma_{J\alpha}} =
-\pdev{\ln Z_{\alpha}}{\psi_{\alpha}}.
\end{equation}
Thus, the mean phase-space density plays the role that energy does in the regular prescription of statistical mechanics, and $\psi_{\alpha}(\v{v})$ that of inverse temperature, which is local in $\v{v}$. In this formalism, therefore,
\begin{equation}
\label{eqn:S3:E17}
\begin{split}
\crl{g_{\alpha}^{2}}(\v{v}) & = \frac{1}{Z_{\alpha}(\psi_{\alpha}(\v{v}))}\sum_{J}\eta_{J\alpha}^{2}e^{-\psi_{\alpha}(\v{v})\eta_{J\alpha} - \gamma_{J\alpha}} - f_{0\alpha}^{2}(\v{v}) \\ & = \frac{1}{Z_{\alpha}}\pdevn{Z_{\alpha}}{\psi_{\alpha}}{2} - \frac{1}{Z_{\alpha}^{2}}\left(\pdev{Z_{\alpha}}{\psi_{\alpha}}\right)^{2} = \pdevn{\ln Z_{\alpha}}{\psi_{\alpha}}{2} = -\pdev{f_{0\alpha}}{\psi_{\alpha}},
\end{split}
\end{equation}
reminiscent of the heat capacity of a system in the Gibbs ensemble. Such a closure appears to have been first proposed by \cite{Chavanis2005}, in the context of geophysical turbulence.\footnote{For a different collision integral, but one can recover an integral similar to the one that we are about to produce if one applies this closure to one of the collision integrals proposed in \cite{Chavanis2004}.}

With $\crl{g_{\alpha}^{2}}(\v{v})$ thus specified, we may substitute (\ref{eqn:S3:E17}) into (\ref{eqn:S1:E35}) to get
\begin{multline}
\label{eqn:S3:E18}
\pdev{f_{0\alpha}}{t} = -\sum_{\alpha''}\frac{16\pi^{3}q_{\alpha}^{2}q_{\alpha''}^{2}}{m_{\alpha}V}\pdev{}{\v{v}}\cdot\sum_{\v{k}}\frac{\v{k}\v{k}}{k^{4}}\cdot\int \dd{\v{v}''}\frac{\delta(\v{k}\cdot(\v{v}-\v{v}''))}{\left|\epsilon_{\v{k},\v{k}\cdot\v{v}} \right|^{2}}\\\left(\frac{\delgam_{\alpha''}}{m_{\alpha}}\pdev{\psi_{\alpha}}{\v{v}} - \frac{\delgam_{\alpha}}{m_{\alpha''}}\pdev{\psi_{\alpha''}}{\v{v}''} \right)\pdev{f_{0\alpha}}{\psi_{\alpha}}(\v{v})\pdev{f_{0\alpha''}}{\psi_{\alpha''}}(\v{v}'').
\end{multline}
This is the collision integral for a multi-waterbag Lynden-Bell plasma. The instantaneous relationship between $f_{0\alpha}(\v{v})$ and $\psi_{\alpha}(\v{v})$ is given by (\ref{eqn:S3:E16}) with $Z_{\alpha}$ defined by (\ref{eqn:S3:E15}) and~$\gamma_{J\alpha}$'s set by (\ref{eqn:S3:E12}) and (\ref{eqn:S3:E14}). The set of constants $n_{J\alpha}$ in (\ref{eqn:S3:E12}) is fixed by the initial condition and cannot change during the evolution of $f_{0\alpha}$. This is the way in which phase-volume conservation in a collisionless plasma imprints a signature of the initial distribution (its `waterbag content') on its otherwise universal evolution towards the Lynden-Bell equilibria.

Note that the closure proposed above amounts to assuming that the system always quickly attains a `local' equilibrium in phase space given by (\ref{eqn:S3:E14}) with a~$\v{v}$-dependent `inverse phase temperature' $\psi_{\alpha}(\v{v})$. The integral (\ref{eqn:S3:E18}) then describes the evolution toward a `global' equilibrium. We shall discuss the plausibility of this assumption in section~\ref{Section:5}.
\subsection{Properties of the multi-waterbag collision integral}
\label{Section:3.3}
Having derived the collision integral (\ref{eqn:S3:E18}), we now proceed to study its properties. First, we will determine its fixed points, then confirm that they are stable attractors by proving an H-theorem for our collision integral.
\subsubsection{Multi-waterbag equilibria}
\label{Section:3.3.1}
Since $\partial f_{0\alpha} / \partial \psi_{\alpha} \leq 0$ (and only zero in pathological cases), the integral on the right-hand side of (\ref{eqn:S3:E18}) can vanish only if its integrand vanishes. Manifestly, it does so if 
\begin{equation}
\label{eqn:S3:E19}
\psi_{\alpha}(\v{v}) = \beta\delgam_{\alpha}\frac{1}{2}m_{\alpha}|\v{v}|^{2} \equiv \beta \delgam_{\alpha}\epsilon_{\alpha}(\v{v}),
\end{equation}
where $\epsilon_{\alpha}(\v{v})$ is the energy of a particle of species $\alpha$ with velocity $\v{v}$. Therefore, from (\ref{eqn:S3:E14}),
\begin{equation}
\label{eqn:S3:E21}
p_{J\alpha}(\v{v}) = \frac{\exp\Big\lbrace-\beta \delgam_{\alpha}\eta_{J\alpha}\left[\epsilon_{\alpha}(\v{v}) - \mu_{J\alpha}\right]\Big\rbrace}{1+ \sum_{J'\neq 0}\exp\Big\lbrace-\beta \delgam_{\alpha}\eta_{J'\alpha}\left[\epsilon_{\alpha}(\v{v}) - \mu_{J'\alpha}\right]\Big\rbrace}.
\end{equation}
We have brought this into a form pleasingly similar to the Fermi-Dirac distribution by rewriting $\gamma_{J\alpha} = -\beta \delgam_{\alpha}\eta_{J\alpha}\mu_{J\alpha}$ to define the chemical potential $\mu_{J\alpha}$ of the waterbag~$J$ of species $\alpha$. These are the equilibria derived by \citet{LyndenBell67} by a statistical-mechanical entropy-maximisation method applied to a multi-waterbag system. In our derivation, they have emerged as fixed points of a plasma's dynamical evolution. In section \ref{Section:3.3.2}, we will prove that these fixed points of our collision integral (\ref{eqn:S3:E18}) are stable and that the system relaxes towards them. First, however, let us discuss the qualitative nature of these solutions.

The similarity between the multi-waterbag equilibria (\ref{eqn:S3:E21}) and the Fermi-Dirac distribution is no accident. Just like the Fermi-Dirac distribution, the Lynden-Bell equilibria maximise an entropy subject to the conservation of energy, momentum and phase volume. Conservation of phase volume in the single-waterbag case meant waterbags excluded each other. In the multi-waterbag case, the same is true and it must be stressed that waterbags of different phase-space densities exclude each other indiscriminately. The result is a very broad class of possible distributions, whose general features, however, can be understood from the relation between the total energy of the system and the phase volume occupied by each waterbag.  

Just like in the case of the Fermi-Dirac distribution, for an initial condition with waterbags of phase-space densities~$\eta_{J\alpha}$ of species~$\alpha$ each of which occupy a phase volume~$\Gamma_{J\alpha}=Vn_{J\alpha}/\eta_{J\alpha}$, there is a minimum possible energy of the system,~$E_{\mathrm{min}}$. A state with this minimum energy is the state that has the waterbag of the highest phase-space density at the lowest energy with waterbags of progressively lower phase-space density forced to higher energies by the exclusion of phase volume. This is analogous to how a suspension of liquids of different densities will arrange themselves in a glass, with the densest taking the lowest energy (cf. \citealt{Gardner63}, \citealt{Lorenz1955}). Let us order~$\eta_{1\alpha}<\eta_{2\alpha}<...<\eta_{N\alpha}$ by increasing waterbag density. The densest waterbag~$\eta_{N\alpha}$ will occupy the sphere in velocity space with maximum energy~$\epsilon_{\mathrm{F}J\alpha}$ given by
\begin{equation}
\frac{\Gamma_{N\alpha}}{V} = \frac{4\pi}{3}\left(\frac{2\epsilon_{\mathrm{F}N\alpha}}{m_{\alpha}} \right)^{3/2},
\end{equation}
while the subsequent waterbags will be forced to higher energies, filling shells in velocity space. The waterbag of density $\eta_{J\alpha}$ will then occupy the shell between energies $\epsilon_{\mathrm{F}(J+1)\alpha}$ and $\epsilon_{\mathrm{F}J\alpha}$ given by
\begin{equation}
\frac{\Gamma_{J\alpha}}{V} = \frac{4\pi}{3}\left[\left(\frac{2\epsilon_{\mathrm{F}J\alpha}}{m_{\alpha}} \right)^{3/2} - \left(\frac{2\epsilon_{\mathrm{F}(J+1)\alpha}}{m_{\alpha}} \right)^{3/2} \right],
\end{equation}
or, equivalently,
\begin{equation}
\frac{4\pi}{3}\left(\frac{2\epsilon_{\mathrm{F}J\alpha}}{m_{\alpha}} \right)^{3/2}  = \sum_{n= J}^{N}\frac{\Gamma_{n\alpha}}{V}.
\end{equation}
The minimum energy is then
\begin{equation}
\begin{split}
E_{\mathrm{min}} & = V\sum_{\alpha}\sum_{J=1}^{N}\frac{2\pi m_{\alpha} \eta_{J\alpha}}{5}\left[\left(\frac{2\epsilon_{\mathrm{F}J\alpha}}{m_{\alpha}} \right)^{5/2} - \left(\frac{2\epsilon_{\mathrm{F}(J+1)\alpha}}{m_{\alpha}}\right)^{5/2} \right] 
\\ & = V\sum_{\alpha}\sum_{J=1}^{N}\frac{2\pi m_{\alpha} \eta_{J\alpha}}{5}\left(\frac{3}{4\pi V}\right)^{5/3}\left[\left(\sum_{n=J}^{N}\Gamma_{n\alpha}\right)^{5/3} - \left(\sum_{n=J+1}^{N}\Gamma_{n\alpha}\right)^{5/3} \right].
\end{split}
\end{equation}
At precisely this energy, the distribution as a function of velocity will look like a ziggurat, with sharp steps corresponding to the different waterbag densities. This state is known generically as a \cite{Gardner63}, or `Gardner-restacked' distribution (normally thought of in the continuous limit of infinitely many infinitesimally separated waterbags--- see section \ref{Section:3.4}). It is a generic `ground state' for a Vlasov plasma (cf. \citealt{helander_2017}).
\begin{figure}
\centering
\includegraphics[width=\textwidth]{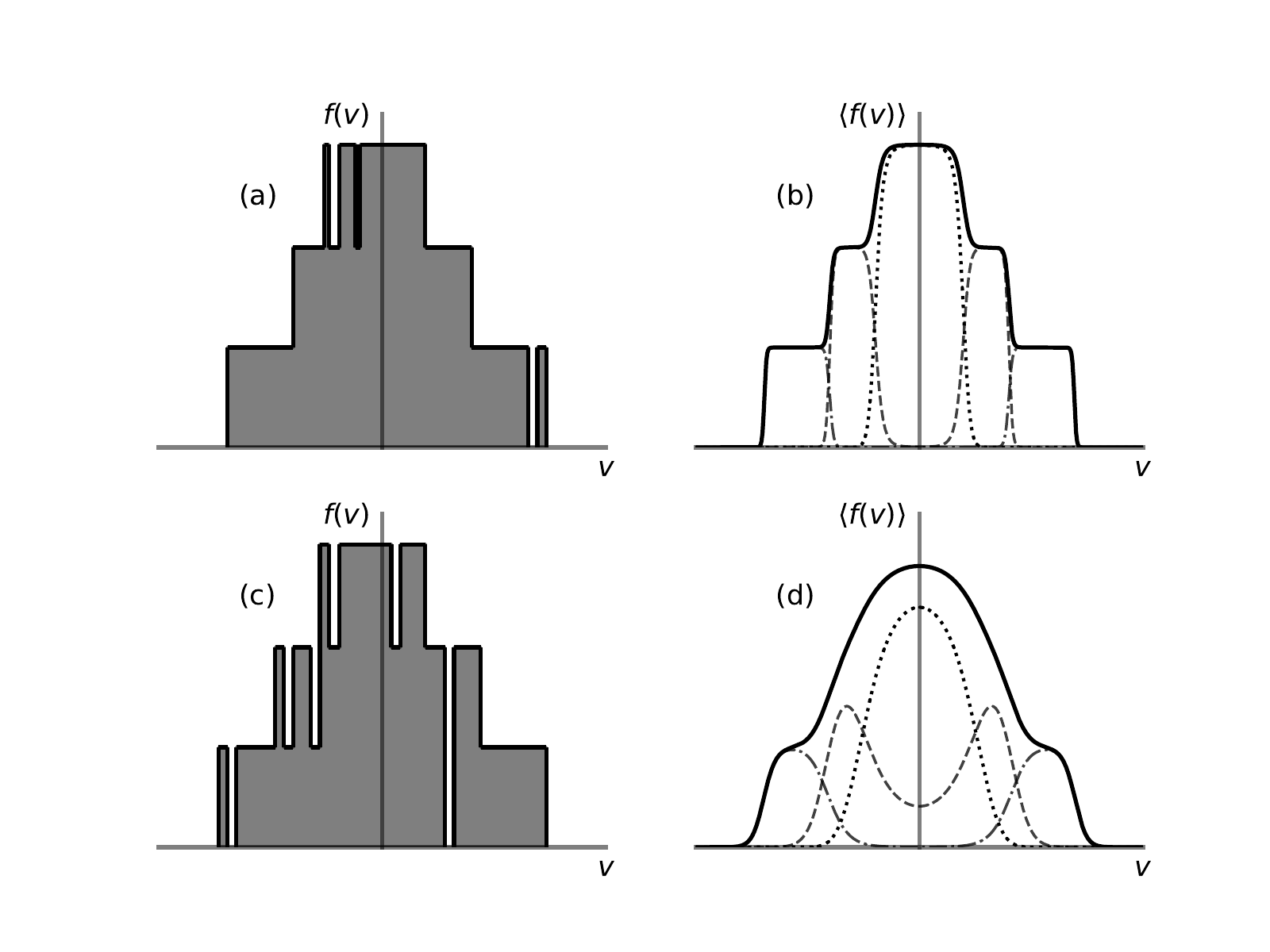}
\caption{A 1D cartoon depicting exact and mean phase-space densities for a three-waterbag system. Panels~{(b)} and~{(d)} show~$f_{0}(v)$ corresponding to the exact~$f(v)$ shown in panels (a) and (c), respectively. The dotted, dashed and dot-dashed lines in panels~{(b)} and~{(d)} show to the contributions~$\eta_{J}p_{J}(v)$ of each of the three waterbag densities to the mean phase-space density. The system in the top two panels has a total energy equal to~$E = 1.003 E_{\mathrm{min}}$, where~$E_{\mathrm{min}}$ is the minimal energy that the system could have subject to phase-volume conservation. The system in the bottom two panels has~$E = 1.05 E_{\mathrm{min}}$. This allows the second most dense waterbag (dashed line) to intermingle with the densest waterbag (dotted line) giving a substantial contribution to the mean phase-space density at $v=0$. In contrast, in the top panels, the little energy that the system has above~$E_{\mathrm{min}}$ permits very little mixing between waterbags, making the mean phase-space density appear very close to a step function (a `ziggurat').}
\label{Figure 3}
\end{figure}

However, this state is not at all representative of the full range of possible Lynden-Bell equilibria. Since the energy of the system remains constant under the collision integrals that we are considering (see section \ref{Section:2.4}), for the Gardner distribution to reachable, the system must have started with the minimum energy possible, but that necessarily implies that it already was in the Gardner state. As we did in section \ref{Section:3.1} for the Fermi-Dirac distribution, let us ask for what range of initial energies the Lynden-Bell distribution~(\ref{eqn:S3:E21}) will be qualitatively similar to the Gardner distribution. To answer this question, we must consider the energy required to allow two waterbags to intermingle freely in velocity space as opposed to being cleanly separated. Should the~$(J+1)$-st and~$J$-th waterbags freely mix, they would have an average phase-space density
\begin{equation}
\bar{\eta} = \frac{\eta_{(J+1)\alpha} \Gamma_{(J+1)\alpha}+ \eta_{J\alpha} \Gamma_{J\alpha}}{\Gamma_{(J+1)\alpha} + \Gamma_{J\alpha}}
\end{equation}
between the energies $\epsilon_{(J+2)\alpha}$ and $\epsilon_{J \alpha}$. The difference between such a state's energy and the original energy is
\begin{align}
\label{eqn:Exchange energy}
  \phantom{\Delta E_{J \alpha}}
  &\begin{aligned}
    \mathllap{\Delta E_{J\alpha}} & = \frac{2\pi m_{\alpha}V}{5}\left(\frac{2}{m_{\alpha}} \right)^{5/2}\Bigg[\bar{\eta} \left(\epsilon_{\mathrm{F}J\alpha}^{5/2} - \epsilon_{\mathrm{F}(J+2)\alpha}^{5/2} \right)\\ & \quad - \eta_{J\alpha} \left(\epsilon_{\mathrm{F}J\alpha}^{5/2} - \epsilon_{\mathrm{F}(J+1)\alpha}^{5/2} \right) - \eta_{(J+1)\alpha} \left(\epsilon_{\mathrm{F}(J+1)\alpha}^{5/2} - \epsilon_{\mathrm{F}(J+2)\alpha}^{5/2} \right)\Bigg]
  \end{aligned} \nonumber \\ 
  &\begin{aligned}
  \mathllap{} & = \frac{2^{7/2}\pi V}{5m_{\alpha}^{3/2}}(\eta_{(J+1)\alpha} - \eta_{J\alpha})\times \\ & \quad \Bigg(\frac{\Gamma_{(J+1)\alpha}}{\Gamma_{(J+1)\alpha}+ \Gamma_{J\alpha}}\epsilon_{\mathrm{F}J\alpha}^{5/2} + \frac{\Gamma_{J\alpha}}{\Gamma_{(J+1)\alpha}+ \Gamma_{J\alpha}}\epsilon_{\mathrm{F}(J+2)\alpha}^{5/2} -\epsilon_{\mathrm{F}(J+1)\alpha}^{5/2} \Bigg).
  \end{aligned}
\end{align}
The somewhat complex form of (\ref{eqn:Exchange energy}) is due to the fact that the Lynden-Bell equilibria allow for a huge number of possible initial conditions. Nevertheless, some important, and simple, features can be gleaned from (\ref{eqn:Exchange energy}). First, $\Delta E_{J\alpha}$ can be very small if either~$\eta_{(J+1)\alpha} - \eta_{J\alpha}$ is small or, more subtly, if one of $\Gamma_{J\alpha}$ or $\Gamma_{(J+1)\alpha}$ is small compared to the other. The first of these possibilities is a manifestation of the fact that, if the phase-space densities of two waterbags are similar, the distribution function does not change significantly by allowing them to intermingle. The second possibility makes the energy required to intermingle the waterbags small because only a small portion of phase space needs to be `excited'.

The corollary of (\ref{eqn:Exchange energy}) is, therefore, that the initial energy of the system must only be greater than its minimum possible energy by an amount~$E-E_{\mathrm{min}} \sim \mathrm{min}_{J}\Delta E_{J\alpha}$ for the Lynden-Bell equilibrium distribution of species $\alpha$ to be considerably distinct from the Gardner distribution. If, for instance, $\Delta E_{J\alpha}$ were minimal for a particular $J = \bar{J}$, then at energies such that~$E-E_{\mathrm{min}} \sim \Delta E_{\bar{J}\alpha}$, the equilibrium would allow the waterbags $\bar{J}$ and $\bar{J}+1$ to be intermingled without the phase space occupied by the other waterbags being `unlocked'. An example of this is shown in figure \ref{Figure 3} for a 1D, three-waterbag system: the two left panels show the exact phase-space densities $f(v)$, the two right panels the corresponding mean ones~$f_{0}(v)$. For the latter, we also show the contributions $\eta_{J}p_{J}(v)$ from each of the three waterbags. In the two upper panels, where the energy of the system was chosen to be very close to $E_{\mathrm{min}}$, the equilibrium Lynden-Bell looks like a ziggurat. More interestingly, when the system has more energy, as it does in the two lower panels, it first `unlocks' mixing between the two densest waterbags. Thus, the second densest waterbag makes a significant contribution to the phase-space density at zero velocity. The least dense waterbag is still very localised in phase space, indicating that more energy would need to be present in the system to make all waterbags fully non-degenerate. What all this means for Lynden-Bell equilibria is that, despite the relative simplicity of solving a set of transcendental equations enforcing waterbag-content and energy conservation, the shape of these equilibria can be significantly varied for different initial conditions.

\subsubsection{The H-theorem}
\label{Section:3.3.2}
To prove the stability of the Lynden-Bell equilibria (\ref{eqn:S3:E21}), we will now prove an H-theorem for the collision integral (\ref{eqn:S3:E18}). Namely, we will prove that there is a functional of $f_{0\alpha}$ that can only be increased by evolution under~(\ref{eqn:S3:E18}). This functional is then the entropy of our system, and if it has a maximum, this is a stable attractor of the evolution, since the system could not depart from this state without lowering the entropy. 

Consider the following obvious candidate for entropy:
\begin{equation}
\label{eqn:S3:E23}
S = -\sum_{\alpha}\frac{V}{\delgam_{\alpha}}\int\dd{\v{v}}\sum_{J}p_{J\alpha}(\v{v})\ln p_{J\alpha}(\v{v}).
\end{equation}
As before, the sum in this definition includes the empty waterbag $J=0$, whose probability is~$p_{0\alpha} = 1 - \sum_{J\neq 0}p_{J\alpha}$. Therefore, (\ref{eqn:S3:E23}) reduces to the well-known Fermi-Dirac entropy in the single-waterbag case. According to our closure scheme, $p_{J\alpha}(\v{v})$ can be written as (\ref{eqn:S3:E14}). The entropy (\ref{eqn:S3:E23}) then becomes
\begin{equation}
\label{eqn:S3:E24}
\begin{split}
S &= \sum_{\alpha}\frac{V}{\delgam_{\alpha}}\int\dd{\v{v}}\sum_{J}p_{J\alpha}(\v{v})\Big[\psi_{\alpha}(\v{v})\eta_{J\alpha}  + \gamma_{J\alpha} +  \ln Z_{\alpha}(\psi_{\alpha}(\v{v}))\Big] \\ & = \sum_{\alpha}\frac{V}{\delgam_{\alpha}}\left\lbrace\int\dd{\v{v}} \Big[f_{0\alpha}(\v{v})\psi_{\alpha}(\v{v}) + \ln Z_{\alpha}(\psi_{\alpha}(\v{v}))\Big] + \sum_{J}\gamma_{J\alpha}\frac{n_{J\alpha}}{\eta_{J\alpha}}\right\rbrace.
\end{split}
\end{equation}
We now take the time derivative of (\ref{eqn:S3:E24}):
\begin{equation}
\label{eqn:S3:E25}
\dev{S}{t}  = \sum_{\alpha}\frac{V}{\delgam_{\alpha}}\left[\int\dd{\v{v}} \Bigg(\pdev{f_{0\alpha}}{t}\psi_{\alpha} + f_{0\alpha}\pdev{\psi_{\alpha}}{t} + \pdev{\ln Z_{\alpha}}{t}\Bigg) + \sum_{J}\pdev{\gamma_{J\alpha}}{t}\frac{n_{J\alpha}}{\eta_{J\alpha}}\right].
\end{equation}
This can be simplified by taking the time derivative of (\ref{eqn:S3:E16}) and noting that 
\begin{equation}
\label{eqn:S3:E26}
\begin{split}
\pdev{\ln Z_{\alpha}}{t} &  = -\frac{1}{Z_{\alpha}}\sum_{J}e^{-\psi_{\alpha}\eta_{J\alpha} - \gamma_{J\alpha}}\Big(\pdev{\psi_{\alpha}}{t}\eta_{J\alpha} + \pdev{\gamma_{J\alpha}}{t} \Big)\\ &    = - f_{0\alpha}\pdev{\psi_{\alpha}}{t} - \sum_{J}p_{J\alpha}\pdev{\gamma_{J\alpha}}{t}.
\end{split}
\end{equation}
Substituting this into (\ref{eqn:S3:E25}) and using (\ref{eqn:S3:E12}), we get
\begin{equation}
\label{eqn:S3:E27}
\dev{S}{t}  = \sum_{\alpha}\frac{V}{\delgam_{\alpha}}\int\dd{\v{v}}\pdev{f_{0\alpha}}{t}\psi_{\alpha}.
\end{equation}
In this form, we may finally use the collision integral (\ref{eqn:S3:E18}) and integrate by parts:
\begin{align}
\label{eqn:S3:E28}
  \phantom{\dev{S}{t}} 
  &\begin{aligned}
  \mathllap{\dev{S}{t}} = \sum_{\alpha\alpha''}\frac{16\pi^{3}q_{\alpha}^{2}q_{\alpha''}^{2}}{\delgam_{\alpha}\delgam_{\alpha''}}\sum_{\v{k}}\iint & \dd{\v{v}}\dd{\v{v}''}\frac{\delta(\v{k}\cdot(\v{v}-\v{v}''))}{k^{4}|\epsilon_{\v{k},\v{k}\cdot\v{v}}|^{2}}\pdev{f_{0\alpha}}{\psi_{\alpha}}(\v{v})\pdev{f_{0\alpha''}}{\psi_{\alpha''}}(\v{v}'') \\ & \left[\frac{\delgam_{\alpha''}^{2}}{m_{\alpha}^{2}}\left(\v{k}\cdot\pdev{\psi_{\alpha}}{\v{v}} \right)^{2}  - \frac{\delgam_{\alpha}\delgam_{\alpha''}}{m_{\alpha}m_{\alpha''}}\v{k}\cdot\pdev{\psi_{\alpha''}}{\v{v}''}\v{k}\cdot\pdev{\psi_{\alpha}}{\v{v}}\right] 
  \end{aligned} \nonumber \\
  &\begin{aligned}
    \mathllap{} = \sum_{\alpha\alpha''}\frac{8\pi^{3}q_{\alpha}^{2}q_{\alpha''}^{2}}{\delgam_{\alpha}\delgam_{\alpha''}}\sum_{\v{k}}\iint & \dd{\v{v}}\dd{\v{v}''} \frac{\delta(\v{k}\cdot(\v{v}-\v{v}''))}{k^{4}|\epsilon_{\v{k},\v{k}\cdot{\v{v}}}|^{2}}\pdev{f_{0\alpha}}{\psi_{\alpha}}(\v{v})\pdev{f_{0\alpha''}}{\psi_{\alpha''}}(\v{v}'') \\  & \Bigg(\frac{\delgam_{\alpha''}}{m_{\alpha}}\v{k}\cdot\pdev{\psi_{\alpha}}{\v{v}}- \frac{\delgam_{\alpha}}{m_{\alpha''}}\v{k}\cdot\pdev{\psi_{\alpha''}}{\v{v}''} \Bigg)^{2} \geq 0.
  \end{aligned}
\end{align}
Here, to get to the second equality, we have symmetrised the expression by swapping~${\alpha \leftrightarrow \alpha''}$ and $\v{v} \leftrightarrow \v{v}''$. From (\ref{eqn:S3:E17}), we note that $\partial f_{0\alpha}/\partial\psi_{\alpha}$ is a negative semi-definite quantity so the integrand in (\ref{eqn:S3:E28}) is certainly non-negative. This proves that the entropy~(\ref{eqn:S3:E24}) is increased by the collision integral (\ref{eqn:S3:E18}) unless the integrand is zero, in which case the entropy is conserved. For the latter possibility to be realised, the squared expression in (\ref{eqn:S3:E28}) must vanish, which it does only for $\psi_{\alpha}(\v{v})$ given by (\ref{eqn:S3:E19}). Thus, the collision integral~(\ref{eqn:S3:E18}) relaxes the mean phase-space density of the system towards the Lynden-Bell multi-waterbag equilibria (\ref{eqn:S3:E21}).
\subsection{Continuous limit of the multi-waterbag formalism}
\label{Section:3.4}
While conceptually enlightening, using a discrete number of waterbag densities is somewhat problematic. At face value, it restricts one to considering initial conditions that are piecewise-constant functions. In fact, a continuous phase-space density $f_{\alpha}(\v{v})$ can easily be accommodated by making the grid of~$\eta$ arbitrarily fine. We do this by choosing some large number of waterbags densities to fill the interval $[0,\eta_{\alpha\max}]$ with spacing $\Delta\eta\to 0$. We can then rewrite what were sums over~$J$ as integrals with respect to the waterbag density~$\eta$: 
\begin{equation}
\label{eqn:S3:E29}
\sum_{J} \to \frac{1}{\Delta\eta}\int_{0}^{\eta_{\alpha \max}}\dd{\eta}.
\end{equation}
Note that the sum above includes the empty waterbag, and so care must be taken to ensure that, when the sums are transformed in this way, the empty waterbag has been included. Now all discrete quantities indexed by $J$ are to be upgraded to continuous functions of~$\eta$. Namely, we let 
\begin{equation}
\gamma_{\alpha}(\eta) \equiv \gamma_{J\alpha}, \quad\quad\quad P_{0\alpha}(\v{v},\eta) \equiv  \frac{p_{J\alpha}(\v{v})}{\Delta\eta}, 
\end{equation}
the latter function being a probability density with respect to $\eta$ (hence the normalisation to $\Delta \eta$). Then (\ref{eqn:S3:E14}) becomes
\begin{equation}
\label{eqn:S3:E30}
P_{0\alpha}(\v{v},\eta) = \frac{1}{Z_{\alpha}(\psi_{\alpha}(\v{v}))}e^{-\psi_{\alpha}(\v{v})\eta - \gamma_{\alpha}(\eta)},
\end{equation}
with the partition function (\ref{eqn:S3:E15}) redefined as
\begin{equation}
\label{eqn:S3:E31}
Z_{\alpha}(\psi_{\alpha}(\v{v})) = \int_{0}^{\eta_{\alpha\mathrm{max}}}\dd{\eta}e^{-\psi_{\alpha}(\v{v})\eta - \gamma_{\alpha}(\eta)}.
\end{equation}
The Lagrange multiplier $\psi_{\alpha}(\v{v})$ is now determined by the continuous version of~(\ref{eqn:S3:E9}) and~(\ref{eqn:S3:E16}): 
\begin{equation}
\label{eqn:S3:E32}
f_{0\alpha}(\v{v}) = \int\dd{\eta} \eta P_{0\alpha}(\v{v},\eta) = -\pdev{\ln Z_{\alpha}}{\psi_{\alpha}}.
\end{equation}
To determine $\gamma_{\alpha}(\eta)$, one must solve the continuous version of the constraint (\ref{eqn:S3:E12}) fixing the `waterbag content' of the distribution: 
\begin{equation}
\label{eqn:S3:E33}
\int \dd{\v{v}} P_{0\alpha}(\v{v},\eta) = \frac{n_{\alpha}(\eta)}{\eta \Delta\eta} \equiv \rho_{\alpha}(\eta),
\end{equation}
where $n_{\alpha}(\eta)$ is the continuous generalisation of~$n_{J\alpha}$, and the newly defined function~$\rho_{\alpha}(\eta)$ encodes all the information about the species $\alpha$ that must be retained from its initial distribution (an infinite set of Casimir invariants).

Thus, the price of the generalisation to a continuous-waterbag model is having to solve the two coupled integral equations~(\ref{eqn:S3:E32}) and~(\ref{eqn:S3:E33}) for the functions~$\psi_{\alpha}(\v{v})$ and~$\gamma_{\alpha}(\eta)$. The collision integral is then again (\ref{eqn:S3:E18}), its fixed points are~(\ref{eqn:S3:E32}) with $\psi_{\alpha}(\v{v})$ given by~(\ref{eqn:S3:E19}), viz.,
\begin{equation}
f_{0\alpha}(\v{v}) = \frac{\int\dd{\eta}\eta e^{-\psi_{\alpha}(\v{v})\eta - \gamma_{\alpha}(\eta)}}{\int \dd{\eta}e^{-\psi_{\alpha}(\v{v})\eta - \gamma_{\alpha}(\eta)}},
\end{equation}
and the H-theorem (\ref{eqn:S3:E28}) continues to hold, with the continuous limit of the entropy~(\ref{eqn:S3:E23}):
\begin{equation}
\label{eqn:S3:E34}
S = -\sum_{\alpha}\frac{V}{\delgam_{\alpha}}\iint\dd{\v{v}}\dd{\eta}P_{0\alpha}(\v{v},\eta)\ln P_{0\alpha}(\v{v},\eta).
\end{equation} 
\section{Hyperkinetics}
\label{Section:4}
We saw in section (\ref{Section:3}) that, to work out the collisionless evolution of the plasma, we needed to know the probability for the phase-space density $f_{\alpha}(\v{r},\v{v})$ to have a certain value~$\eta$ at a velocity $\v{v}$. We called that probability $p_{J\alpha}(\v{v})$, or, in the continuous limit,~$P_{0\alpha}(\v{v},\eta)$. This led to a closure problem because these probabilities could not be uniquely determined from $f_{0\alpha}(\v{v})$ alone. There was an analogy between the resolution of this closure problem proposed in section \ref{Section:3.2} and the resolution of closure problems in fluid theories: it was done by appealing to a local maximisation of entropy---in our case, local in $\v{v}$, so the functional form of $P_{0\alpha}(\v{v},\eta)$ with respect to $\eta$ was fixed by~(\ref{eqn:S3:E30}), whereas its dependence on $\v{v}$ remained undetermined and encoded by $\psi_{\alpha}(\v{v})$. This fluid analogy is further strengthened by noting that the mean phase-space density was then written in~(\ref{eqn:S3:E32}) as a fluid quantity: the first moment of $P_{0\alpha}(\v{v},\eta)$ with respect to~$\eta$.
 
All this points to an alternative route to describing collisionless relaxation. The closure problem in fluid theories is resolved by recognising that the system can be described kinetically. Following this logic, instead of considering $f_{0\alpha}(\v{v})$ as the core object of our theory, we will consider $P_{0\alpha}(\v{v},\eta)$, from which $f_{0\alpha}(\v{v})$ can be derived. Thus, we are extending our phase space from 6D, $(\v{r},\v{v})$ to 7D, $(\v{r},\v{v},\eta)$, in such a way that the kinetics of $f_{0\alpha}(\v{r},\v{v})$ will be derivable by taking moments of the new kinetics of $P_{\alpha}(\v{r},\v{v},\eta)$ and $P_{0\alpha}(\v{v},\eta)$. Like Lynden-Bell's statistical mechanics, this approach, which we shall call `hyperkinetics', originates from galactic dynamics (as well as fluid mechanics; see \citealt{Chavanis_1996} and references therein). Its logical conclusion is the hyperkinetic collision integral that we shall now derive---it is a version of the collision integral first derived, in the context of galactic dynamics and for a model with discrete multiple waterbags, by \cite{Severne1980} (who used a somewhat different method). 
\subsection{Hyperkinetic collision integral}
\label{Section:4.1}
To construct the kinetics of the individual waterbags, we first define the `waterbag distribution function' 
\begin{equation}
\label{eqn:S4:E1}
P_{\alpha}(\v{r},\v{v},\eta) = \delta(f_{\alpha}(\v{r},\v{v}) - \eta),
\end{equation}
which is the probability density of finding the exact phase-space density $f_{\alpha}(\v{r},\v{v})$ to have the value $\eta$ at the phase-space position $(\v{r},\v{v})$. The evolution equation for $P_{\alpha}$ takes the same form as the Vlasov equation:
\begin{equation}
\label{eqn:S4:E2}
\begin{split}
\pdev{P_{\alpha}}{t} &= \pdev{f_{\alpha}}{t}\delta'(f_{\alpha}(\v{r},\v{v}) - \eta) \\ & = \left(-\v{v}\cdot\nabla f_{\alpha} + \frac{q_{\alpha}}{m_{\alpha}}\nabla \varphi \cdot \pdev{f_{\alpha}}{\v{v}}\right)\delta'(f_{\alpha}(\v{r},\v{v}) - \eta) \\ & = -\v{v}\cdot\nabla P_{\alpha} + \frac{q_{\alpha}}{m_{\alpha}}\nabla \varphi \cdot \pdev{P_{\alpha}}{\v{v}}.
\end{split}
\end{equation}
Since we are still concerned with the relaxation to equilibrium, we now wish to find the collision integral for the evolution of $P_{\alpha}$ in much the same way as we did for $f_{\alpha}$ in \ref{Section:2}. 

As before, we Fourier-decompose 
\begin{equation}
\label{eqn:S4:E3}
P_{\alpha}(\v{r},\v{v},\eta) = P_{0\alpha}(\v{v},\eta) + \sum_{\v{k}}e^{i\v{k}\cdot\v{r}}P_{\v{k}\alpha}(\v{v},\eta),
\end{equation}
and expect the ensemble average of the homogeneous part of the hyperkinetic distribution, $P_{0\alpha}$, to be much greater in size than the fluctuating part, sanctioning the quasilinear approach. Therefore, we again linearise our hyperkinetic equation (\ref{eqn:S4:E2}) to get the fluctuating part of the distribution:
\begin{equation}
\label{eqn:S4:E4}
\pdev{P_{\v{k}\alpha}}{t} + i\v{k}\cdot\v{v}P_{0\alpha} = i\frac{q_{\alpha}}{m_{\alpha}}\varphi_{\v{k}}\v{k}\cdot\pdev{P_{0\alpha}}{\v{v}},
\end{equation}
which in turn determines the evolution of the homogeneous part:
\begin{equation}
\label{eqn:S4:E5}
\pdev{P_{0\alpha}}{t} = \pdev{}{\v{v}}\cdot\left(\frac{q_{\alpha}}{m_{\alpha}}\sum_{\v{k}}\v{k}\mathrm{Im}\crl{\varphi_{\v{k}}^{*}P_{\v{k}\alpha}} \right).
\end{equation}
The only difference with the calculation in section \ref{Section:2} comes from the fact that electric-field perturbations are still made by charge-density perturbations, which are a cumulative effect of all the waterbags. In other words, Poisson's equation (\ref{eqn:S1:E6}) is 
\begin{equation}
\label{eqn:S4:E7}
\varphi_{\v{k}} = \sum_{\alpha}\frac{4\pi q_{\alpha}}{k^{2}}\int \dd{\v{v}}f_{\v{k}\alpha}(\v{v}) =\sum_{\alpha}\frac{4\pi q_{\alpha}}{k^{2}}\iint \dd{\v{v}}\dd{\eta} \eta P_{\v{k}\alpha}(\v{v},\eta).
\end{equation}

The derivation of the collision integral for $P_{0\alpha}$ can now be ported over from section~\ref{Section:2} near verbatim. The only difference is comes from Poisson's equation (\ref{eqn:S4:E7}): instead of velocity integrals, we now have integrals over both velocities and waterbag densities:
\begin{equation}
\label{eqn:S4:E8}
\int\dd{\v{v}}\big(...\big) \to \iint\dd{\v{v}}\dd{\eta} \eta\big(...\big).
\end{equation}
To avoid repetition, we will just note the key equations that are reached throughout the derivation, with reference to the corresponding equations in section~\ref{Section:2}. 

The dielectric function, previously (\ref{eqn:S1:E10}), is now 
\begin{equation}
\label{eqn:S4:E9}
\epsilon_{\v{k}}(p) = 1- i\sum_{\alpha'}\frac{4\pi q_{\alpha'}^{2}}{m_{\alpha'}k^{2}}\iint\dd{\v{v}'}\dd{\eta'}\frac{\eta'}{p+i\v{k}\cdot\v{v}'}\v{k}\cdot\pdev{P_{0\alpha'}}{\v{v}'}.
\end{equation}
The resultant general form of the collision integral, previously (\ref{eqn:S1:E18}), becomes
\begin{equation}
\label{eqn:S4:E10}
\pdev{P_{0\alpha}}{t} = \pdev{}{\v{v}}\cdot\sum_{\alpha''}\!\iint\!\!\dd{\v{v}''}\!\!\dd{\eta''}\eta'' \!\left[\mathsf{D}^{\alpha}_{\alpha\alpha''}(\v{v},\v{v}'',\eta'')\!\cdot\!\!\left.\pdev{P_{0\alpha}}{\v{v}}\right|_{\eta} - \mathsf{D}^{\alpha}_{\alpha''\alpha}(\v{v}'',\v{v},\eta)\!\cdot\!\!\left. \pdev{P_{0\alpha''}}{\v{v}''}\right|_{\eta''} \right].
\end{equation}
The diffusion kernel, via an expression analogous to (\ref{eqn:S1:E19}), can again be simplified by assuming that the fluctuations evolve much more rapidly than the mean: a calculation identical to that done in section \ref{Section:2.3} returns the following analogue of (\ref{eqn:S1:E24}):
\begin{multline}
\label{eqn:S4:E12}
\mathsf{D}^{\alpha}_{\mu\nu}(\v{w},\v{v},\eta) =  \sum_{\nu'}\frac{16\pi^{3}q_{\mu}^{2}q_{\nu}q_{\nu'}}{m_{\alpha}m_{\mu}}\\\mathrm{Re}\sum_{\v{k}}\frac{\v{k}\v{k}}{k^{4}}\delta(\v{k}\cdot(\v{w} - \v{v}))\iint\dd{\v{v}'}\dd{\eta'}\eta'e^{-i\v{k}\cdot(\v{v}-\v{v}')t}\frac{\crl{g_{\v{k}\nu}(\v{v},\eta)g^{*}_{\v{k}\nu'}(\v{v}',\eta')}}{\epsilon_{\v{k},\v{k}\cdot\v{v}}\epsilon^{*}_{\v{k},\v{k}\cdot\v{v}'}},
\end{multline}
where $g_{\v{k}\nu}(\v{v},\eta) = P_{\v{k}\nu}(t=0,\v{v},\eta)$ is the initial distribution and, as in section~\ref{Section:2},~{${\epsilon_{\v{k},\v{k}\cdot\v{v}} \equiv \epsilon_{\v{k}}(-i\v{k}\cdot\v{v})}$}. 

Now we need a closure for the correlation function $\crl{g_{\v{k}\nu}(\v{v},\eta)g^{*}_{\v{k}\nu'}(\v{v}',\eta')}$ in terms of~$P_{0\nu}(\v{v},\eta)$. The first step is again the microgranulation ansatz introduced in section~\ref{Section:2.5}, viz., the assumption that only very near points in phase space are correlated with each other, over a phase-space volume~$\delgam_{\nu}$:
\begin{equation}
\label{eqn:S4:E13}
\crl{g_{\nu}(\v{r},\v{v},\eta)g_{\nu'}(\v{r}',\v{v}',\eta')} = \delgam_{\nu}\delta_{\nu\nu'}\crl{g_{\nu}(\eta)g_{\nu}(\eta')}(\v{v})\delta(\v{v} - \v{v}')\delta(\v{r} - \v{r}'),
\end{equation} 
or, in Fourier space,
\begin{equation}
\label{eqn:S4:E14}
\crl{g_{\v{k}\nu}(\v{v},\eta)g^{*}_{\v{k}\nu'}(\v{v}',\eta')} = \frac{\delgam_{\nu}}{V}\delta_{\nu\nu'}\crl{g_{\nu}(\eta)g_{\nu}(\eta')}(\v{v})\delta(\v{v} - \v{v}').
\end{equation}
This is the generalisation of (\ref{eqn:S1:E32}). Note that, while localising the correlator in the $(\v{r},\v{v})$ space, we allow for correlations between different values of $\eta$. To determine this remaining correlator in terms of $P_{0\nu}$, we use (\ref{eqn:S4:E1}) to find  
\begin{equation}
\label{eqn:S4:E15}
\begin{split}
\crl{g_{\nu}(\eta)g_{\nu}(\eta')}(\v{v}) & = \crl{[\delta(f_{\nu}(\v{r},\v{v}) - \eta) - P_{0\nu}(\v{v},\eta)][\delta(f_{\nu}(\v{r},\v{v}) - \eta') - P_{0\nu}(\v{v},\eta')]} \\  &= \crl{\delta(f_{\nu}(\v{r},\v{v}) - \eta)\delta(f_{\nu}(\v{r},\v{v}) - \eta')} - P_{0\nu}(\v{v},\eta)P_{0\nu}(\v{v},\eta')\\ & = \delta(\eta - \eta')P_{0\nu}(\v{v},\eta) - P_{0\nu}(\v{v},\eta)P_{0\nu}(\v{v},\eta').
\end{split}  
\end{equation}
This formula is a direct generalisation of the single-waterbag closure (\ref{eqn:S3:E5}), but of course there is no longer any need to assume a single-waterbag distribution. Neither is there any need to assume a local maximisation of entropy, as we did for our multi-waterbag distribution in section \ref{Section:3.2}: working in the extend phase space $(\v{r},\v{v},\eta)$ has resolved the closure problem for $\crl{f_{\alpha}^{2}}$ vs. $f_{0\alpha}$ automatically.

Using (\ref{eqn:S4:E14}) and (\ref{eqn:S4:E15}), we can now write the diffusion kernel (\ref{eqn:S4:E12}) in a closed form:
\begin{equation}
\label{eqn:S4:E16}
  \mathsf{D}^{\alpha}_{\mu\nu}(\v{w},\v{v},\eta) =  \frac{16\pi^{3}q_{\mu}^{2}q_{\nu}^{2}\delgam_{\nu}}{m_{\alpha}m_{\mu}V} \sum_{\v{k}}\frac{\v{k}\v{k}}{k^{4}}\frac{\delta(\v{k}\cdot(\v{w}- \v{v}))}{|\epsilon_{\v{k},\v{k}\cdot\v{v}}|^{2}} \Big[\eta - f_{0\nu}(\v{v}) \Big]P_{0\nu}(\v{v},\eta),
\end{equation}
where $f_{0\nu}(\v{v}) = \int\dd{\eta'}\eta'P_{0\nu}(\v{v},\eta')$. Finally, substituting (\ref{eqn:S4:E16}) into (\ref{eqn:S4:E10}) gives us the generalised hyperkinetic collision integral:
\begin{multline}
\label{eqn:S4:E17}
\pdev{P_{0\alpha}}{t} = \sum_{\alpha''}\frac{16\pi^{3}q_{\alpha}^{2}q_{\alpha''}^{2}}{m_{\alpha}V}\pdev{}{\v{v}}\cdot\sum_{\v{k}}\frac{\v{k}\v{k}}{k^{4}}\cdot\int\dd{\v{v}''}\frac{\delta(\v{k}\cdot(\v{v}-\v{v}''))}{|\epsilon_{\v{k},\v{k}\cdot\v{v}}|^{2}}\int\dd{\eta''}\eta''\\ \Bigg\lbrace \frac{\delgam_{\alpha''}}{m_{\alpha}}\Big[ \eta'' - f_{0\alpha''}(\v{v}'') \Big]P_{0\alpha''}(\v{v}'',\eta'')\left.\pdev{P_{0\alpha}}{\v{v}}\right|_{\eta} - \frac{\delgam_{\alpha}}{m_{\alpha''}}\Big[ \eta - f_{0\alpha}(\v{v}) \Big]P_{0\alpha}(\v{v},\eta)\left.\pdev{P_{0\alpha''}}{\v{v}''}\right|_{\eta''}\Bigg\rbrace.
\end{multline}
To reiterate, the only closure required in the derivation of this collision integral was the microgranulation ansatz (\ref{eqn:S4:E13}). No need to break higher-order correlators has arisen, because, within the hyperkinetic formalism, all moments of the phase-space density can be derived from the distribution function $P_{0\alpha}$: $\crl{f_{\alpha}^{n}} = \int\dd{\eta}\eta^{n}P_{0\alpha}(\v{v},\eta)$. Indeed, if one takes the first moment of (\ref{eqn:S4:E17}) with respect to $\eta$, it is easy to show that one recovers~(\ref{eqn:S1:E35}), which was the general collision integral before a choice of waterbag closure was made in section \ref{Section:3}.

As well as conserving total energy, momentum, and particle number, (\ref{eqn:S4:E17}) has two further invariants, which represent the conservation of phase volume, 
\begin{equation}
\label{eqn:S4:E18}
\int \dd{\v{v}} P_{0\alpha}(\v{v},\eta) = \rho_{\alpha}(\eta),
\end{equation}
and the conservation of probability
\begin{equation}
\label{eqn:S4:E19}
\int\dd{\eta}P_{0\alpha}(\v{v},\eta) = 1.
\end{equation}
As in (\ref{eqn:S3:E33}), (\ref{eqn:S4:E18}) distils to a single function $\rho_{\alpha}(\eta)$ (the `waterbag content' of the distribution, or the infinite set of its Casimir invariants) all the information from the initial condition that must be preserved by collisionless evolution. Given these invariants, it is unsurprising that the fixed points of the collision integral (\ref{eqn:S4:E17}) will again be the Lynden-Bell equilibria, viz.,
\begin{equation}
\label{eqn:S4:E20}
P_{0\alpha}(\v{v},\eta) = \frac{e^{-\beta\delgam_{\alpha}\eta\left[\epsilon_{\alpha}  - \mu_{\alpha}(\eta)\right]}}{\int\dd{\eta}e^{-\beta\delgam_{\alpha}\eta\left[\epsilon_{\alpha}  - \mu_{\alpha}(\eta)\right]}},
\end{equation}
where $\epsilon_{\alpha} = m_{\alpha}|\v{v}|^{2}/2$, and the normalisation (\ref{eqn:S4:E19}) has been enforced. The parameters~$\beta$ and $\mu(\eta)$ are determined from (\ref{eqn:S4:E18}) and the energy conservation,
\begin{equation}
V\iint\dd{\v{v}}\dd{\eta}\eta \epsilon_{\alpha}(\v{v})P_{0\alpha}(\v{v},\eta) = E.
\end{equation}
That (\ref{eqn:S4:E20}) are indeed fixed points of (\ref{eqn:S4:E17}) can be confirmed by direct substitution. Just as in section \ref{Section:3.3.2}, we must provide an H-theorem to prove their stability. 
\subsection{Hyperkinetic H-theorem}
\label{Section:4.2} 
We define the Shannon entropy as before but now upgraded to its continuous variant~(\ref{eqn:S3:E34}). Taking the time derivative of $S$ using (\ref{eqn:S4:E17}) and integrating by parts in $\v{v}$, we get
\begin{align}
\label{eqn:S4:E22}
  \phantom{\dev{S}{t}}
  &\begin{aligned}
    \mathllap{\dev{S}{t}} = -\sum_{\alpha}\frac{V}{\delgam_{\alpha}}\iint\dd{\v{v}}\dd{\eta}\Big(1 + \ln P_{0\alpha} \Big)\pdev{P_{0\alpha}}{t}
  \end{aligned} \nonumber \\ 
  &\begin{aligned}
  \mathllap{} =  & \sum_{\alpha\alpha''}\frac{16\pi^{3}q_{\alpha}^{2}q_{\alpha''}^{2}}{\delgam_{\alpha}\delgam_{\alpha''}}\iint\dd{\v{v}}\dd{\v{v}''}\sum_{\v{k}}\frac{\delta(\v{k}\cdot(\v{v}-\v{v}''))}{k^{4}|\epsilon_{\v{k},\v{k}\cdot\v{v}}|^{2}}\\ & \iint\dd{\eta}\dd{\eta''}\eta'' \Bigg\lbrace \frac{\delgam_{\alpha''}^{2}}{m_{\alpha}^{2}}\Big[\eta''- f_{0\alpha''}(\v{v}'')\Big]\frac{P_{0\alpha''}(\v{v}'',\eta'')}{P_{0\alpha}(\v{v},\eta)}\left(\v{k}\cdot\left.\pdev{P_{0\alpha}}{\v{v}}\right|_{\eta} \right)^{2}  \\& - \frac{\delgam_{\alpha}\delgam_{\alpha''}}{m_{\alpha}m_{\alpha''}}\Big[\eta - f_{0\alpha}(\v{v})\Big]\v{k}\cdot\left. \pdev{P_{0\alpha}}{\v{v}} \right|_{\eta}\v{k}\cdot \left. \pdev{P_{0\alpha}}{\v{v}''}\right|_{\eta''} \Bigg\rbrace.
  \end{aligned}
\end{align}
First, we notice that the $\eta''$ integral of the first bracketed term is
\begin{equation}
\label{eqn:S4:E23}
\int \dd{\eta''}\eta'' \Big[\eta'' - f_{0\alpha''}(\v{v}'')\Big]P_{0\alpha''}(\v{v}'',\eta'') = \int\dd{\eta''}\Big[\eta'' - f_{0\alpha''}(\v{v}'') \Big]^{2}P_{0\alpha''}(\v{v}'',\eta'').
\end{equation}
Secondly, in the second bracketed term, anything that is not multiplied by both~$\eta$ and~$\eta''$ integrates to zero by (\ref{eqn:S4:E19}):
\begin{equation}
\label{eqn:S4:E24}
\int\dd{\eta}\left.\pdev{P_{0\alpha}}{\v{v}}\right|_{\eta} = \pdev{}{\v{v}}\int \dd{\eta}P_{0\alpha}(\v{v},\eta) = 0.
\end{equation}
With these insights, the rate of change of entropy becomes
\begin{align}
\label{eqn:S4:E25}
  \phantom{\dev{S}{t}}
  &\begin{aligned}
    \mathllap{\dev{S}{t}} = & \sum_{\alpha\alpha''}\frac{16\pi^{3}q_{\alpha}^{2}q_{\alpha''}^{2}}{\delgam_{\alpha}\delgam_{\alpha''}}\iint\dd{\v{v}}\dd{\v{v}''}\sum_{\v{k}}\frac{\delta(\v{k}\cdot(\v{v}-\v{v}''))}{k^{4}|\epsilon_{\v{k},\v{k}\cdot\v{v}}|^{2}}\\ & \iint\dd{\eta}\dd{\eta''} \Bigg\lbrace \frac{\delgam_{\alpha''}^{2}}{m_{\alpha}^{2}}\Big[\eta''- f_{0\alpha''}(\v{v}'')\Big]^{2}\frac{P_{0\alpha''}(\v{v}'',\eta'')}{P_{0\alpha}(\v{v},\eta)}\left(\v{k}\cdot\left.\pdev{P_{0\alpha}}{\v{v}}\right|_{\eta} \right)^{2} \\& - \frac{\delgam_{\alpha}\delgam_{\alpha''}}{m_{\alpha}m_{\alpha''}}\eta \eta''\v{k}\cdot\left. \pdev{P_{0\alpha}}{\v{v}} \right|_{\eta}\v{k}\cdot \left. \pdev{P_{0\alpha''}}{\v{v}''}\right|_{\eta''} \Bigg\rbrace.
  \end{aligned}
\end{align}
Finally, we symmetrise the entire expression by swapping $\alpha \leftrightarrow \alpha''$, $\eta \leftrightarrow \eta''$, $\v{v} \leftrightarrow \v{v}''$, which allows us to write (\ref{eqn:S4:E25}) in an explicitly positive-semidefinite form:
\begin{align}
\label{eqn:S4:E26}
\phantom{\dev{S}{t}}
  &\begin{aligned}
    \mathllap{\dev{S}{t}} = & \sum_{\alpha\alpha''}\frac{8\pi^{3}q_{\alpha}^{2}q_{\alpha''}^{2}}{\delgam_{\alpha}\delgam_{\alpha''}}\iint\dd{\v{v}}\dd{\v{v}''}\sum_{\v{k}}\frac{\delta(\v{k}\cdot(\v{v}-\v{v}''))}{k^{4}|\epsilon_{\v{k},\v{k}\cdot\v{v}}|^{2}}\\ & \iint\dd{\eta}\dd{\eta''}\Bigg\lbrace\frac{\delgam_{\alpha''}}{m_{\alpha}}\Big[\eta'' - f_{0\alpha''}(\v{v}'')\Big]\sqrt{\frac{P_{0\alpha''}(\v{v}'',\eta'')}{P_{0\alpha}(\v{v},\eta)}}\v{k}\cdot \left.\pdev{P_{0\alpha}}{\v{v}}\right|_{\eta}  \\ & -  \frac{\delgam_{\alpha}}{m_{\alpha''}}\Big[\eta - f_{0\alpha}(\v{v})\Big]\sqrt{\frac{P_{0\alpha}(\v{v},\eta)}{P_{0\alpha''}(\v{v}'',\eta'')}}\v{k}\cdot \left.\pdev{P_{0\alpha''}}{\v{v}''}\right|_{\eta''} \Bigg\rbrace^{2} \geq 0.
  \end{aligned}
\end{align}
Expanding the squared expression in (\ref{eqn:S4:E26}) does indeed recover (\ref{eqn:S4:E25}) as all excess terms cancel. The collision integral (\ref{eqn:S4:E17}) therefore never decreases the Shannon entropy~(\ref{eqn:S3:E34}). 

To prove that all initial conditions will eventually reach their corresponding Lynden-Bell equilibrium~(\ref{eqn:S4:E20}), we must show that the rate of entropy growth (\ref{eqn:S4:E26}) will equal zero if and only if the Lynden-Bell equilibrium has been reached. Owing to the squared expression in (\ref{eqn:S4:E26}), the entropy growth will equal zero when 
\begin{multline}
\label{eqn:S4:E27}
\frac{1}{\delgam_{\alpha}m_{\alpha}\Big[\eta - f_{0\alpha}(\v{v}) \Big]P_{0\alpha}(\v{v},\eta)}\v{k}\cdot\left.\pdev{P_{0\alpha}}{\v{v}}\right|_{\eta} =\\ \frac{1}{\delgam_{\alpha''}m_{\alpha''}\Big[\eta'' - f_{0\alpha''}(\v{v}'') \Big]P_{0\alpha''}(\v{v}'',\eta'')}\v{k}\cdot\left.\pdev{P_{0\alpha''}}{\v{v}''}\right|_{\eta''},
\end{multline} 
for all $\alpha$, $\alpha''$, $\eta$ and $\eta''$ when $\v{k}\cdot\left(\v{v}- \v{v}'' \right) = 0$. Save for $\v{v}$ and $\v{v}''$, the left- and right-hand sides of (\ref{eqn:S4:E27}) are functions of distinct, independent variables. The only solution to (\ref{eqn:S4:E27}) must, therefore, satisfy
\begin{equation}
\label{eqn:S4:E28}
\v{k}\cdot\left.\pdev{P_{0\alpha}}{\v{v}}\right|_{\eta} = -\beta\delgam_{\alpha}m_{\alpha}\Big[\eta - f_{0\alpha}(\v{v}) \Big]P_{0\alpha}(\v{v},\eta)\v{k}\cdot\v{v},
\end{equation}
where $\beta$ is an as yet undetermined constant, although it is clear it will come to mean the thermodynamic beta shortly. Note that by writing $\v{k}\cdot\v{v}$ we have assumed, without loss of generality, that we are in the zero-net-momentum frame. It is clear that the Lynden-Bell equilibria (\ref{eqn:S4:E20}) satisfy the condition (\ref{eqn:S4:E28}) [as they must, being maximisers of the entropy (\ref{eqn:S3:E34})]. 

To prove that these are the only solutions, we rewrite (\ref{eqn:S4:E28}) as
\begin{equation}
\label{eqn:S4:E29}
\left.\pdev{\ln P_{0\alpha}}{\v{v}}\right|_{\eta} = -\beta\delgam_{\alpha}\eta m_{\alpha}\v{v} + \beta m_{\alpha}\delgam_{\alpha}\int\dd{\eta}\eta P_{0\alpha}(\v{v},\eta)\v{v}.
\end{equation}
Then, without loss of generality, 
\begin{equation}
\label{eqn:S4:E30}
P_{0\alpha}(\v{v},\eta) = \frac{C_{\alpha}(\v{v},\eta)}{\int \dd{\eta}C_{\alpha}(\v{v},\eta)},
\end{equation}
for which (\ref{eqn:S4:E29}) becomes
\begin{equation}
\label{eqn:S4:E31}
\begin{split}
\left.\pdev{\ln C_{\alpha}}{\v{v}}\right|_{\v{v}} + \beta \delgam_{\alpha}\eta m_{\alpha}\v{v} & = \beta\delgam_{\alpha}m_{\alpha}\frac{\int\dd{\eta}\eta C_{\alpha}(\v{v},\eta)}{\int\dd{\eta}C_{\alpha}(\v{v},\eta)}\v{v} + \pdev{}{\v{v}}\ln\int\dd{\eta}C_{\alpha}(\v{v},\eta)\v{v}\\ & = \pdev{}{\v{v}}\Phi_{\alpha}(\v{v}).
\end{split}
\end{equation}
In (\ref{eqn:S4:E31}), we have collected all $\eta$-dependent terms on the right-hand side and further declared that, since the curl of the left-hand is zero, these terms can be written as the gradient of some function $\Phi_{\alpha}(\v{v})$, which will be determined self-consistently after determining $C_{\alpha}(\v{v})$. With this sleight of hand, (\ref{eqn:S4:E31}) is solved by 
\begin{equation}
\label{eqn:S4:E32}
C_{\alpha}(\v{v},\eta)  = e^{\Phi_{\alpha}(\v{v})}e^{-\beta\delgam_{\alpha}\eta\left[\epsilon_{\alpha}(\v{v}) -  \mu_{\alpha}(\eta)\right]},
\end{equation}
where $\epsilon_{\alpha}(\v{v}) = m_{\alpha}|\v{v}|^{2}/2$ and the chemical potential $\mu_{\alpha}(\eta)$ has emerged as an integration constant. Now, substituting this into (\ref{eqn:S4:E30}), we find that $P_{0\alpha}(\v{v},\eta)$ is the Lynden-Bell equilibrium (\ref{eqn:S4:E20}), proving that it is the only solution for which the entropy does not grow. Note that the determination of $\Phi_{\alpha}(\v{v})$ is unimportant for the calculation of $P_{0\alpha}(\v{v},\eta)$ because $C_{\alpha}(\v{v})$ can be freely multiplied by any function of $\v{v}$ without changing $P_{0\alpha}(\v{v},\eta)$; also, given (\ref{eqn:S4:E32}), the solvability condition 
\begin{equation}
\pdev{}{\v{v}}\Phi_{\alpha}(\v{v}) = \beta\delgam_{\alpha}m_{\alpha}\frac{\int\dd{\eta}\eta C_{\alpha}(\v{v},\eta)}{\int\dd{\eta}C_{\alpha}(\v{v},\eta)}\v{v} + \pdev{}{\v{v}}\ln\int\dd{\eta}C_{\alpha}(\v{v},\eta)\v{v}
\end{equation}
is satisfied for all functions $\Phi_{\alpha}(\v{v})$. This completes the proof that all initial conditions will reach their Lynden-Bell equilibria.

\subsection{Relation between multi-waterbag and hyperkinetic collision integrals}
\label{Section:5}
In this section, we aim to draw a relation between the two collision integrals derived above: the multi-waterbag collision integral~(\ref{eqn:S3:E18}) and the hyperkinetic collision integral~(\ref{eqn:S4:E17}). The apparent distinction between these collision integrals is the need for an artificial closure in the multi-waterbag collision integral that is not present in the hyperkinetic collision integral. We will begin to shed light on this by stating the following exact relation between the two collision integrals:

\textit{1) The hyperkinetic collision integral recovers the multi-waterbag collision integral if the waterbag distribution function $P_{0\alpha}(\v{v},\eta)$ satisfies the closure relation (\ref{eqn:S3:E30})}. To prove this, we observe, as we did already in section (\ref{Section:4.1}), that the first moment of the hyperkinetic collision integral (\ref{eqn:S4:E17}) with respect to $\eta$ is (\ref{eqn:S1:E35}) with $\crl{g_{\alpha}^{2}}(\v{v})$ given by 
\begin{equation}
\label{eqn:S5:E2}
\crl{g_{\alpha}^{2}}(\v{v}) = \int\!\!\dd{\eta}\Big[\eta - f_{0\alpha}(\v{v})\Big]^{2}P_{0\alpha}(\v{v},\eta) = \int \!\!\dd{\eta} \eta^{2}P_{0\alpha}(\v{v},\eta) - \left[\int\!\! \dd{\eta} \eta P_{0\alpha}(\v{v},\eta) \right]^{2}.
\end{equation}
Substituting $P_{0\alpha}(\v{v},\eta)$ from (\ref{eqn:S3:E30}) into (\ref{eqn:S5:E2}) gives 
\begin{equation}
\crl{g_{\alpha}^{2}}(\v{v}) = \frac{1}{Z_{\alpha}(\psi_{\alpha})}\int\dd{\eta}\eta^{2}e^{-\psi_{\alpha}(\v{v})\eta - \gamma_{\alpha}(\eta)}  - f_{0\alpha}^{2}(\v{v}),
\end{equation}  
which is the continuous version of (\ref{eqn:S3:E17}) used to arrive at the multi-waterbag collision integral. Therefore, if the waterbag distribution function satisfies (\ref{eqn:S3:E30}), the hyperkinetic collision integral reduces to the multi-waterbag collision integral. However, this is not an automatic guarantee that the waterbag distribution function $P_{0\alpha}$ will `stay on the closure'. In the event, however, it is possible to prove that it will. 

\textit{2) If the waterbag distribution function $P_{0\alpha}(\v{v},\eta)$ evolving under the hyperkinetic collision integral (\ref{eqn:S4:E17}) ever satisfies the closure (\ref{eqn:S3:E30}), then it will continue to satisfy it for all future times.} For a given waterbag distribution function $P_{0\alpha}(\v{v},\eta)$, let us define~$\bar{P}_{0\alpha}(\v{v},\eta)$ to be the waterbag distribution function that maximises the entropy~(\ref{eqn:S3:E34}) subject to having the same mean phase-space density~$f_{0\alpha}(\v{v})$ and waterbag content~$\rho_{\alpha}(\eta)$ as~$P_{0\alpha}(\v{v},\eta)$, i.e.,~$\bar{P}_{0\alpha}(\v{v},\eta)$ is the waterbag distribution function~(\ref{eqn:S3:E30}). Now, we examine the difference between the entropies of~$\bar{P}_{0\alpha}(\v{v},\eta)$ and~$P_{0\alpha}(\v{v},\eta)$:
\begin{equation}
\label{eqn:S5:E3}
\bar{S} - S = \sum_{\alpha}\frac{V}{\delgam_{\alpha}}\iint\dd{\v{v}}\dd{\eta} \Big(P_{0\alpha}\ln P_{0\alpha} - \bar{P}_{0\alpha}\ln\bar{P}_{0\alpha}\Big).
\end{equation}
We first note that this relative entropy is, by definition of $\bar{P}_{0\alpha}$, non-negative and only zero if $P_{0\alpha} = \bar{P}_{0\alpha}$ for all $\alpha$. Furthermore, since $\bar{P}_{0\alpha}$ is determined by $P_{0\alpha}$ and $\bar{S}$ by $\bar{P}_{0\alpha}$, we may calculate this relative entropy at each time, and take its time derivative. Taking the time derivative of $S$ is straightforward given the collision integral (\ref{eqn:S4:E17}). The time derivative of $\bar{S}$ can be determined in terms of only $\psi_{\alpha}$ and $\partial f_{0\alpha}/ \partial t$, as the entropy has the form (\ref{eqn:S3:E25}), the only difference being that the evolution of $f_{0\alpha}$ is now governed by the hyperkinetic collision integral, not the multi-waterbag one. Using (\ref{eqn:S3:E27}), we may, therefore, write the evolution of the relative entropy as 
\begin{equation}
\label{eqn:S5:E4}
\begin{split}
\dev{}{t}\left(\bar{S} - S \right) & = \sum_{\alpha}\frac{V}{\delgam_{\alpha}}\int\dd{\v{v}}\Bigg[\psi_{\alpha}\pdev{f_{0\alpha}}{t} + \int\dd{\eta}\left(1 + \ln P_{0\alpha} \right)\pdev{P_{0\alpha}}{t}\Bigg]\\ & = \sum_{\alpha}\frac{V}{\delgam_{\alpha}}\iint\dd{\v{v}}\dd{\eta}  \Big(\eta\psi_{\alpha} + \ln P_{0\alpha}\Big)\pdev{P_{0\alpha}}{t}.
\end{split}
\end{equation}
Consequently, when the waterbag distribution function satisfies the closure (\ref{eqn:S3:E30}), the relative entropy will be zero by definition of $\bar{P}_{0\alpha}$ and, by (\ref{eqn:S5:E4}), its time derivative will also be zero, so the relative entropy will stay zero for all future times. Since the relative entropy is zero only for $P_{0\alpha} = \bar{P}_{0\alpha}$ this proves our second statement. Therefore, it is possible to construct initial conditions for which the multi-waterbag collision integral~(\ref{eqn:S3:E18}) correctly describes the evolution due to the hyperkinetic collision integral for all future times. 

Given that the distribution function $P_{0\alpha}(\v{v},\eta)$ remains on the closure if it begins exactly on the closure, it is natural to ask if the evolution of the distribution function is, in fact, forced towards the closure by the hyperkinetic collision integral. If this was true, it would make the multi-waterbag collision integral (\ref{eqn:S3:E18}) a valid approximation for the hyperkinetic collision integral after an initial transient period. However, for this to be true, there would have to be two timescales hidden within the hyperkinetic collision operator: a shorter timescale on which the hyperkinetic distribution function approached the closure and a longer timescale over which the closure evolved according to the multi-waterbag collision integral. It is trivially clear that this separation of timescales exists in a one-dimensional system, since the 1D hyperkinetic collision integral forbids the mean phase-space density to change. In this case, the only evolution is towards satisfying the closure (\ref{eqn:S3:E30}). In a higher number of dimensions, it may therefore be useful to consider two types of collisions: those which do and do not alter the mean phase-space density~$f_{0\alpha}(\v{v})$. This artificial divide would then describe two processes by which the system raises its entropy: one by making $f_{0\alpha}(\v{v})$ approach a Lynden-Bell equilibrium and the other by reordering waterbags without altering $f_{0\alpha}(\v{v})$. The latter process is one in which entropy is maximised locally in phase space (i.e., for each $\v{v}$). A useful analogy might be the conventional collisional dynamics of gases, which relax quickly, at the collision rate, to a local Maxwellian, and slowly, at the diffusion rate, to the global one.

\section{Collisionless vs. collisional relaxation}
\label{Section:6}
Thus far we have not discussed the absence of true collisions within this formalism. In this context, `true collisions' are any  type of relaxation to equilibrium that does not conserve phase volume and thus releases the stranglehold that the invariants (\ref{eqn:S3:E1}) had on the evolution. It is for this reason that \cite{LyndenBell67} originally proposed the idea of a violent relaxation: so that steady states could be reached long before the conservation of phase volume was broken. The collision integrals derived above do not describe such a violent, highly nonlinear, regime but rather a quasilinear one, where the mean distribution function evolves slowly compared to the fluctuations. Nevertheless, for our collision integrals to be valid, the rate of relaxation due to them must be much greater than the rate of relaxation due to true collisions. We shall estimate the collisionless relaxation rate in a moment, but first let us show how to recover the collision integral of \cite{Balescu60} and \cite{Lenard60} from the Kadomtsev-Pogutse collision integral (\ref{eqn:S3:E6}), in order to have a `true collisionality' with which to compare our `collisionless collision rate'. 
\subsection{Balescu-Lenard collision integral}
\label{Section:BL}
In reality, a plasma is not a phase fluid but a collection of $N$ particles. The true, exact phase-space density of these particles is the Klimontovich distribution (see, e.g., \citealt{KlimontovichBook}):
\begin{equation}
f(\v{r},\v{v}) = \sum_{i=1}^{N}\delta(\v{r}-\v{r}_{i})\delta(\v{v}-\v{v}_{i}),
\end{equation}
where $\v{r}_{i}$ and $\v{v}_{i}$ are the particles' instantaneous positions and velocities, respectively. Since each particle thus occupies precisely zero phase volume, it might seem as though phase-volume conservation were a meaningless idea. The reason one can talk about phase-volume conservation at all is that one assumes that particles that are neighbours in phase space move in a similar way, implying that replacing the Klimontovich distribution with a smoothed phase-space density is a reasonable approximation. It is then the phase volumes associated with this smoothed function that are conserved. This is closely related to the microgranulation ansatz (\ref{eqn:S1:E32}), which posits that, within a phase volume $\delgam_{\alpha}$, the fluctuations of the phase-space density are correlated, implying that particles are moving collectively. A `true collision' occurs when a single particle is not correlated at all with its neighbouring particles. This can be accommodated within the microgranulation ansatz by assuming that each particle is its own waterbag. Mathematically this is just the single-waterbag model of section \ref{Section:3.1} in the limit where the correlation volume $\delgam_{\alpha}$ is made smaller that the inter-particle separation in phase space. Then, by assumption there is only one particle in the correlation volume, so 
\begin{equation}
\label{eqn:S6:E2}
\delgam_{\alpha}\eta_{\alpha} = 1. 
\end{equation}
The smoothed distribution function is given by the average occurrence of single particles, hence $f_{0\alpha}(\v{v}) \ll \eta_{\alpha}$. Under these assumptions, the Kadomtsev-Pogutse collision integral~(\ref{eqn:S3:E6}) becomes the Balescu-Lenard collision integral:
\begin{multline}
\label{eqn:S6:E1}
\pdev{f_{0\alpha}}{t} =  \sum_{\alpha''}\frac{16\pi^{3}q_{\alpha}^{2}q_{\alpha''}^{2}}{m_{\alpha}V}\pdev{}{\v{v}}\cdot\sum_{\v{k}}\frac{\v{k}\v{k}}{k^{4}}\cdot\int\dd{\v{v}''}\frac{\delta(\v{k}\cdot(\v{v}-\v{v}''))}{\left|\epsilon_{\v{k},\v{k}\cdot\v{v}} \right|^{2}}\\ \Bigg[\frac{1}{m_{\alpha}}f_{0\alpha''}(\v{v}'')\pdev{f_{0\alpha}}{\v{v}} - \frac{1}{m_{\alpha''}}f_{0\alpha}(\v{v})\pdev{f_{0\alpha''}}{\v{v}''}\Bigg].
\end{multline}
This reduction allows one to make a useful comparison between the `true' collision rate, which is the typical collision rate associated with the Balescu-Lenard integral (\ref{eqn:S6:E1}), and the effective collision rate associated with the hyperkinetic `collisionless collision integral'~(\ref{eqn:S4:E17}). Before proceeding to do that, we observe in passing that the formalism developed in sections~\ref{Section:2} and~\ref{Section:3} also allows one to derive very efficiently true collision operators for quantum plasmas consisting of fermionic and bosonic species. This is done in appendix \ref{Section:Quantum relaxation}.

\subsection{Effective collision rates}
\label{Section:6.1}
To keep this discussion as transparent as possible, let us consider only the like-particle effective collision rates. Consider the generic quasilinear collision integral (\ref{eqn:S1:E35}), with~$\alpha'' = \alpha$ and 
\begin{equation}
\label{eqn:Correlation ordering}
\crl{g_{\alpha}^{2}}(\v{v}) = \int \dd{\eta}\left[\eta - f_{0\alpha}(\v{v}) \right]^{2} P_{0\alpha}(\v{v},\eta), \quad \quad f_{0\alpha}(\v{v}) = \int\dd{\eta}\eta P_{0\alpha}(\v{v},\eta),
\end{equation}
where $P_{0\alpha}(\v{v},\eta)$ is evolved by (\ref{eqn:S4:E17}). This integral should be compared to its Balescu-Lenard counterpart (\ref{eqn:S6:E1}), also with $\alpha'' = \alpha$. It is immediately apparent that the difference in the rates of the true and effective collisions is
\begin{equation}
\label{eqn:S6:E8}
\frac{\nu_{\alpha\alpha}^{\mathrm{eff}}}{\nu_{\alpha\alpha}^{\mathrm{true}}} \sim \frac{\delgam_{\alpha}\crl{g_{\alpha}^{2}}}{f_{0\alpha}} \equiv \delgam_{\alpha}\eta_{\alpha}^{\mathrm{eff}},
\end{equation} 
where $\eta_{\alpha}^{\mathrm{eff}}$ is the typical deviation of the phase-space density from its mean. The quantity on the right-hand side of (\ref{eqn:S6:E8}) is the typical variation in the number of particles contained in the correlation volume $\delgam_{\alpha}$. Since this collisionless theory is built upon the assumption that the correlation volume is sufficiently large for its mean phase-space density $\eta$ to be meaningfully specified, we have inherently assumed that the number of particles contained in a correlation volume is large. However, this does not tell us immediately about the typical deviation of this number from its mean. We therefore aim to compare $\eta_{\alpha}^{\mathrm{eff}}$ to the typical value of the phase-space density,~$n_{\alpha}/v_{\mathrm{th}\alpha}^{3}$. To make this comparison quantitative, we estimate~(\ref{eqn:S6:E8}) close to a Lynden-Bell equilibrium. In this case, using~(\ref{eqn:S3:E17}) with~{${\psi_{\alpha} = \beta \delgam_{\alpha}\epsilon_{\alpha}}$}, we find 
\begin{equation}
\label{eqn:S6:Eta scale}
\delgam_{\alpha}\eta_{\alpha}^{\mathrm{eff}} = -\frac{1}{\beta}\pdev{\ln f_{0\alpha}}{\epsilon_{\alpha}} \sim \frac{1}{\beta m_{\alpha} v_{\mathrm{th}\alpha}^{2}},
\end{equation}
where $v_{\mathrm{th}\alpha}$ is the typical velocity scale of~$f_{0\mathrm{\alpha}}$. Since~$\beta$ must be determined from the constraints of energy conservation~(\ref{eqn:S1:E29}) and phase-volume conservation~(\ref{eqn:S3:E33}),~(\ref{eqn:S6:Eta scale}) provides an implicit expression for the~$\eta_{\alpha}^{\mathrm{eff}}$ in terms of~$E$ and~$\rho_{\alpha}(\eta)$. For the purposes of order-of-magnitude estimates, the exact calculation of $\beta$ is unnecessary and it will suffice to consider two relevant limits. In the degenerate case, when the initial condition is very close to the minimum-energy state (the Gardner state; see section~\ref{Section:3.3.1}),~$\beta$ will be very large and can be computed by a Sommerfeld-like expansion. In this case, naively ordering~{${\eta_{\mathrm{\alpha}}^{\mathrm{eff}} \sim n_{\alpha}/v_{\mathrm{th}\alpha}^{3}}$} would be an overestimation. This is because deviations from the mean phase-space density require the system to have sufficient energy to allow two species of waterbag to intermingle (see discussion around~\ref{Section:3.3.1}). Thankfully, as discussed in section~\ref{Section:3.3.1}, the system does not need to be energetically very far from its Gardner distribution to become, at least partially, non-degenerate. In the non-degenerate limit, which is by far the most common, waterbags of different phase-space density can freely intermingle, and so we can estimate~$\eta_{\mathrm{\alpha}}^{\mathrm{eff}} \sim n_{\alpha}/v_{\mathrm{th}\alpha}^{3}$, implying~$\delgam_{\alpha}\eta_{\alpha}^{\mathrm{eff}}$ is on the order of the number of particles in a correlation volume, which we have assumed to be large.

Thus, the effective collision rate is generally much larger than the true-collision rate. Note however, that this fast collisionless relaxation must still be slower that the evolution of the fluctuations. The most straightforward estimate of the typical rate of the latter is the plasma frequency,~$\omega_{\mathrm{pe}} = (4\pi e^{2}n_{\mathrm{e}}/m_{\mathrm{e}})^{1/2}$ (specialising to electrons for the purposes of this estimate). Thus, we require (and expect) the ordering
\begin{equation}
\label{eqn:ordering requirement}
\nu_{\mathrm{ee}}^{\mathrm{true}} \sim \frac{\omega_{\mathrm{pe}}}{n_{\mathrm{e}}\lambda_{\mathrm{De}}^{3}} \ll \nu_{\mathrm{ee}}^{\mathrm{eff}} \ll \omega_{\mathrm{pe}},
\end{equation}
where $\lambda_{\mathrm{De}}\sim \omega_{\mathrm{pe}}v_{\mathrm{the}}$ is the Debye length. This places a constraint on the correlation volume:
\begin{equation}
\label{eqn:resulting constraint}
1 \ll \delgam_{\mathrm{e}}\eta_{\mathrm{e}}^{\mathrm{eff}} \ll n_{\mathrm{e}}\lambda_{\mathrm{De}}^{3}.
\end{equation}
If the plasma parameter $n_{\mathrm{e}}\lambda_{\mathrm{De}}^{3}$ is large (i.e., if the ideal-gas approximation applies), this constraint is not very stringent and still allows the effective collision frequency to be much larger than $\nu_{\mathrm{ee}}^{\mathrm{true}}$. 

A more stringent constraint emerges if one works out the time $\tau_{\mathrm{c}}$ that it takes for exact phase-volume conservation to be broken. This time-scale is substantially less well understood as it depends on the exact rate at which the fluctuating part $\delta f_{\alpha}$ of the exact phase-space density is altered irrevocably by collisions. A typical estimate (see, e.g., \citealt{SuOberman1968}) is 
\begin{equation}
\label{eqn:S6:E9}
\tau_{\mathrm{c}}^{-1} \sim (\nu_{ee}^\mathrm{true})^{1/3}\omega_{\mathrm{pe}}^{2/3} \sim \frac{\omega_{\mathrm{pe}}}{(n_{\mathrm{e}}\lambda_{\mathrm{De}}^{3})^{1/3}}.
\end{equation}
We may then argue that, for the collisionless relaxation to the Lynden-Bell equilibria to be of any importance it must happen long before phase-volume conservation is broken:~(\ref{eqn:resulting constraint}) is then revised to 
\begin{equation}
\label{eqn:upgraded constraint}
(n_{\mathrm{e}}\lambda_{\mathrm{De}}^{3})^{2/3}\ll \delgam_{\mathrm{e}}\eta_{\mathrm{e}}^{\mathrm{eff}} \ll n_{\mathrm{e}}\lambda_{\mathrm{De}}^{3}.
\end{equation}  
Depending on just how large the plasma parameter is, this could be a more difficult ordering to satisfy. However, if satisfied, it would make the collisionless relaxation rate far larger than the rate of relaxation due to `true' collisions. To know just how much larger, we must be able to calculate $\delgam_{\alpha}$ independently. Obviously this, along with a quantitative assessment of the validity of the microgranulation ansatz, requires a full theory of the two-point correlation function of the phase-space density. Without such a theory, certain revealing estimates can, nevertheless, be made.
\subsection{Energy of fluctuations and the correlation volume}
In all the above, the major shortcoming of the theory is the lack of clarity about the size of $\delgam_{\alpha}$. In order to relate $\delgam_{\alpha}$ to something measurable within a plasma, let us calculate the energy $E_{\varphi}$ stored in the electric-field fluctuations under the microgranulation ansatz~(\ref{eqn:S4:E14}). This energy is given by 
\begin{equation}
\label{eqn:Electric field energy}
E_{\varphi} = \int\dd{\v{r}}\frac{|\nabla\varphi|^{2}}{8\pi} = \frac{1}{2}V\sum_{\alpha}q_{\alpha}\sum_{\v{k}}\iint\dd{\v{v}}\dd{\eta}\eta\mathrm{Re}\crl{\varphi_{\v{k}}^{*}P_{\v{k}\alpha}(\v{v},\eta)}.
\end{equation}
where we have integrated by parts and used Poisson's equation (\ref{eqn:S4:E7}) in the second equality. Only the real part has survived because the summation in $\v{k}$ is even. While one could again apply Poisson's equation to the remaining $\varphi_{\v{k}}$ in (\ref{eqn:Electric field energy}) and then naively utilise the microgranlation ansatz (\ref{eqn:S4:E14}), it is worth noting that the correlator in (\ref{eqn:Electric field energy}) is the real part of the same correlator the imaginary part of which appeared in (\ref{eqn:S4:E5}). The real part of this correlator is equivalent to the hyperkinetic generalisation of the discarded term of~(\ref{eqn:S1:E16}). Recovering the hyperkinetic generalisation of this term, from~(\ref{eqn:S1:E16}) via~(\ref{eqn:S1:E37}), we get
\begin{multline}
\mathrm{Re}\crl{\varphi_{\v{k}}^{*}P_{\v{k}\alpha}(\v{v},\eta)} = \mathrm{Re}\sum_{\alpha}\frac{4\pi q_{\alpha'}}{k^{2}}\iint \frac{\dd{p}\dd{p'}}{(2\pi)^{2}}\frac{e^{(p+p')t}}{\epsilon_{\v{k}}(p)\epsilon_{\v{k}}^{*}(p'^{*})} \\ \times\iint\dd{\v{v}'}\dd{\eta'}\eta' \frac{\crl{g_{\v{k}\alpha}(\v{v},\eta)g_{\v{k}\alpha'}^{*}(\v{v}',\eta')}}{(p+i\v{k}\cdot\v{v})(p'-i\v{k}\cdot\v{v}')},
\end{multline}
with the understanding that, after the microgranulation ansatz, the real part of the other contributions will vanish. Carrying out the $p$ and $p'$ contour integration (see section \ref{Section:2.3} for details) and applying the microgranulation ansatz (\ref{eqn:S4:E14}), we find 
\begin{equation}
\label{eqn:Correlator real part}
\mathrm{Re}\crl{\varphi_{\v{k}}^{*}P_{\v{k}\alpha}(\v{v},\eta)} = \frac{4\pi q_{\alpha}\delgam_{\alpha}}{k^{2}V|\epsilon_{\v{k},\v{k}\cdot\v{v}}|^{2}}\int \dd{\eta'}\eta' \crl{g_{\v{k}\alpha}(\v{v},\eta)g_{\v{k}\alpha}^{*}(\v{v},\eta')}.
\end{equation}
Substituting (\ref{eqn:Correlator real part}) into (\ref{eqn:Electric field energy}) and using the formula (\ref{eqn:S4:E15}) for the correlation function of $g_{\v{k}\alpha}(\v{v},\eta)$, we get
\begin{align}
\label{eqn:Electric field energy 2}
\phantom{E_{\varphi}}
&\begin{aligned}
	\mathllap{E_{\varphi}} = \sum_{\alpha}\sum_{\v{k}}\frac{2\pi q_{\alpha}^{2}\delgam_{\alpha}}{k^{2}}\int\dd{\v{v}}\frac{1}{{|\epsilon_{\v{k},\v{k}\cdot\v{v}}|^{2}}}\iint\dd{\eta}\dd{\eta'}\eta\eta' & \Big[\delta(\eta-\eta')P_{0\alpha}(\v{v},\eta)\\ & \quad - P_{0\alpha}(\v{v},\eta)P_{0\alpha}(\v{v},\eta') \Big]
\end{aligned} \nonumber \\
&\begin{aligned}
	\mathllap{} = \sum_{\alpha}\sum_{\v{k}}\frac{2\pi q_{\alpha}^{2}\delgam_{\alpha}}{k^{2}}\int\dd{\v{v}}\frac{\crl{g_{\alpha}^{2}}(\v{v})}{|\epsilon_{\v{k},\v{k}\cdot\v{v}}|^{2}},
\end{aligned}
\end{align}
where $\crl{g_{\alpha}^{2}}(\v{v})$ is given by~(\ref{eqn:Correlation ordering}). This formula describes the energy in the fluctuating electric field for a plasma obeying the microgranulation ansatz.  

In a Lynden-Bell equilibrium, (\ref{eqn:Electric field energy 2}) can be further simplified by calculating $\delgam_{\alpha}\crl{g_{\alpha}^{2}}$ via~(\ref{eqn:S6:E8}) and~(\ref{eqn:S6:Eta scale}). Approximating also~{${|\epsilon_{\v{k},\v{k}\cdot\v{v}}|^{2}\approx 1}$}, we have
\begin{equation}
\label{eqn:star estimate}
E_{\varphi} \approx  \sum_{\alpha}\sum_{\v{k}}\frac{2\pi q_{\alpha}^{2}}{k^{2}}\int \dd{\v{v}}\left(-\frac{1}{\beta}\pdev{f_{0\alpha}}{\epsilon_{\alpha}} \right)  = \sum_{\alpha}\frac{4Vq_{\alpha}^{2}k_{\mathrm{max}}}{\beta m_{\alpha}}\int_{0}^{\infty} \dd{v}f_{0\alpha}(v),
\end{equation}
where~$k_{\mathrm{max}}$ is the UV cutoff for the wavenumber integal (in 3D). Using~(\ref{eqn:S6:Eta scale}) and making a rough estimate of everything, we have
\begin{equation}
\label{eqn:starstar estimate}
E_{\varphi} \sim \sum_{\alpha}\delgam_{\alpha}\eta_{\alpha}^{\mathrm{eff}}q_{\alpha}^{2}n_{\alpha}k_{\mathrm{max}}V.
\end{equation}

Let us compare these estimates with the energy stored in electric fluctuations associated with true, Coulomb collisions (i.e., with discrete particle noise). The latter can be calculated in a manner analogous to the above but using the Balescu-Lenard collision operator~(\ref{eqn:S6:E1}). To do this, we set
\begin{equation}
\label{eqn:Single waterbag limit}
P_{0\alpha}(\v{v},\eta) = \left[1-\frac{f_{0\alpha}(\v{v})}{\eta_{\alpha}}\right]\delta(\eta) + \frac{f_{0\alpha}(\v{v})}{\eta_{\alpha}}\delta(\eta - \eta_{\alpha}),
\end{equation}
where $\eta_{\alpha} = \delgam_{\alpha}^{-1}$ according to~(\ref{eqn:S6:E2}). Inserting~(\ref{eqn:Single waterbag limit}) into~(\ref{eqn:Electric field energy 2}) and collecting only the lowest-order terms in the limit of $f_{0\alpha}/\eta_{\alpha} \to 0$ gives
\begin{equation}
\label{eqn:Electric field energy 3}
E_{\varphi,\mathrm{BL}} = \sum_{\alpha}\sum_{\v{k}}\frac{2\pi q_{\alpha}^{2}}{k^{2}}\int\dd{\v{v}}\frac{f_{0\alpha}(\v{v})}{|\epsilon_{\v{k},\v{k}\cdot\v{v}}|^{2}}\approx \sum_{\alpha}4V q_{\alpha}^{2}k_{\mathrm{max}}n_{\alpha},
\end{equation}
where the last estimate has been obtained in the same manner as~(\ref{eqn:star estimate}). From a direct comparison of~(\ref{eqn:Electric field energy 2}) and~(\ref{eqn:starstar estimate}) with~(\ref{eqn:Electric field energy 3}), it is clear that, both~$\v{k}$ by~$\v{k}$ and overall, a plasma obeying the microgranulation ansatz has more energy stored in the electric field than a collisional plasma, by a factor of~$\delgam_{\alpha}\eta_{\alpha}^{\mathrm{eff}} \gg 1$, the typical number of particles in a correlation volume.\footnote{As with the estimates of the collision timescale, we do not expect this factor to be large for distributions sufficiently close to their ground states (Gardner distributions) when~{${\eta_{\alpha}^{\mathrm{eff}} \ll n_{\alpha}/v_{\mathrm{th}\alpha}^{3}}$}, but such systems will barely evolve anyway.}

Note that, despite the fluctuation energy being large compared to particle noise, it will still be small compared to the kinetic energy of the particles
\begin{equation}
K = V\sum_{\alpha}\int\dd{\v{v}}\frac{m_{\alpha}|\v{v}|^{2}}{2}f_{0\alpha} \sim V\sum_{\alpha}m_{\alpha}v_{\mathrm{th}\alpha}^{2}n_{\alpha}.
\end{equation}
Therefore, using~(\ref{eqn:starstar estimate}), we find
\begin{equation}
\label{eqn:Electric field comparison}
\frac{E_{\varphi}}{K} \sim \frac{\sum_{\alpha}\delgam_{\alpha}^{\mathrm{eff}}\eta_{\alpha}^{\mathrm{eff}}q_{\alpha}^{2}n_{\alpha}k_{\mathrm{max}}}{\sum_{\alpha}m_{\alpha}v_{\mathrm{th}\alpha}^{2}n_{\alpha}} \sim \frac{\delgam_{\mathrm{e}}\eta_{\mathrm{e}}^{\mathrm{eff}}}{n_{\mathrm{e}}\lambda_{\mathrm{De}}^{3}}k_{\mathrm{max}}\lambda_{\mathrm{De}} \ll 1,
\end{equation}
ignoring at the last step any potential disparities between contributions from different species. Taking cue from our discussion in section~\ref{Section:6.1}, we conclude that this ratio will be small in all conceivable cases of interest (at least as long as this theory is valid). 

Since the distributions evolved by the quasilinear collision integrals are assumed to be linearly stable (see section~\ref{Section:2.3}), their kinetic energy does not change [see (\ref{eqn:S1:E30})] and, therefore, neither can the fluctuation energy~$E_{\varphi}$ change. It is then useful to think of~$E_{\varphi}$ as a feature of the system that tells us about the size of the correlation volume $\delgam_{\alpha}$ and of the applicability of the collisionless relaxation, via the estimate~(\ref{eqn:starstar estimate}).

This calculation does not determine $\delgam_{\alpha}$, however, together with the discussion of the collision timescales, it sets feasible limits on what the allowed values of $\delgam_{\alpha}$ can be. Crucially, we see that a larger value of $\delgam_{\alpha}$ corresponds to a larger energy in the fluctuating electric field. This further suggests that the validity of the Balescu-Lenard collision integral should be called into question when the fluctuations in the electric field are anomalously large compared to (\ref{eqn:Electric field energy 3}). The Balescu-Lenard collision integral is designed to describe slow relaxation mediated by discrete particle noise. The hyperkinetic collision integral, on the other hand, describes a faster relaxation mediated by correlated volumes of phase space. 

\subsection{Caveats on the existence of a Lynden-Bell plasma}
\label{Section:5.4}
We are now equipped with a better understanding of how the microgranulation ansatz has affected not only the collision integral, but also the fluctuations of the electric field. It is therefore now prudent, in spite of the comfort provided by the estimates~(\ref{eqn:upgraded constraint}) and~(\ref{eqn:starstar estimate}), to question what has been lost and, therefore, what the validity of this closure is. 

The defining, but also the most fragile, feature of the collisionless collision integrals derived above is phase-volume conservation by the evolution of the exact phase-space density $f_{\alpha}$. We argued in section \ref{Section:6.1} that, to be in any way important, our relaxation had to be faster than phase-volume conservation could be broken. While estimates (\ref{eqn:resulting constraint}) and (\ref{eqn:S6:E9}) serve as guesses for how fast phase-volume conservation is broken, there is numerical evidence \citep{Zhdankin2021} that in turbulent systems Casimir invariants can be broken on timescales that are independent of the true collision rate -- analogously perhaps to the way in which `dissipative anomalies' arise generically in turbulent environments (cf. \citealt{Eyink2018}, \citealt{Schekochihin2009}). If this is the case, then one would not be justified in writing a collisionless Vlasov equation (\ref{eqn:S1:E1}) for $f_{\alpha}$ or (\ref{eqn:S4:E2}) for $P_{\alpha}$, rendering the subsequent calculations formally invalid. In such systems, it may nevertheless be a meaningful question to ask whether the waterbag density $\rho(\eta)$ (equivalently the Casimir invariants) will evolve towards anything universal. If so, then despite its non-conservation, waterbag content could still be a relevant constraint.

Should phase-volume conservation survive, we must further ask if it is reasonable to expect that the correlation volume $\delgam_{\alpha}$ should be a constant independent of time. Naturally one might assume that, as fluctuations travel through phase space, they gradually break up and form smaller and smaller structures. Another factor that indicates that~$\delgam_{\alpha}$ should be dependent on time is~(\ref{eqn:Electric field energy 2}) itself. Since the kinetic energy of the mean distribution function is also constant, then to conserve the total energy, the electric-field energy must also be constant. By~(\ref{eqn:Electric field energy 2}), this implies that~$\delgam_{\alpha}$ must vary in concert with the integral of~$\crl{g_{\alpha}^{2}}$ over velocities. Alternatively one could take the view that the mean energy of the distribution function is only approximately constant, justified by the smallness of the electric-field energy compared to the kinetic energy as computed in~(\ref{eqn:Electric field comparison}).

Only one of our results requires that $\delgam_{\alpha}$ not be a function of time: the H-theorem. If $\delgam_{\alpha}$ evolves with time, then the H-theorem is broken because our entropy (\ref{eqn:S3:E34}) has a prefactor proportional to $\delgam_{\alpha}$. This could be remedied if~$\delgam_{\alpha}$ were independent of the species, because it would be an overall multiplicative prefactor that the entropy need not include. This would result in a working, but weaker, H-theorem under which any state could be a steady state if~$\delgam_{\alpha}$ decayed sufficiently fast to halt its evolution. Such relaxation has been called `incomplete violent relaxation', indicating that the system tried to reach its Lynden-Bell equilibrium, but stalled before the relaxation could be completed \citep{Chavanis2006}. 

Thus, for the phase-volume conservation and the correlation volume to be meaningful features of any complete theory, they must earn their place in it. Since the microgranulation ansatz grants them a privileged position without question, it cannot be trusted without question. Nevertheless, it provides an insight into what interesting effects could be contained in a theory where these features are valid and meaningful. We give one such interesting example in the next section.

\section{Strange relaxation in multispecies Lynden-Bell plasma}
\label{Section:6}
Having derived the hyperkinetic collision integral, its conservation laws, steady states and H-theorem, it would seem that relaxation to equilibrium has successfully been turned into an app. Like a figure of Greek myth, the collision integral is then destined to be made redundant by its own offspring: the H-theorem. This is because the H-theorem prescribes for each initial condition a steady-state distribution function, which makes evolving the hyperkinetic collision integral in time seem moot. In this section, however, we will show that there are certain fairly general initial conditions for which the process of relaxation has interesting and non-trivial properties. 

For this to be the case, it is clear that there must be multiple timescales within the problem. Based on our estimate (\ref{eqn:S6:E8}) of effective collision frequency relative to the true collision frequency, it is clear that one way to make this possible is to consider two species with a large mass ratio: electrons and ions. Borrowing intuition from the conventional collisional theory, this will allow for a separation of timescales whereby particles of the lighter, faster, species will collide most frequently while the heavier, slower, species will collide, and therefore evolve, at a lower rate. This intuitive picture is, however, complicated by the fact that, from our knowledge of the steady states (\ref{eqn:S4:E20}), we expect the thermal velocities of the particles to be such that
\begin{equation}
\label{eqn:S7:E1}
\beta \delgam_{\alpha}\eta_{\alpha}^{\mathrm{eff}} \frac{1}{2}m_{\alpha}v_{\mathrm{th}\alpha}^{2}\sim 1 \implies v_{\mathrm{th}\alpha}\sim \sqrt{\frac{2}{\beta \delgam_{\alpha}^{\mathrm{eff}}\eta_{\alpha}^{\mathrm{eff}}m_{\alpha}}}.
\end{equation}
This is to say that, relative to other species, particles of a given species behave as though they had an effective mass $\delgam_{\alpha}\eta_{\alpha}^{\mathrm{eff}}$ times greater than their true mass. To avoid this complexity, we will restrict ourselves to the case where the typical number of particles within a correlation volume is comparable for electrons and ions, viz.,
\begin{equation}
\label{eqn:S7:E2}
\delgam_{\mathrm{e}}\eta_{\mathrm{e}}^{\mathrm{eff}} \sim \delgam_{\mathrm{i}}\eta_{\mathrm{i}}^{\mathrm{eff}}.
\end{equation}
This amounts to assuming that collisionless effects do not override the scale separation imposed by the mass ratio, which is what creates the interesting effects in the conventional collisional theory. From (\ref{eqn:S4:E17}) and the orderings (\ref{eqn:S7:E1}) and (\ref{eqn:S7:E2}), it is apparent that the same-species collision operators only contain a single timescale. This is simply a statement that the only purpose of the same-species collision operator is to increase the entropy of that species independent of the others, and that it can only do that on a single timescale. In contrast, owing to the mass ratio, the inter-species collisions can have multiple timescales. We know from standard collisional theory that these give rise to the rates at which momentum and temperature are equalised between species. To extract similar, but distinct, features from our new collision integral, we would therefore have to expand it in small mass ratio. Before we do this, however, it is possible to anticipate from simple observations what the interesting new physics will be.
\subsection{Preview of strange relaxation}
\label{Section:Prophesy}
\begin{figure}
\centering
\includegraphics[width=0.75\textwidth]{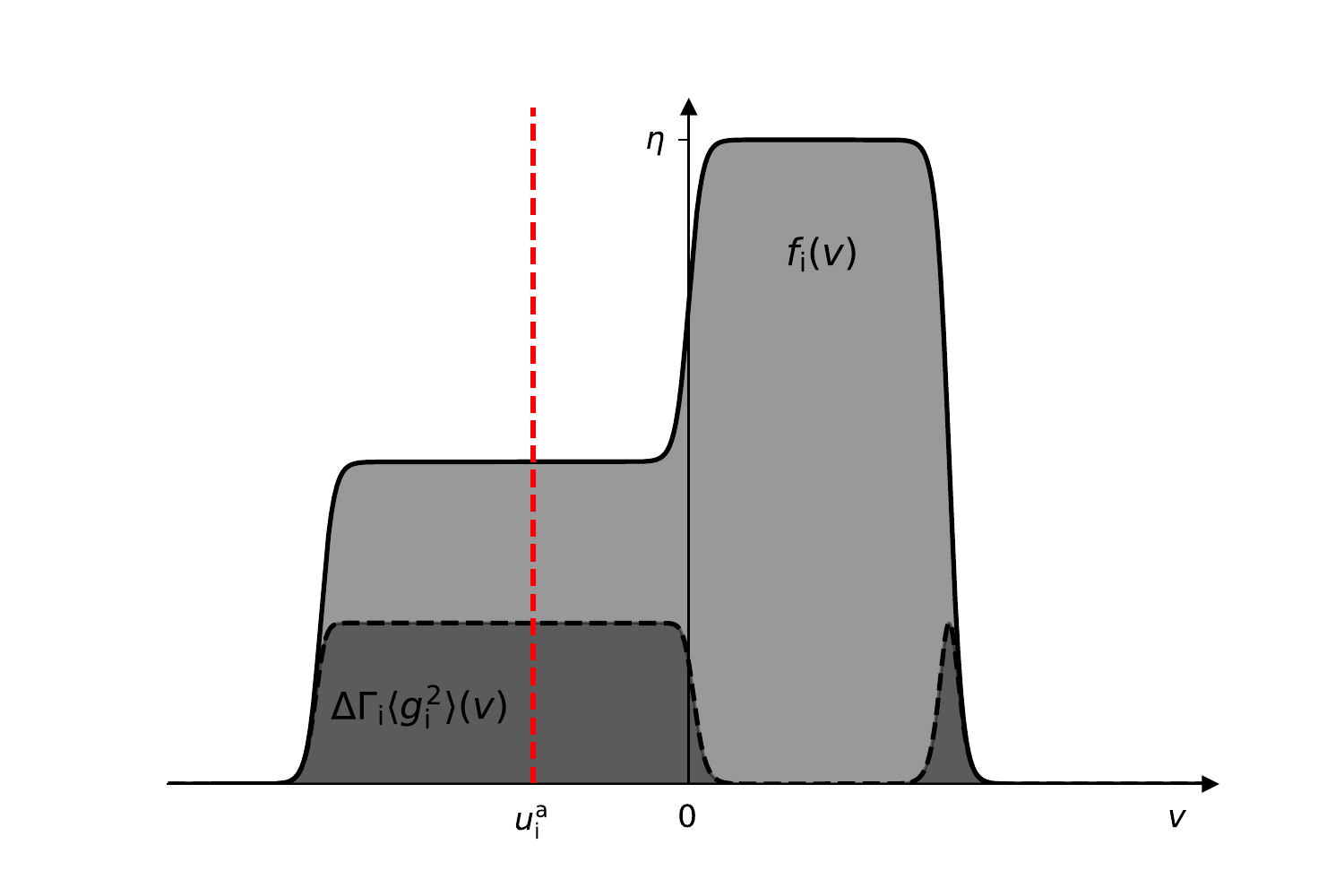}
\caption{A cartoon of a possible anisotropic ion distribution function that possesses an anomalous velocity $\v{u}_{\ii}^{\mathrm{a}}$. The solid line represents the mean ion distribution function. If the exact ion phase-space density is viewed as a single waterbag [as in the Kadomtsev-Pogutse collision integral (\ref{eqn:S3:E6})], then the `effective distribution' of ions with which the electrons will interact is shown by the dashed black line. The mean ion velocity is zero, but the anomalous velocity is not, as shown by the dashed red line. Note that the height of the effective distribution would be rescaled if $\eta$ and $\delgam_{\ii}$ took different values, but this would only affect the rate of relaxation to the anomalous velocity, not the anomalous velocity itself.}
\label{Figure 4}
\end{figure}

To study the interaction between species we will write our collision operator~(\ref{eqn:S4:E17}) as
\begin{equation}
\label{eqn:S7:E6}
\pdev{P_{\alpha}}{t} = \sum_{\alpha'}C_{\alpha\alpha'}[P_{\alpha},P_{\alpha'}],
\end{equation}
in terms of the interspecies collision operator $C_{\alpha\alpha'}$, which can be easily read from elements of the species sum in (\ref{eqn:S4:E17}). For compactness of notation, we shall henceforth drop 0's from the subscripts of the mean distribution function. For electrons interacting with ions, the dominant effect is the diffusion-like first term of (\ref{eqn:S4:E17}). Comparing this to the true-collision operator~(\ref{eqn:S6:E1}), we see that (aside from acting on~$P_{\ee}$ instead of~$f_{\ee}$) the new, `collisionless' feature in this diffusion term is that~$f_{\ii}$ is replaced by~$\delgam_{\ii}\crl{g_{\ii}^{2}}$ with~$\crl{g_{\ii}^{2}}$ defined by (\ref{eqn:S5:E2}). Therefore, any effects on the electrons relating to $f_{\ii}$ in the standard collisional theory will now be replaced by the analogous effects relating to $\delgam_{\ii}\crl{g_{\ii}^{2}}$ (see figure \ref{Figure 4} for an illustration of this in a single-waterbag distribution). In particular, instead of being isotropised by the ions of density~$n_{\ii}$ and dragged towards the ion velocity $\v{u}_{\ii}$, the electrons will see an `anomalous density' $n_{\ii}^{\mathrm{a}}$ and find themselves dragged towards an `anomalous velocity' $\v{u}_{\ii}^{\mathrm{a}}$, given by 
\begin{equation}
\label{eqn:S7:Anomalous}
n_{\ii}^{\mathrm{a}} = \delgam_{\ii}\int\dd{\v{v}}'\crl{g_{\ii}^{2}}(\v{v}'), \quad \quad \v{u}_{\ii}^{\mathrm{a}} = \frac{\int\dd{\v{v}'}\v{v}'\crl{g_{\ii}^{2}}(\v{v}')}{\int\dd{\v{v}'} \crl{g_{\ii}^{2}}(\v{v}')}.
\end{equation}
Since ion relaxation is slow compared to the electron one, the ion distribution need not be isotropic, so, in general,~$\v{u}_{\ii}^{\mathrm{a}} \neq 0$. Furthermore, there is no need for~$\v{u}_{\ii}^{\mathrm{a}}$ to point in the same direction as the ions' mean velocity~$\v{u}_{\ii}$, and indeed it is even possible that~$\v{u}_{\ii} = 0$ while~$\v{u}_{\ii}^{\mathrm{a}} \neq 0 $. This means that collisionless relaxation can lead to spontaneous generation of current -- and, therefore, of magnetic field -- from ion anisotropies (e.g., from an ion heat flux). 

The physical significance of this becomes especially clear for a single-waterbag example shown in figure \ref{Figure 4}. The phase-space exclusion effect makes the ions behave as though they were fermions. Therefore, the probability for an electron to have a `collision' with an ion whose velocity is $\v{v}$ is not just the probability $f_{\ii}(\v{v})/\eta$ that an ion can be found at that velocity but the probability that an ion can be found at that velocity $\v{v}$ and that the velocity $\v{v} + \Delta\v{v}$ into which that ion will be scattered by the interaction is not already occupied. Since an ion is deflected by a very small amount in a collision with an electron~($\Delta\v{v} \ll \v{v}$), this probability is approximately proportional to $f_{\ii}(\v{v})\left[\eta - f_{\ii}(\v{v})\right]$, which is precisely $\crl{g_{\ii}^{2}}(\v{v})$ from which the anomalous ion velocity and density are defined in (\ref{eqn:S7:Anomalous}). This means that densely packed portions of the ion phase space become effectively invisible to the electrons. Thus, the mean velocity towards which the electrons are dragged is not the true ion mean velocity but the mean velocity of those ions that move in the less densely occupied regions of the phase space. 

After an initial period of this `strange relaxation' the electrons will converge to their own Lynden-Bell equilibrium moving at the anomalous ion velocity $\v{u}_{\ii}^{\mathrm{a}}$. The ions, however, will not thus far have relaxed significantly except to ensure the conservation of total momentum (and thus alter the mean velocity by a mass-ratio-small amount~{${\sim -\v{u}_{\ii}^{\mathrm{a}}m_{\ee}n_{\ee}/m_{\ii}n_{\ii}}$}). Once the ion-ion relaxation timescale is reached, the ions will begin to erase any anisotropy in their distribution function as they proceed towards their own Lynden-Bell equilibrium. As their anisotropy vanishes, the anomalous mean ion velocity will tend to the true mean ion velocity, $\v{u}_{\ii}^{\mathrm{a}}\to \v{u}_{\ii}$, and the electrons' mean velocity will doggedly follow it, relaxing towards two Lynden-Bell equilibria of equal velocities but distinct temperatures. Finally, on the longest (ion-electron) interaction timescale, the distributions will equalise their temperatures, reaching the overall maximum-entropy state and completing the relaxation.
   
This is the physics of the strange relaxation process. In the remainder of this section, we demonstrate formally that this is indeed what happens by carrying out the mass-ratio expansion of the electron-ion (section \ref{Section:First order}) and the ion-electron (section \ref{Section:ion-electron}) collision operators. This calculation follows the standard path well trodden for true collisions, but with a few important adjustments. 
\subsection{Landau form of the hyperkinetic collision operator}
\label{Section:Landau Collision operator}
We will first make a further simplification by reducing our hyperkinetic collision operator (\ref{eqn:S4:E17}) to the so-called `Landau form'. Doing so amounts to finding an approximate expression for the $\v{k}$ sum
\begin{equation}
\label{eqn:S7:E3}
\frac{1}{V}\sum_{\v{k}}\frac{\v{k}\v{k}}{k^{4}}\frac{\delta\left(\v{k}\cdot(\v{v}-\v{v}') \right)}{|\epsilon_{\v{k},\v{k}\cdot\v{v}}|^{2}},
\end{equation}
which is complicated by the presence of the dielectric function
\begin{equation}
\label{eqn:S7:E4}
\epsilon_{\v{k},\v{k}\cdot\v{v}} = 1 + \sum_{\alpha'}\frac{4\pi q_{\alpha'}^{2}}{m_{\alpha'}k^{2}}\iint \dd{\v{v}'}\dd{\eta'}\frac{\eta'}{\v{k}\cdot\left(\v{v}-\v{v}' \right)}\v{k}\cdot\pdev{P_{\alpha'}}{\v{v}}.
\end{equation}
Here the $\v{v}'$ integral is taken along the Landau contour. To simplify (\ref{eqn:S7:E3}), the important feature to note is that, for length scales shorter than the Debye length and in the absence of instabilities, the second term in (\ref{eqn:S7:E4}) will be small and $\epsilon_{\v{k},\v{k}\cdot\v{v}}$ can be approximated by unity. For scales significantly longer than the Debye length,~$\epsilon_{\v{k},\v{k}\cdot\v{v}} \sim (\lambda_{\mathrm{De}}k)^{-2}$, which will make their contribution to the sum in (\ref{eqn:S7:E3}) small in $(\lambda_{\mathrm{De}}k)^{4}$. Therefore, following \cite{Landau1936}, we truncate the $\v{k}$ sum at $(k\lambda_{\mathrm{De}})\gtrsim 1$, and approximate the dielectric function by unity. Denoting $\v{w} = \v{v}-\v{v}'$ and integrating in cylindrical coordinates such that $\v{k} = k_{\parallel}\hat{\v{w}} + \v{k}_{\perp}$, we get
\begin{equation}
\label{eqn:S7:E5}
\frac{1}{V}\sum_{\v{k}}\frac{\v{k}\v{k}}{k^{4}}\frac{\delta(\v{k}\cdot\v{w})}{|\epsilon_{\v{k},\v{k}\cdot\v{v}}|^{2}} \approx \iiint\frac{\dd{\v{k}}}{\left(2\pi \right)^{3}}\frac{\v{k}_{\perp}\v{k}_{\perp}}{k_{\perp}^{4}}\delta(k_{\parallel}w) = \frac{1}{8\pi^{2}}\frac{1}{w}\left(\iden -\frac{\v{w}\v{w}}{w^{2}} \right)\ln\left(\frac{k_{\mathrm{max}}}{k_{\mathrm{min}}}\right).
\end{equation}
Here we have truncated the $k_{\perp}$ integral, not only at $k_{\mathrm{min}}^{-1} \sim \lambda_{\mathrm{De}}$ as discussed above, but also at scales shorter than $k_{\mathrm{max}}^{-1}$. In the standard collisional theory, this corresponds to cutting off the integral at the distance of closest approach of two particles and gives rise to the Coulomb logarithm. In our treatment, such a high-$k$ cutoff should instead be associated with the physical length scale of the correlations, which may be thought of as the distance of closest approach for two waterbags. Henceforth, we will denote this generalised Coulomb-logarithm-like quantity by $\Lambda_{\alpha\alpha'}$.

With the approximation (\ref{eqn:S7:E5}), the interspecies collision operator in (\ref{eqn:S4:E17}) becomes
\begin{multline}
\label{eqn:S7:E7}
C_{\alpha\alpha'}[P_{\alpha},P_{\alpha'}] = \frac{\gamma_{\alpha\alpha'}}{m_{\alpha}}\pdev{}{\v{v}}\cdot\int\frac{\dd{\v{v}}'}{w}\left( \iden -\frac{\v{w}\v{w}}{w^{2}}\right)\cdot\int\dd{\eta'}\eta' \\ \Bigg\lbrace \frac{\delgam_{\alpha'}}{m_{\alpha}}\left[\eta' - f_{\alpha'}(\v{v}') \right]P_{\alpha'}(\v{v}',\eta')\left.\pdev{P_{\alpha}}{\v{v}}\right|_{\eta} - \frac{\delgam_{\alpha}}{m_{\alpha'}}\left[\eta - f_{\alpha}(\v{v}) \right]P_{\alpha}(\v{v},\eta)\left.\pdev{P_{\alpha'}}{\v{v}'}\right|_{\eta'}\Bigg\rbrace,
\end{multline}
where $\gamma_{\alpha\alpha'} = 2\pi q_{\alpha}^{2}q_{\alpha'}^{2}\Lambda_{\alpha\alpha'}$.
\subsection{Electron-ion relaxation}
\subsubsection{Zeroth order: istropisation of electrons}
\label{Section:Zeroth Order}
To lowest order in $\sqrt{m_{\ee}/m_{\ii}}$, the second term in brackets in (\ref{eqn:S7:E7}) vanishes entirely and the velocity difference between the electrons and ions, $\v{w}$, becomes approximately equal to the electron velocity $\v{v}$. This leaves the collision operator in the immensely simple form
\begin{equation}
C_{\eci} = \frac{\gamma_{\eci}n_{\ii}^{\mathrm{a}}}{m_{\ee}^{2}}\pdev{}{\v{v}}\cdot\left[\frac{1}{v}\left(\iden - \frac{\v{v}\v{v}}{v^{2}}\right)\cdot \left. \pdev{P_{\ee}}{\v{v}}\right|_{\eta} \right],
\end{equation}
where $n_{\ii}^{\mathrm{a}}$ is the anomalous ion density given by the first expression in (\ref{eqn:S7:Anomalous}). This operator is a pitch-angle-scattering (Lorentz) operator (see, e.g., \citealt{Helander}), which causes relaxation on timescales comparable to that of electron-electron interactions. Its effect is to isotropise $P_{\ee}$. Physically this is a consequence of the ions' high mass. Like ping-pong balls bouncing off bowling balls, the electrons bounce off the ions without exchanging any energy and the only effect is to isotropise their distribution. However, to this lowest order, we have only retained terms ordered with the electron thermal velocity, losing any speeds ordered with the ion velocity. To retain velocities of that size, viz., the electrons' mean velocity, we must go to next order.
\subsubsection{First order: anomalous drag}
\label{Section:First order}
To next order, determining the collision integral becomes moderately less trivial. First, we will expand the electron hyperkinetic distribution function in mass ratio and assume that its lowest-order part, denoted by~$P_{\ee}^{\ISO}(v,\eta)$, is isotropic. We further assume that any deviations from isotropy are due to velocities ordered with the ion thermal velocity, so the anisotropic correction, denoted~$P_{\ee}^{\ANI}(\v{v},\eta)$, will be smaller than~$P_{\ee}^{\ISO}(v,\eta)$ by a factor of~$\sqrt{m_{\ee}/m_{\ii}}$. As well as this, since the $\v{v}'$ integral in (\ref{eqn:S7:E7}) is over ion velocities,~$\v{v}'$ will be small in~$\sqrt{m_{\ee}/m_{\ii}}$ compared to~$\v{v}$. Therefore, we can now expand the $\v{w}$-dependent tensor in (\ref{eqn:S7:E7}) to first order in $\sqrt{m_{\ee}/m_{\ii}}$, giving
\begin{equation}
\label{eqn:S7:tensor expansion}
\begin{split}
\frac{1}{w}\left(\iden - \frac{\v{w}\v{w}}{w^{2}} \right)  & \approx \frac{1}{v}\left( \iden - \frac{\v{v}\v{v}}{v^{2}} \right) - \v{v}'\cdot\pdev{}{\v{v}}\left[\frac{1}{v}\left(\iden - \frac{\v{v}\v{v}}{v^{2}} \right) \right] \\ & = \frac{1}{v}\left(1 + \frac{3\v{v}\cdot\v{v}'}{v^{2}} \right)\left(\iden - \frac{\v{v}\v{v}}{v^{2}} \right) + \frac{\v{v}'\v{v} + \v{v}\v{v}' - 2(\v{v}\cdot\v{v}')\iden}{v^{3}}.
\end{split}
\end{equation}
Since the anisotropic part of the hyperkinetic distribution function is already small, we only need the lowest-order contribution from (\ref{eqn:S7:tensor expansion}) to act on $P_{\ee}^{\ANI}(\v{v},\eta)$. For the isotropic part $P_{\ee}^{\ISO}(\v{v})$, the lowest-order contribution vanishes (as it must, since any isotropic distribution is a solution of the lowest-order problem) and we must keep the next-order terms. These terms are
\begin{equation}
\begin{split}
\frac{1}{w}\left(\iden - \frac{\v{w}\v{w}}{w^{2}} \right)\cdot \pdev{P_{\ee}^{\ISO}}{\v{v}} &\approx \frac{\v{v}'\v{v} + \v{v}\v{v}' - 2(\v{v}\cdot\v{v}')\iden}{v^{3}}\cdot \frac{\v{v}}{v}\pdev{P_{\ee}^{\ISO}}{v} \\ & = \frac{1}{v}\left(\iden -\frac{\v{v}\v{v}}{v^{2}} \right)\cdot\frac{\v{v}'}{v}\pdev{P_{\ee}^{\ISO}}{v} = \frac{1}{v}\left(\iden - \frac{\v{v}\v{v}}{v^{2}} \right)\cdot \pdev{}{\v{v}}\left(\frac{\v{v}\cdot\v{v}'}{v}\pdev{P_{\ee}^{\ISO}}{v} \right).
\end{split}
\end{equation}
In the final equality, we made use of the isotropy of $P_{\ee}^{\ISO}$ in order to insert an additional derivative into the product. The effect of this is to cast this term in the form of a pitch-angle-scattering operator. The resulting collision operator is
\begin{equation}
\label{eqn:S7:E11}
C_{\eci}[P_{\ee},P_{\ii}] = \frac{\gamma_{\eci}}{m_{\ee}^{2}}n_{\ii}^{\mathrm{a}}\pdev{}{\v{v}}\cdot \left\lbrace\frac{1}{v}\left(\iden -\frac{\v{v}\v{v}}{v^{2}} \right)\cdot \pdev{}{\v{v}}\left[P_{\ee}^{\ANI}(\v{v},\eta) + \frac{\v{v}\cdot\v{u}_{\ii}^{\mathrm{a}}}{v}\pdev{P_{\ee}^{\ISO}}{v} \right]\right\rbrace,
\end{equation}
where $n_{\ii}^{\mathrm{a}}$ and $\v{u}_{\ii}^{\mathrm{a}}$ are given by (\ref{eqn:S7:Anomalous}).

Despite not knowing the exact evolution due to (\ref{eqn:S7:E11}), we may use the fact that it is a pitch-angle-scattering operator to read off the steady state. This will occur when the expression in the square brackets is isotropic. Without loss of generality, we may define~$P_{\ee}^{\ANI}$ to have zero spherical average (isotropic part), in which case the only way for the term in square brackets in (\ref{eqn:S7:E11}) to be isotropic is for it to be identically zero. Therefore the fixed point of the collision operator (\ref{eqn:S7:E11}) is
\begin{equation}
\begin{split}
P_{\ee}(\v{v},\eta) & = P_{\ee}^{\ISO}(v,\eta) + P_{\ee}^{\ANI}(\v{v},\eta) = P_{\ee}^{\ISO}(v,\eta) - \frac{\v{v}\cdot\v{u}_{\ii}^{\mathrm{a}}}{v}\pdev{P_{\ee}^{\ISO}}{v} \approx P_{\ee}^{\ISO}(|\v{v}-\v{u}_{\ii}^{\mathrm{a}}|,\eta),
\end{split}
\end{equation}
the last equality holding with $O(m_{\ee}/m_{\ii})$ precision. 

Thus, as prophesied in section \ref{Section:Prophesy}, the electrons become isotropic around the anomalous ion velocity $\v{u}_{\ii}^{\mathrm{a}}$ rather than the true mean ion velocity. Dynamically, this manifests itself as a drag on the electrons with the rate of change of the electron momentum~$m_{\ee}n_{\ee}\v{u}_{\ee}$ being
\begin{equation}
\label{eqn:S7:E12}
\begin{split}
m_{\ee}n_{\ee}\dev{\v{u}_{\ee}}{t}& = \dev{}{t}\int \dd{\v{v}}m_{\ee}\v{v}\int \dd{\eta}\eta P_{\ee}(\v{v},\eta) = \int\dd{\v{v}}m_{\ee}\v{v}\int \dd{\eta}\eta C_{\eci}[P_{\ee},P_{\ii}] \\ & =  -\frac{\gamma_{\eci}n_{\ii}^{\mathrm{a}}}{m_{\ee}}\int\frac{\dd{\v{v}}}{v}\left(\iden -\frac{\v{v}\v{v}}{v^{2}} \right)\cdot\pdev{}{\v{v}}\left(f_{\ee}^{\ANI} + \frac{\v{v}\cdot\v{u}_{\ii}^{\mathrm{a}}}{v}\pdev{f_{\ee}^{\ISO}}{v} \right) \\ & = -\frac{\gamma_{\eci}n_{\ii}^{\mathrm{a}}}{m_{\ee}}\left[\int \frac{\dd{\v{v}}}{v} \left(\iden -\frac{\v{v}\v{v}}{v^{2}} \right)\pdev{f_{\ee}^{\ISO}}{v} \right]\v{u}_{\ii}^{\mathrm{a}} - \frac{2\gamma_{\eci}n_{\ii}^{\mathrm{a}}}{m_{\ee}}\int \dd{\v{v}}\frac{\v{v}}{v^{3}}f_{\ee}^{\mathrm{\ANI}} \\ & = \frac{8\pi \gamma_{\eci}n_{\ii}^{\mathrm{a}}f_{\ee}^{\ISO}(0)}{3m_{\ee}}\v{u}_{\ii}^{\mathrm{a}} - \frac{2\gamma_{\eci}n_{\ii}^{\mathrm{a}}}{m_{\ee}}\int\dd{\v{v}}\frac{\v{v}}{v}f_{\ee}^{\ANI},
\end{split}
\end{equation}
where~$f_{\ee}^{\ISO}(v)$ and~$f_{\ee}^{\ANI}(\v{v})$ are, respectively, the isotropic and anisotropic parts of the electron phase-space density. Going from the third to the fourth line we used the isotropy of~$f_{\ee}^{\ISO}$ to compute the angle integral explicitly. The same cannot be done for the integral over the anisotropic part of the phase-space density. However, if we assume that, to lowest order, the electron phase-space density is isotropic around the mean electron velocity~$\v{u}_{\ee}$, then
\begin{equation}
\label{eqn:S7:E13B}
f_{\ee}(\v{v}) = f_{\ee}^{\ISO}(|\v{v}-\v{u}_{\ee}|) \approx f_{\ee}^{\ISO} - \frac{\v{v}\cdot\v{u}_{\ee}}{v}\pdev{f_{\ee}^{\ISO}}{v} \implies f_{\ee}^{\ANI}(\v{v}) = -\frac{\v{v}\cdot\v{u}_{\ee}}{v}\pdev{f_{\ee}^{\ISO}}{v},
\end{equation}
which allows the final integral to be computed, giving the anomalous drag force
\begin{equation}
\v{F}_{\eci} = m_{\ee}n_{\ee}\dev{\v{u}_{\ee}}{t} = -\frac{8\pi \gamma_{\eci}n_{\ii}^{\mathrm{a}}f_{\ee}^{\ISO}(0)}{3m_{\ee}}\left(\v{u}_{\ee} - \v{u}_{\ii}^{\mathrm{a}} \right).
\end{equation}
This is the same as the standard expression for the collisional drag \citep{spitzer1967physics}, but replacing the ion density and mean velocity with the corresponding anomalous variants~(\ref{eqn:S7:Anomalous}).
 
We note that neither the lowest- nor the first-order approximations of the electron-ion collision operator yet describe the transfer of energy between the two species. To achieve this, and obtain the expected relaxation of temperature, one must go to higher order, which is easiest to do via the ion-electron collision operator.
\subsection{Ion-electron relaxation: temperature equilibration}
\label{Section:ion-electron}
The ion-electron collision operator (\ref{eqn:S7:E7}) is 
\begin{multline}
\label{eqn:S7:E18}
C_{\ice} = \frac{\gamma_{\eci}}{m_{\ii}}\pdev{}{\v{v}}\int \frac{\dd{\v{v}'}}{w}\left(\iden -\frac{\v{w}\v{w}}{w^{2}} \right)\cdot \int\dd{\eta'}\eta'\\ \Bigg \lbrace \underbrace{\frac{\delgam_{\ee}}{m_{\ii}}\left[\eta' - f_{\ee}(\v{v}') \right]P_{\ee}(\v{v}',\eta')\pdev{P_{\ii}}{\v{v}}}_{\sim \frac{\delgam_{\ee}\eta_{\ee}^{\mathrm{eff}}}{m_{\ii}v_{\mathrm{thi}}}P_{\ee}P_{\ii}}  - \underbrace{\frac{\delgam_{\ii}}{m_{\ee}}\left[\eta - f_{\ii}(\v{v}) \right]P_{\ii}(\v{v},\eta)\pdev{P_{\ee}}{\v{v}'}}_{\sim \frac{\delgam_{\ii}\eta_{\ii}^{\mathrm{eff}}}{m_{\ee}v_{\mathrm{the}}}P_{\ee}P_{\ii}} \Bigg\rbrace.
\end{multline}
 Naively, this gives an ion-electron relaxation rate that is comparable to the ion-ion relaxation rate, which is smaller than the electron-ion one by a factor of $\sqrt{m_{\ee}/m_{\ii}}$. However, this neglects the fact that the electron distribution will isotropise itself on the electron-ion relaxation timescale. Thus, the lowest-order term in (\ref{eqn:S7:E18}) will vanish before it ever gets a chance to participate. This renders the ion-electron relaxation rate smaller than the ion-ion one by another factor of $\sqrt{m_{\ee}/m_{\ii}}$. As in the case of true collisions, this does not preclude the conservation of total momentum because an order-unity velocity change of the electrons only requires an order-mass-ratio adjustment to the ion velocity in order to conserve momentum, and the ion-electron collision operator is still fully capable of achieving a change this small on the electron-ion collision timescale.

Using the fact that the electron distribution will long since have become isotropic to lowest order, we can again carry out a mass-ratio expansion of the ion-electron collision operator. Since the first bracketed term in (\ref{eqn:S7:E18}) is nominally smaller, the lowest-order contribution only requires the isotropic part of the electron distribution function. This becomes (noting that now the integration is over electron velocities so,~$\v{v}'' \gg \v{v}$)
\begin{equation}
\label{eqn:S7:E16}
\int\frac{\dd{\v{v}'}}{v'}\left(\iden - \frac{\v{v}'\v{v}'}{v'^{2}} \right)\int \dd{\eta'}\eta'\left[\eta' - f_{\ee}(\v{v}') \right]P_{\ee}(\v{v}',\eta) = \frac{8\pi}{3}\iden \int_{0}^{\infty}\crl{(g_{\ee}^{\ISO})^{2}}(\v{v}')v'\dd{v'}.
\end{equation}
The second bracketed term in (\ref{eqn:S7:E18}) requires both the isotropic and anisotropic contributions, leading us to evaluate the integral
\begin{equation}
\label{eqn:S7:E17}
\begin{split}
\int\frac{\dd{\v{v}'}}{w}\left(\iden - \frac{\v{w}\v{w}}{w^{2}} \right)\cdot \pdev{f_{\ee}}{\v{v}'} & = \int \frac{\dd{\v{v}'}}{w}\left(\iden - \frac{\v{w}\v{w}}{w^{2}} \right)\cdot \left(\pdev{f_{\ee}^{\ANI}}{\v{v}'} + \frac{\v{v}'}{v'}\pdev{f_{\ii}^{\ISO}}{v'} \right) \\ & = \int \frac{\dd{\v{v}'}}{v'}\left(\iden - \frac{\v{v}'\v{v}'}{v'^{2}} \right)\cdot \pdev{f_{\ee}^{\ANI}}{\v{v}'} + \left[\int\frac{\v{v}'}{v'^{2}}\left(\iden  -\frac{\v{v}'\v{v}'}{v'^{2}} \right)\pdev{f_{\ee}^{\ISO}}{v'} \right]\cdot\v{v} \\ & = 2\int \dd{\v{v}'}\frac{\v{v}'}{v'^{2}}f_{\ee}^{\ANI}(\v{v}') - \frac{8\pi f_{\ee}^{\ISO}(0)}{3}\v{v}.
\end{split}
\end{equation}
Collecting the contributions (\ref{eqn:S7:E16}) and (\ref{eqn:S7:E17}) together, we find that the electron-ion collision operator is, to lowest order,
\begin{multline}
\label{eqn:S7:E21}
C_{\ice} = -\frac{\gamma_{\eci}\delgam_{\ii}}{m_{\ee}m_{\ii}}\pdev{}{\v{v}}\cdot\left\lbrace \left[\eta - f_{\ii}(\v{v}) \right]P_{\ii}(\v{v},\eta)\left[2\int \dd{\v{v}'}\frac{\v{v}'}{v'^{2}}f_{\ee}^{\ANI}(\v{v}') - \frac{8\pi f_{\ee}^{\ISO}(0)}{3}\v{v} \right]\right\rbrace  \\ + \frac{8\pi \gamma_{\eci}\delgam_{\ee}}{3m_{\ii}^{2}}\int\crl{(g_{\ee}^{\ISO})^{2}}(v')v'\dd{v'}\pdevn{P_{\ii}}{\v{v}}{2}.
\end{multline}
While not amazingly insightful, it is at least obvious how to verify that, together with the electron-ion collision operator~(\ref{eqn:S7:E11}), this conserves the total momentum of the system, giving the opposite momentum change to (\ref{eqn:S7:E12}), as it must. 

Of course, we need not have stopped at the simplest assumption that the electron distribution is, to lowest order, isotropic. The electron-electron interactions, which are just as frequent as the electron-ion interactions, will push the electron distribution function towards a Lynden-Bell equilibrium, which will have some mean velocity $\v{u}_{\ee}$ that lies close to the anomalous ion velocity $\v{u}_{\ii}^{\mathrm{a}}$ given by (\ref{eqn:S7:Anomalous}). Such a Lynden-Bell equilibrium for the electrons has the form
\begin{equation}
f_{\ee}(\v{v}) = \frac{\int \dd{\eta}\eta e^{-\beta_{\ee}\delgam_{\ee}\eta\left[\frac{1}{2}m_{\ee}|\v{v}- \v{u}_{\ee}|^{2} - \mu_{\ee}(\eta) \right]}}{\int \dd{\eta} e^{-\beta_{\ee}\delgam_{\ee}\eta\left[\frac{1}{2}m_{\ee}|\v{v}- \v{u}_{\ee}|^{2} - \mu_{\ee}(\eta) \right]}}.
\end{equation}
This immediately allows $f_{\ee}^{\ANI}$ in~(\ref{eqn:S7:E21}) to be computed as in~(\ref{eqn:S7:E13B}). Using the property of Lynden-Bell equilibria expressed by~(\ref{eqn:S6:E8}) and~(\ref{eqn:S6:Eta scale}), we also find
\begin{equation}
\label{eqn:S7:Property}
\crl{(g_{\ee}^{\ISO})^{2}} = -\frac{1}{\beta_{\ee}\delgam_{\ee}m_{\ee}v}\pdev{f_{\ee}^{\ISO}}{v}.
\end{equation}
The ion-electron collision operator (\ref{eqn:S7:E21}) then reduces to
\begin{equation}
C_{\ice}[P_{\ii},P_{\ee}] = \frac{8\pi \gamma_{\eci}\delgam_{\ii}f_{\ee}^{\ISO}(0)}{3m_{\ii}m_{\ee}}\pdev{}{\v{v}}\cdot\left\lbrace \frac{1}{\beta_{\ee}\delgam_{\ii}m_{\ii}}\pdev{P_{\ii}}{\v{v}} + \left(\v{v}-\v{u}_{\ee} \right)\left[\eta - f_{\ii}(\v{v}) \right]P_{\ii}(\v{v},\eta)\right\rbrace.
\end{equation}
Clearly, this will be stationary for a Lynden-Bell equilibrium if the mean electron velocity~$\v{u}_{\ee}$ is equal to the mean ion velocity (i.e., if neither distribution has a mean velocity in the frame moving with $\v{u}_{\ee}$), and the thermodynamic beta $\beta_{\ii}$ of the ions' Lynden-Bell equilibrium is equal to that of the electrons, $\beta_{\ee}$.

Because the ion-ion interactions increase the entropy of the ion hyperkinetic distribution function and are faster than the ion-electron interactions, it is certainly true that the ion distribution will be a Lynden-Bell equilibrium. Furthermore, because the anomalous ion velocity $\v{u}_{\ii}^{\mathrm{a}}$ is equal to the mean ion velocity for isotropic distributions like the Lynden-Bell equilibria, the electron-ion collision operator will have seen to it that the mean ion and electron velocities match. However, since the electron-electron, electron-ion and ion-ion collision operators to lowest order do not alter the energies of the two distributions, it is not guaranteed that the electrons and ions will have the same thermodynamic beta. Thus, as anticipated, the final piece of the relaxation puzzle, the relaxation of temperatures, occurs on the ion-electron interaction timescale. By assuming a Lynden-Bell equilibrium for ions with mean velocity equal to the electron mean velocity (i.e., both species stationary in the zero momentum frame), we can calculate the rate of change of the kinetic energy of the ion distribution function:
\begin{equation}
\begin{split}
\dev{E_{\ii}}{t} &= \dev{}{t}\int\dd{\v{v}}\frac{1}{2}m_{\ii}v^{2}\int \dd{\eta}\eta P_{\ii}(\v{v},\eta) \\ & = \frac{8\pi \gamma_{\eci}f_{\ee}^{\ISO}(0)}{3m_{\ii}m_{\ee}}\left[ \frac{3}{\beta_{\ee}}\int \dd{\v{v}}f_{\ii}(\v{v}) - m_{\ii}\delgam_{\ii}\int \dd{\v{v}}|\v{v}|^{2}\crl{g_{\ii}^{2}}(\v{v}) \right] \\ & = \frac{8\pi \gamma_{\eci}f_{\ee}^{\ISO}(0)}{3m_{\ii}m_{\ee}}\left(\frac{3n_{\ii}}{\beta_{\ee}} + \frac{1}{\beta_{\ii}}\int \dd{\v{v}}\v{v}\cdot\pdev{f_{\ii}}{\v{v}} \right) \\ & =  \frac{8\pi \gamma_{\eci}f_{\ee}^{\ISO}(0)n_{\ii}}{3m_{\ii}m_{\ee}} \left(\frac{1}{\beta_{\ee}} - \frac{1}{\beta_{\ii}} \right),
\end{split}
\end{equation}
where, in going to the second line we integrated by parts, in going to the third line we exploited the relationship (\ref{eqn:S7:Property}) between $\crl{g_{\ii}^{2}}$ and $f_{\ii}$, and finally integrated by parts again in going to the last line. This is simply another manifestation of the second law of thermodynamics (guaranteed by the H-theorem): energy will flow from the thermodynamically hotter species to the thermodynamically colder one until temperatures equalise. Note, however, that this energy flow need not produce the equilibration of kinetic energies, which are no longer directly tied to the thermodynamic temperatures, the equilibria being non-Maxwellian.

\section{Summary and discussion}
\label{Section:7}
The existence of collision integrals that have the \cite{LyndenBell67} equilibria as their steady-state solutions implies that these equilibria can be reached dynamically. Two collision integrals have arisen: the multi-waterbag collision integral~(\ref{eqn:S3:E18}), and the hyperkinetic collision integral~(\ref{eqn:S4:E17}), both generalising the single-waterbag collision integral first derived by \cite{Kadomtsev_Pogutse70}. It is a key ingredient that these collision integrals are equipped with H-theorems, confirming that the dynamical evolution towards the Lynden-Bell equilibria is made inevitable by the requirement to increase the Lynden-Bell entropy. The derivation of both collision integrals had to contend with a closure problem: the evolution of the mean phase-space density~$f_{0\alpha}$ requires knowledge of the correlation function of the exact phase-space density between separate parts of phase space:~${\crl{f_{\alpha}(\v{r},\v{v})f_{\alpha'}(\v{r}',\v{v}')}}$, which is not, in general, known. For both collision integrals, this problem is partially resolved by the microgranulation ansatz (see section~\ref{Section:2.5}), a version of which was first used by \cite{Kadomtsev_Pogutse70} to derive their collision integral. The microgranulation ansatz posits that the exact phase-space density of particles of species $\alpha$ is correlated over a small, but non-zero, volume in phase space~$\delgam_{\alpha}$, but is perfectly mixed over larger volumes. In contrast, one can derive the Balescu-Lenard integral~(\ref{eqn:S6:E1}) describing `true' particle collisions by assuming that all particles are statistically independent, i.e., that a particle is only correlated with itself. The effect of the microgranulation ansatz is to reduce the problem of calculating the worrisome two point correlator~$\crl{f_{\alpha}(\v{r},\v{v})f_{\alpha'}(\v{r}',\v{v}')}$ to one of calculating a more manageable one-point correlator~$\crl{f_{\alpha}^{2}}(\v{v})$. 

A closure is still required because the variance of a random quantity cannot, in general, be determined by its mean. \cite{Kadomtsev_Pogutse70} restricted the exact phase-space density~$f_{0\alpha}(\v{r},\v{v})$ to only two possible values, $\eta_{\alpha}$ or $0$, which immediately implied~{${\crl{f_{\alpha}^{2}} = \eta_{\alpha}f_{0\alpha}}$}, removing the closure problem (see section \ref{Section:3.1}). However, this single-waterbag model is obviously extremely non-general. To move past it, in section \ref{Section:3.2}, a scheme motivated by statistical mechanics is employed, which can be traced back to the treatment of geophysical turbulence by \cite{Chavanis2005}. By maximising the Shannon entropy~(\ref{eqn:S3:E11}) subject to a continuum of constraints~(\ref{eqn:S3:E33}) (the `waterbag content' of the distribution function, or its Casimir invariants), one finds that the correlator~$\crl{f_{\alpha}^{2}}$ can be written implicitly in terms of~$f_{0\alpha}$. This leads to the multi-waterbag collision integral~(\ref{eqn:S3:E18}), which grows the Lynden-Bell entropy~(\ref{eqn:S3:E23}) and has the Lynden-Bell equilibria~(\ref{eqn:S3:E21}) as its only fixed points. Notably, if laterally to our main purpose here, this formalism allows one to recover very efficiently the `true' collision integrals: Balescu-Lenard (section \ref{Section:BL}) and the collision integrals for fermionic and bosonic quantum plasmas (Appendix \ref{Section:Quantum relaxation}).

An alternative route to solving the closure problem for~$\crl{f_{\alpha}^{2}}$ is to dodge it entirely. In section~\ref{Section:4}, instead of trying to determine the evolution of the mean phase-space density~$f_{0\alpha}(\v{v})$, a `hyperkinetic' approach is introduced, treating the exact phase-space density~$f_{\alpha}(\v{r},\v{v})$ as a random field and asking for the probability of finding it to have a particular value $\eta$ at a particular position in phase space. This method, pioneered by \cite{Severne1980} (for discrete waterbags) and having its origins in vortex kinetics and galactic dynamics \citep{Chavanis_1996}, resolves the closure problem by calculating~$f_{0\alpha}$ and~$\crl{f_{\alpha}^{2}}$ as the first and second moments, respectively, of the `hyperkinetic distribution function'~$P_{0\alpha}(\v{v},\eta) = \crl{\delta(f_{\alpha}(\v{r},\v{v})- \eta)}$. This function is evolved by the hyperkinetic collision integral~(\ref{eqn:S4:E17}), which also grows the Lynden-Bell entropy~(\ref{eqn:S4:E17}) and has the Lynden-Bell equilibria~(\ref{eqn:S4:E20}) as its only fixed points.

It is immediately apparent that the statistical mechanical closure~(\ref{eqn:S3:E30}) used to derive the multi-waterbag collision integral~(\ref{eqn:S3:E18}) is simply a special case of the hyperkinetic distribution function. In section~\ref{Section:5}, we show that in fact their connection is deeper: we prove that, should the hyperkinetic distribution be placed precisely on the closure~(\ref{eqn:S3:E30}), the two collision integrals would then be equivalent for all future times. This could imply that entropy maximisation local in phase space is an inherent feature of the hyperkinetic collision integral~(\ref{eqn:S4:E17}). For this to be so, there would have to be a shorter effective collision timescale on which the hyperkinetic distribution function relaxed towards the closure~(\ref{eqn:S3:E30}) before continuing its relaxation towards a Lynden-Bell equilibrium on a longer timescale.

While conceptually fascinating, to be put on solid ground, the `collisionless collision integals' must be verified. While there is substantial evidence of the validity (within certain regimes) of the Balescu-Lenard integral~(\ref{eqn:S6:E1}) describing `true' Coulomb collisions, it remains to be seen whether the hyperkinetic collision integral, and, more generally and especially, the microgranulation ansatz, are valid. We therefore outline a number potential observable quantities that would be indicative of the collisionless relaxation. Most obviously, the verification of Lynden-Bell equilibria themselves would vindicate the assumption of phase-volume conservation despite the fairly stringent formal limitation on it imposed by~(\ref{eqn:upgraded constraint}) (the requirement that collisionless relaxation occur before the exact phase-space density is filamented down to collisional scales), and potentially an even more stringent limitation in systems where a dissipative anomaly is present and the phase-volume conservation is broken on a time scale independent of `true' collisionality (see discussion and references in section~\ref{Section:5.4}). As discussed in section~\ref{Section:3.3.1}, the Lynden-Bell equilibria can be extremely varied, but there is a direct correspondence between the waterbag content of the initial condition and the ultimate equilibrium, which can clearly be tested. The relaxation process itself would bear telltale signs, should it obey the microgranulation ansatz. In section~\ref{Section:6}, we showed that a plasma obeying the microgranulation ansatz has an effective collision rate much higher than the true-collision rate associated with the Balescu-Lenard collision integral, and that the typical energy stored in the electric fluctuations in such a plasma is much greater than the fluctuation level arising from Coulomb collisions. This creates a picture of collisionless relaxation mediated not by single-particle collisions but by the effective collisions of larger correlated volumes of phase space, which seed larger perturbations to the electric field. 

In section~\ref{Section:6}, this collisionless relaxation is shown to have some curious consequences due to inter-species interactions. Most interestingly, electrons are dragged towards a mean velocity that is not necessarily the mean velocity of the ions, as the case of `true' collisions, but a certain anomalous velocity associated with an anisotropic ion distribution, which can be non-zero even if the mean ion velocity is zero (e.g., for an ion distribution that carries a heat flux but no net momentum). A current and, therefore, magnetic field, could be spontaneously generated by this mechanism.

Despite accommodating larger fluctuations in the electric field than the collisional theory, the collisionless relaxation described above is still assumed to take place in the quasilinear regime. It seems likely that there exist regimes where the fluctuations are larger still, and therefore a full nonlinear solution to the two-point correlation function of the phase-space density is required. What these regimes are, and whether there is a limit in which they recover the microgranulation ansatz, will be the subject of future work.

{\em Note.} Shortly before this paper was submitted, a preprint by \cite{Chavanis2021} appeared, where he independently develops a theory of collisionless relaxation along the lines that are, in certain respects, parallel to our reasoning, and reaches some of the same (or similar) conclusions. We refer the reader to his paper also for a detailed historical review of his and others' work on collisionless relaxation in a range of physical systems---primarily gravitating and fluid-dynamical.

\section*{Acknowledgements}
We would like to thank Michael Barnes, Ben Chandran, Ilya Dodin, Bill Dorland, Jean-Baptiste Fouvry, Chris Hamilton, Per Helander, Sid Parameswaran, Luis Silva, and Dmitri Uzdensky  for illuminating discussions. RJE would like to further thank Felix Parra, whose lectures on collisional plasma physics inspired the treatment of strange relaxation in section~\ref{Section:6}.  We also thank Pierre-Henri Chavanis for  bringing to our attention his early work on statistical-mechanical closures for collision integrals \citep{Chavanis2004,Chavanis2005}. RJE's and TA's work was supported by UK EPSRC studentships. TA's work was also supported by the Euratom research and training programme within the framework of the EUROfusion consortium (grant agreement No.~633053) and by the UKRI Energy Programme (grant EP/T012250/1). The views and opinions expressed herein do not necessarily reflect those of the European Commission. AB was supported by a Marshall Scholarship. The work of AAS was supported in part by a UK EPSRC grant EP/R034737/1.

\appendix
\section{Collision integrals in quantum plasmas}
\label{Section:Quantum relaxation}
In this appendix, we show how the scheme used in sections~\ref{Section:2} and~\ref{Section:3} can be applied to work out swiftly the `true' collision integrals for fermions and bosons. It is perhaps unsruprising that this is possible given the close analogy between phase-volume conservation and the exclusion principle. However, the procedure is non-rigorous because sections~\ref{Section:2} and~\ref{Section:3} assume point particles occupying definite positions in the ($\v{r},\v{v}$) phase space which is not the case for quantum gases. This issue can be handled rigorously by the derivation and expansion of the \cite{UehlingUhlenbeck} collision operator leading to the quantum Balescu-Lenard collision operator and its Landau siblings (see, e.g., \citealt{DANIELEWICZ}). We instead opt for the cheap and fast route to arrive at the (not-yet-closed) collision operator~(\ref{eqn:S1:E35}). This is written in terms of the unknown correlation volume~$\delgam_{\alpha}$ and the correlator~$\crl{g_{\alpha}^{2}}(\v{v})$, which we can resolve by falling back on the quantum nature of our particles. To consider true collisions, we assume that particles are uncorrelated wherever this is compatible with quantum constraints (the exclusion principle). Crucially, however, due to the quantisation of their positions and momenta, particles occupy -- and, therefore, must be correlated over -- a finite phase space volume,~$\delgam_{\alpha}$. We can infer this volume by considering particles in a box of volume~$V$, for which the possible momenta are
\begin{equation}
\v{p} = \frac{2\pi\hbar}{V^{1/3}}\left(i_{x}, i_{y}, i_{z} \right),
\end{equation}
where $i_{x}$, $i_{y}$, $i_{z}$ are integers. The phase-volume per particle can then be read off from the spacing of their momenta:
\begin{equation}
\label{eqn:crit density}
\delgam_{\alpha} = \left(\frac{2\pi \hbar}{m_{\alpha}}\right)^{3} \equiv \eta_{0\alpha}^{-1},
\end{equation}
where we have defined the `quantum density' $\eta_{0\alpha}$ (equivalent in spirit to the `waterbag density' considered in section~\ref{Section:3}) to be the inverse of this correlation volume. The possible values of the `exact phase-space density' $f_{\alpha}$ are then simply $\eta_{0\alpha}$ multiplied by the possible occupation numbers. 

To calculate the correlator $\crl{g_{\alpha}^{2}}(\v{v})$ (the variance of the initial condition, understood in the sense discussed in section~\ref{Section:2.5}), we will appeal to the closure scheme in section~\ref{Section:3.2}, which will account for the particle statistics. We consider all possible occupation numbers at velocity $\v{v}$, and maximise entropy subject to knowing the mean phase-space density~$f_{0\alpha}$ (equal to $\eta_{0\alpha}$ times the mean occupation number). The correlator $\crl{g_{\alpha}^{2}}(\v{v})$ is then $\eta_{0\alpha}^{2}$ times the variance of the occupation number given this maximum-entropy assignment. We will consider the case where the bosons or fermions have $\sigma_{\alpha}$ possible spins (or indeed any quantum number giving rise to degeneracy). Then the partition function becomes
\begin{equation}
\label{eqn:A1:E3}
Z_{\alpha} = \sum_{n_{1},n_{2}...n_{\sigma_{\alpha}}}e^{-\psi_{\alpha}(\v{v})\eta_{0\alpha}\sum_{j}n_{j}} = \left[\sum_{n}e^{-\psi_{\alpha}(\v{v})\eta_{0\alpha}n} \right]^{\sigma_{\alpha}},
\end{equation}
where $\psi_{\alpha}(\v{v})$ is a Lagrange-multiplier function chosen as in (\ref{eqn:S3:E13}) to guarantee the correct mean phase-space density~$f_{0\alpha}(\v{v})$. The first sum in (\ref{eqn:A1:E3}) is over all possible occupation numbers $n_{j}$ of each of the possible spins. Since the effect on the phase-space density due to occupation numbers from different spins is additive, this is then split into the product of $\sigma_{\alpha}$ identical sums. The remaining sum is taken over the allowed values of the occupation number at each spin: $n \in \lbrace 0,1 \rbrace$ for fermions and $n \in \lbrace 0,1,2,... \rbrace$ for bosons. The result is
\begin{equation}
Z_{\alpha} = \left[1 \pm  e^{-\psi_{\alpha}(\v{v})\eta_{0\alpha}}\right]^{\pm \sigma_{\alpha}},
\end{equation}
with the `$+$' sign for fermions and the `$-$' sign for bosons. As in (\ref{eqn:S3:E17}), $\crl{g_{\alpha}^{2}}$ can now be computed thus: 
\begin{equation}
\label{eqn:A1:E5}
\crl{g_{\alpha}^{2}} =\pdevn{\ln Z_{\alpha}}{\psi_{\alpha}}{2} =  \left(\eta_{0\alpha} \mp \frac{f_{0\alpha}}{\sigma_{\alpha}} \right)f_{0\alpha},
\end{equation}
with the `$-$' sign for fermions and the `$+$' sign for bosons. 

Using (\ref{eqn:A1:E5}) in (\ref{eqn:S1:E35}) gives the desired collision operator for charged particles obeying Bose-Einstein or Fermi-Dirac statistics:
\begin{multline}
\label{eqn:A1:E6}
\pdev{f_{0\alpha}}{t} = \sum_{\alpha'}\frac{16\pi^{3}q_{\alpha}^{2}q_{\alpha'}^{2}}{m_{\alpha}V}\pdev{}{\v{v}}\cdot\sum_{\v{k}}\frac{\v{k}\v{k}}{k^{4}}\cdot\int \dd{\v{v}'}\frac{\delta(\v{k}\cdot(\v{v}-\v{v}'))}{\left|\epsilon_{\v{k},\v{k}\cdot\v{v}} \right|^{2}} \\ \left\lbrace \frac{1}{m_{\alpha}}\left[1 \mp \frac{f_{0\alpha'}(\v{v}')}{\sigma_{\alpha'}\eta_{0\alpha'}} \right]f_{0\alpha'}(\v{v}')\pdev{f_{0\alpha}}{\v{v}} - \frac{1}{m_{\alpha'}}\left[1 \mp \frac{f_{0\alpha}(\v{v})}{\sigma_{\alpha}\eta_{0\alpha}} \right]f_{0\alpha}(\v{v})\pdev{f_{0\alpha'}}{\v{v}'}\right\rbrace.
\end{multline}
This clearly recovers the Balescu-Lenard collision operator (\ref{eqn:S6:E1}) when the system is nowhere near degeneracy (i.e., when $f_{\alpha} \ll  \sigma_{\alpha}\eta_{0\alpha}$). We further note that the results regarding strange relaxation in section \ref{Section:7} are equally applicable (and easier to interpret physically) for the collision operators (\ref{eqn:A1:E6}). This implies an anomalous resistivity of quantum plasmas resulting from degeneracy prohibiting certain collisions. The modification to the electron-ion friction force in quantum plasmas has been studied before (e.g., see \citealt{Daligault2016,Rightley2021}) in systems close to the relevant equilibrium (i.e., after isotropisation has removed the possibility of $\v{u}_{\ii}^{\mathrm{a}}\neq \v{u}_{\ii}$). We note, however, that, owing to the scaling of (\ref{eqn:crit density}) with species mass, ions are conventionally less degenerate than electrons, so achieving $\v{u}_{\ii}^{\mathrm{a}} \neq \v{u}_{\ii}$ requires extreme densities.

\bibliographystyle{jpp}

\bibliography{KPS}

\end{document}